\documentclass[aps,prd,reprint,superscriptaddress,nofootinbib, amsfonts]{revtex4-2}

\usepackage[all]{xy}
\usepackage{amsmath,amsthm,amssymb}
\usepackage[dvips]{graphicx}
\usepackage{comment}
\usepackage{array}
\usepackage{bm}
\allowdisplaybreaks[1]
\usepackage{xcolor}
\usepackage{hyperref}
\usepackage{tcolorbox}
\usepackage{mathrsfs}
\usepackage{float}
\usepackage[caption=false]{subfig}
\usepackage{orcidlink}
\usepackage{mathtools}
\usepackage{booktabs} 
\usepackage{multirow}
\usepackage{makecell}

\newcommand{\lm}{\ell m}

\newcommand{\WigD}[3]{\mathcal{D}_{#1,#2}^{#3}}
\newcommand{\swsh}[2]{\prescript{}{#1}Y_{#2}}

\newcommand{\tref}{t_{\rm ref}}
\newcommand{\tstart}{t_{\rm start}}
\newcommand{\tend}{t_{\rm end}}

\newcommand{\phiorb}{\phi_{\rm orb}}

\newcommand{\lamvec}{\vec{\lambda}}
\newcommand{\chivec}{\vec{\chi}}

\newcommand{\Mthr}{M_{\rm thr}}

\newcommand{\modelname}{SEOBNRv5PHM\_NNSur7dq10}

\newcommand{\chris}[1]{}

\begin{document}

\title{Fast neural network surrogate for multimodal effective-one-body gravitational waveforms from generically precessing compact binaries}

\def\Birmingham{Institute for Gravitational Wave Astronomy and School of Physics and Astronomy, University of Birmingham, Edgbaston, Birmingham, B15 2TT, United Kingdom}

\author{Christopher Whittall \orcidlink{0000-0003-2152-6004}}
\affiliation{\Birmingham}

\author{Geraint Pratten
\orcidlink{0000-0003-4984-0775}}
\affiliation{\Birmingham}

\date{\today}
\begin{abstract} 
Gravitational waveform templates are a key ingredient for the detection and characterization of gravitational waves emitted by compact binary mergers in the universe. These templates must be physically accurate and extensive, but also highly computationally efficient, two requirements that are often in tension. One solution to this problem is the development of surrogate models, which are fast, data-driven models trained to predict the output of a slower, physically realistic waveform model. In this article we build on existing work to incorporate machine learning techniques into the conventional reduced order surrogate framework, with a focus on extending coverage to waveform models that describe generically precessing quasicircular binaries. In particular, we present \modelname, a reduced order neural network surrogate of the SEOBNRv5PHM waveform model, valid up to mass ratios 1:10 for precessing quasicircular binary black hole systems with arbitrary spin magnitudes and orientations. The faithfulness of the surrogate to SEOBNRv5PHM is validated, and the surrogate is successfully applied to Bayesian parameter inference using both real and injected gravitational wave data. The surrogate is approximately 5 times faster than SEOBNRv5PHM when evaluating a single waveform on a CPU, and nearly 1000 times faster per-waveform when amortizing the cost over large waveform batches on a GPU. 
\end{abstract}

\maketitle


\section{Introduction}
Ten years after the announcement of the first direct detection of gravitational waves \cite{LIGOScientific:2016aoc}, the size of the most recent edition of the Gravitational Wave Transient Catalog stands at $218$ events \cite{LIGOScientific:2025slb}, with more expected to be announced with the full upcoming results of the final parts of the 4th LIGO-Virgo-KAGRA (LVK) observing run \cite{LVK:2026obs}. This number is expected to continue growing with future observing runs, and to increase particularly sharply in the 2030s with the proposed introduction of next-generation ground-based interferometers such as the Einstein Telescope \cite{Punturo:2010zz, Branchesi:2023mws, ET:2025xjr} and Cosmic Explorer \cite{Reitze:2019iox, Evans:2023euw}, and the planned launch of the space-based Laser Interferometer Space Antenna (LISA) mission \cite{LISA:2024hlh}.

Our ability to detect gravitational waves from compact binary mergers and precisely infer their source properties relies on the provision of accurate waveform templates which, given the parameters of a speculated binary configuration, predict the gravitational waveform that would be emitted. State-of-the-art waveform models now contain a highly faithful description of many important physical features, of which the inclusion of higher-order multipole modes and the effects of spin-induced precession of the orbital plane \cite{Apostolatos:1994mx} are the focus of this work. Although this improved physics is crucial for accurately identifying source properties and breaking the degeneracies between certain parameters \cite{Vecchio:2003tn, Chatziioannou:2014coa, Pratten:2020igi}, the downside is a significant increase in the computational cost of evaluating a waveform. This is a major challenge for parameter estimation using Bayesian inference, which typically requires $\sim 10^7$ likelihood evaluations per signal, each of which requires the generation of one waveform template. Different approaches have been developed to reduce the cost of waveform evaluation, most significantly the use of fast phenomenological models (see Refs.~\cite{Pratten:2020ceb, Thompson:2023ase, Hamilton:2025xru} for recent examples) and reduced-order surrogate models \cite{Field:2011mf, Field:2013cfa, Purrer:2014fza, Blackman:2015pia, Blackman:2017dfb, Galley:2016mvy, Varma:2019csw}. The likelihood calculation can also be sped up using techniques such as relative binning \cite{Cornish:2010kf, Zackay:2018qdy, Leslie:2021ssu}, multi-banding \cite{Vinciguerra:2017ngf, Morisaki:2021ngj}, or reduced order quadratures \cite{Canizares:2013ywa, Canizares:2014fya, Smith:2016qas, Tissino:2022thn}, which reduce the number of frequencies at which the waveform must be evaluated and stored. Methods for estimating the posterior distribution directly using machine learning, without on-the-fly likelihood evaluations, have also been demonstrated \cite{Chua:2019wwt, Dax:2021tsq, Dax:2024mcn, Hu:2024lrj, Hu:2025vlp}.

The purpose of this work is to build a reduced order neural network surrogate of SEOBNRv5PHM \cite{Ramos-Buades:2023ehm}, a state-of-the-art effective-one-body model describing the inspiral, merger and ringdown of quasicircular, precessing, black hole binaries, including higher-order multipole modes. Surrogate models are fast, data-driven models designed to accurately predict the output of a slower base model across a fixed region of intrinsic parameter space. Reduced-order modelling techniques \cite{Field:2013cfa, Field:2011mf, Barrault2004, maday2009general} are used to reduce the complexity of this task, efficiently representing long time-series data in terms of the minimal number of parameters that must be interpolated across parameter space. Neural network surrogates are a subclass of reduced order surrogate models that make use of artificial neural networks to perform this interpolation \cite{Chua:2018woh, Khan:2020fso, Thomas:2022rmc, Tissino:2022thn}. Our surrogate will span the full $7$-dimensional quasicircular intrinsic parameter space, covering mass ratios up to 1:10 with fully generic spin orientations and spin magnitudes up to the extremal limit for both binary components. This builds on previous results demonstrating the effectiveness of the neural network surrogate approach for quadrupolar aligned-spin waveforms \cite{Khan:2020fso, Fragkouli:2022lpt, GramaxoFreitas:2024bpk}, and multimodal waveforms from precessing binaries with a non-spinning secondary black hole \cite{Thomas:2022rmc}. Our work confirms that the neural network approach scales well to the $7$d interpolation problem, offering hope of future extensibility to problems with even higher dimensionality, such as the inclusion of orbital eccentricity. The neural network framework is also highly computationally efficient, and naturally amenable to batched waveform evaluation and hardware acceleration. 

We commence in Sec.~\ref{sec:methodology} by discussing the phenomonology of precessing waveforms and their decomposition into simpler pieces, provide a broad outline of how we will construct neural network surrogates for each of these pieces, and introduce the waveform mismatch metrics that we will use to quantify the faithfulness of our final surrogate model. Section~\ref{sec:model_construction} provides a detailed summary of the construction of the surrogate for each piece of the waveform, and describes the implementation of the final waveform model on CPU and GPU architectures. Our final surrogate returns waveforms on a fixed geometric time grid, so in Sec.~\ref{sec:domain_of_validity} we consider how the starting frequency of our waveform changes with the total mass, and deduce the mass range for which our model is valid as a function of the smallest frequency used for analysis and the largest mass ratio under consideration. The remainder of Sec.~\ref{sec:model_performance} is used to demonstrate the performance of our surrogate, computing the mismatch of our surrogate waveforms against SEOBNRv5PHM (Sec.~\ref{sec:faithfulness}), and timing the computational cost of single- and batched waveform evaluation using different computational architectures (Sec.~\ref{sec:model_timings}). We explore the application of our model to Bayesian inference on real and injected gravitational wave signals in Sec.~\ref{sec:bayesian_inference}. We conclude in Sec.~\ref{sec:conclusions} by summarizing the construction of our model, its performance and limitations, and the directions for future development. 

We consistently use geometric units $G = 1 = c$ throughout this work, unless otherwise stated. 

\section{Methodology}\label{sec:methodology}
In this section we will review our methodology for constructing surrogate waveform models for precessing binary black holes. The first step in this process, described in Sec.~\ref{sec:wvform_decomp}, is to appropriately decompose the complex waveforms into simpler data pieces which can be modelled more easily on their own. We follow this with a discussion of our general approach for building the individual surrogate models, starting with a review of how reduced-order modelling techniques can be used to provide compressed representations of time-series data (Sec.~\ref{sec:data_compression}), and then describe the use of artificial neural networks to interpolate the offline training data across the binary parameter space (Sec.~\ref{sec:ANN_review}). We conclude in Sec.~\ref{sec:mismatch_defs} by introducing the different measures of faithfulness by which we we will test our final surrogate model.

\subsection{Waveform decomposition}\label{sec:wvform_decomp}

Binary black holes moving along quasi-spherical orbits are described by seven intrinsic parameters $\lamvec = (q, \chivec_1, \chivec_2)$, where $q := m_1/m_2 \geq 1$ is the mass ratio and $\chivec_1$, $\chivec_2$ are the dimensionless spins of the two black holes. The total mass $M := m_1 + m_2$ sets the overall time and length scales of the binary, but otherwise factors out of the waveform completely. When the black hole spins are not aligned with the instantaneous orbital angular momentum, $\vec{L}(t)$, general-relativistic spin-spin and spin-orbit coupling causes the spatial orientation of both spins and the orbital angular momentum to change continuously \cite{Apostolatos:1994mx, Kidder:1995zr}. This most commonly takes the form of {\it simple precession} around the (approximately fixed) direction of the total angular momentum of the system, $\vec{J}(t)$. In rare cases, if the spins are nearly anti-aligned with $\vec{L}$, there may be a time during the inspiral when cancellation between spin and orbital angular momenta gives $\vec{J} \approx 0$, leading to {\it transitional precession}~\cite{Apostolatos:1994mx}. When this occurs, the orientation of $\vec{J}$ can change significantly, causing $\vec{J}$, $\vec{L}$ and $\chivec_i$ to ``tumble" for a period of time until GW emission reduces the magnitude of $\vec{L}$ and breaks the cancellation in $\vec{J}$. In either case, precession modulates the phase and amplitude of the GW signal and induces mixing among the $m$-modes at a given $\ell$ in the waveform.

\begin{figure*}[tb]
  \centering
  \includegraphics[width=\linewidth]{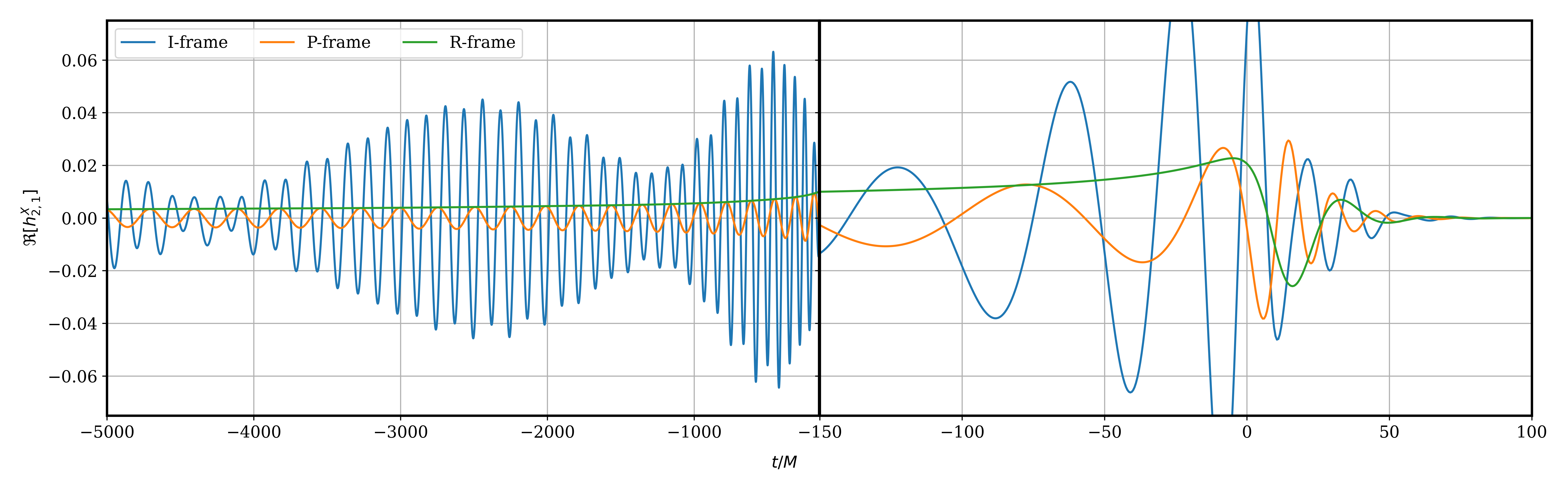}~\\~\\
  \caption{\label{fig:IPR_frames} Real part of the $(\ell, m) = (2,1)$ waveform mode in the I (blue), P (orange) and R (green) frames for the precessing binary with intrinsic parameters $q = 8$, $\chivec_1 = (0.8, 0, 0)$ and $\chivec_2 = (-0.28, 0, 0.28)$ at reference orbital frequency $\Omega_{\rm ref} = 0.007/M$, using the SEOBNRv5PHM model. Transforming from the inertial I-frame to the co-precessing P-frame removes the distinctive precession-induced amplitude modulation of the modes, while transforming from the P-frame to the co-rotating R-frame removes the dominant oscillatory phase from each of the modes.}
\end{figure*}

\begin{figure}[tb]
  \centering
  \includegraphics[width=\linewidth]{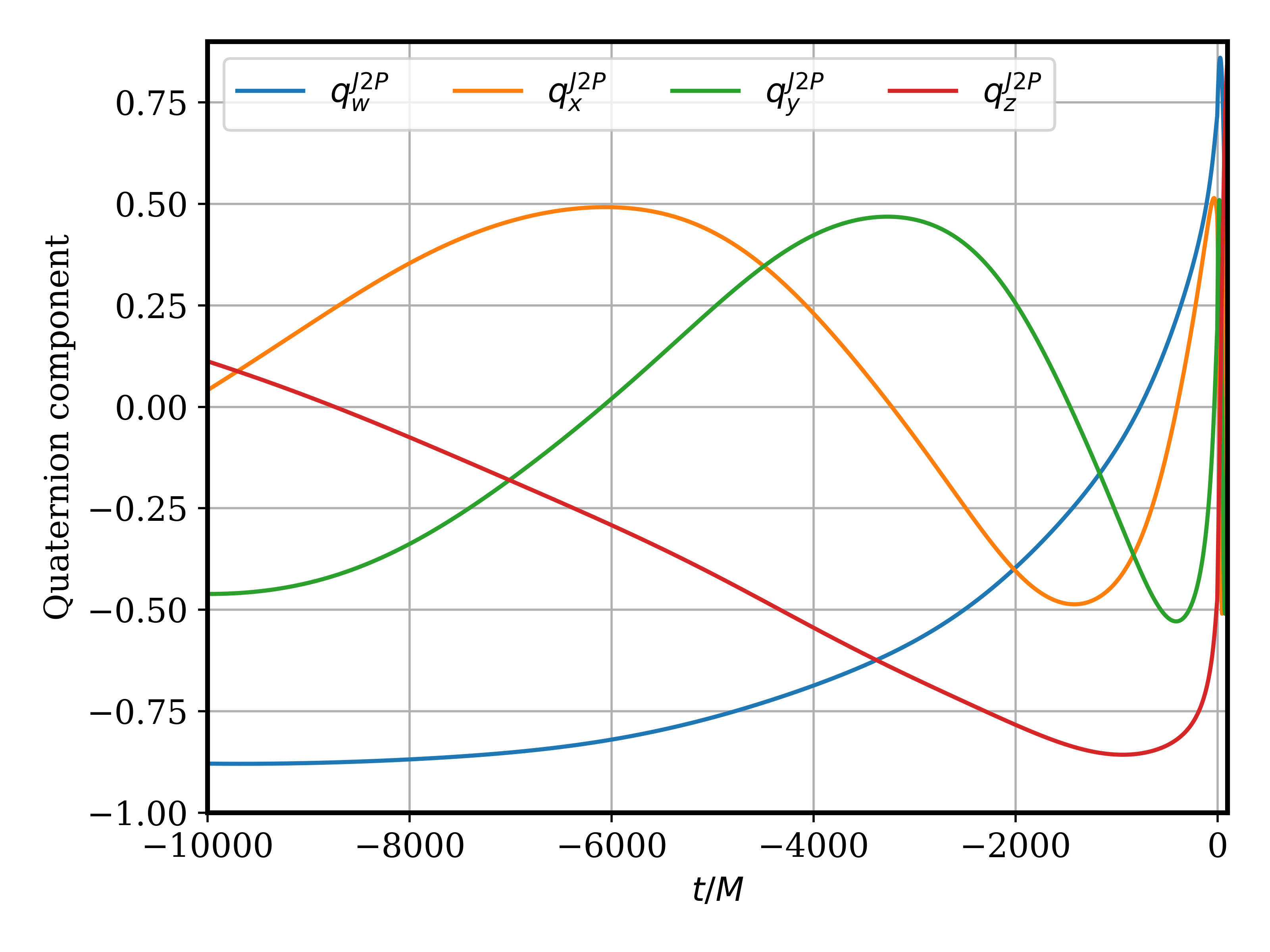}~\\~\\
  \caption{\label{fig:qJ2P_illustration} Components of the quaternion $q_{J2P}(t)$ for the precessing binary with intrinsic parameters $q = 8$, $\chivec_1 = (0.8, 0, 0)$ and $\chivec_2 = (-0.28, 0, 0.28)$ at reference orbital frequency $\Omega_{\rm ref} = 0.007/M$, using the SEOBNRv5PHM model.}
\end{figure}

Our analysis of precessing waveforms involves several frames of reference. The first frame, which we shall refer to as the I-frame, is an inertial frame which has its $z$-axis aligned with the instantaneous orbital angular momentum $\vec{L}(t_{\rm ref})$ at the ($\lamvec$ dependent) reference time $t_{\rm ref}$ at which the orbital frequency $\Omega$ is equal to some reference value $\Omega_{\rm ref}$~\cite{Schmidt:2017btt}. The I-frame defines the inertial source frame in which the GW polarizations $h_{+,\times}^I$ are specified in the SEOBNRv5PHM model \cite{Ramos-Buades:2023ehm}. The polarizations at sky location $(\iota, \phi_0)$ are given in terms of spin-weighted spherical harmonic modes,
\begin{align}
    h_+^I(t; \lamvec; \iota, \phi_0) - &i h_{\times}^I(t; \lamvec; \iota, \phi_0) \nonumber \\&= \sum_{\ell=2}^{\infty}\sum_{m=-\ell}^{\ell} h_{\lm}^I(t; \lamvec)\swsh{-2}{\lm}(\iota, \phi_0), \label{eq:Iframe_mode_decomp}
\end{align}
The quantities $\iota$ and $\phi_0$ represent the inclination of the binary with respect to the observer and the reference phase respectively. Note that we shall specify the intrinsic parameters $\lamvec$  at the reference frequency $\Omega_{\rm ref}$, i.e. $\lamvec = (q, \chivec_1(t_{\rm ref}), \chivec_2(t_{\rm ref}))$, with the spin components $\chivec_i$ given in the I-frame. An illustration of $h_{2,1}^I(t)$ for an example precessing orbit is shown in blue in Fig.~\ref{fig:IPR_frames}, exhibiting the characteristic modulation of the amplitude.

The second frame of interest is the P-frame, corresponding to a non-inertial observer who co-precesses with the orbital plane of the binary \cite{Schmidt:2010it, Schmidt:2012rh, Boyle:2011gg}. To be precise, the P-frame is constructed such that its $z$-axis is aligned with the instantaneous orbital angular momentum $\vec{L}(t)$ at each instant. In this frame, the waveform modes $h_{\lm}^P$ take on a simpler form, illustrated by the orange curve in Fig.~\ref{fig:IPR_frames}, and appear approximately like those of a non-precessing binary \cite{Schmidt:2010it, Schmidt:2012rh}. SEOBNRv5PHM includes by default the $(2,2)$, $(2,1)$, $(3,3)$, $(3,2)$, $(4,4)$ and $(4,3)$ modes in the P-frame~\cite{Ramos-Buades:2023ehm, Mihaylov:2023bkc}, with the corresponding negative $m$-modes given by the assumed symmetry \footnote{This conjugate symmetry provides a convenient approximation, but does not hold exactly for precessing binaries \cite{Schmidt:2012rh, Boyle:2014ioa, Ramos-Buades:2020noq}. The most recent version of SEOBNRv5PHM \cite{Estelles:2025zah} supports optional mode-asymmetries in the P-frame, but we do not include this effect in our present work.}~\cite{Thorne:1980ru}:
\begin{align}
    h_{\ell, -m}^P = (-1)^\ell h_{\lm}^{P*}. \label{eq:Pmode_symmetry}
\end{align}
We will therefore only need to model those P-frame modes with $m \geq 0$.

The I-frame modes $h_{\lm}^I$ are related to the P-frame modes $h_{\lm}^P$ by a time-dependent rotation, \cite{Schmidt:2010it, Schmidt:2012rh, OShaughnessy:2011pmr, Boyle:2014ioa}
\begin{align}
    h^I_{\lm}(t) = \sum_{m'=-\ell}^{\ell}\WigD{m}{m'}{\ell*}(q_{P2I}(t))h_{\lm}^P(t), \label{eq:P2I_mode_rotation}
\end{align}
where $\WigD{m}{m'}{\ell}$ are the Wigner D-matrices, which depend on the unit quaternion $q_{P2I}(t)$ describing the rotation from the P-frame to the I-frame. Note that we have suppressed the dependencies on $\lamvec$ for brevity, and will continue to do so for the remainder of this section. The complex strain in the I-frame can be directly related to the P-frame modes following the algorithm outlined in \cite{Boyle:2014ioa}
\begin{align}
    h_+^I(t; \iota, &\phi_0) - i h_{\times}^I(t; \iota, \phi_0) \nonumber \\ &= \sum_{\ell,m}\sqrt{\frac{2\ell+1}{4\pi}} \WigD{m}{2}{\ell}\left(q_f(t; \iota, \phi_0)\right)h_{\lm}^P(t), \label{eq:Iframe_strain_opt}
\end{align}
where $q_f(t; \iota, \phi_0) := q_{P2I}(t)\cdot q_N(\iota, \phi_0)$ and $q_N(\iota, \phi_0)$ is the quaternion describing the rotation onto the line-of-sight of the observer. Equation~\eqref{eq:Iframe_strain_opt} is the most efficient way to evaluate the polarizations, requiring fewer $\WigD{m}{m'}{\ell}$ evaluations compared to rotating the modes using Eq.~\eqref{eq:P2I_mode_rotation} and then evaluating Eq.~\eqref{eq:Iframe_mode_decomp} directly \cite{Boyle:2014ioa}. 

For the purposes of building our surrogate model, we find it convenient to perform one further frame transformation. Defining the orbital phase in terms of the $(2,2)$ mode in the P-frame,
\begin{align}
    \phi_{\rm orb}(t) := \frac{1}{2}\arg\left[h_{2,2}^P(t)\right], \label{eq:orb_phase_def}
\end{align}
the modes in the so-called R-frame are defined by 
\begin{align}
    h_{\lm}^R(t) = h_{\lm}^P(t) e^{-im\phi_{\rm orb}(t)}. \label{eq:Rmode_def}
\end{align}
The R-frame represents the frame of an observer that is co-rotating with the binary \cite{Boyle:2013nka}. In this frame, the modes grow steadily during the inspiral, with minimal oscillation until the merger-ringdown phase, as illustrated by the green curve in Fig.~\ref{fig:IPR_frames}. This simple morphology makes the R-frame modes significantly easier to model than either the I or P frame modes. 

Finally, we note that the SEOBNRv5PHM model uses a second inertial frame to assist with the construction of the merger-ringdown signal \cite{Ramos-Buades:2023ehm}. This frame, known as the J-frame, has its $z$-axis aligned with the final total angular momentum, $J_f$. The \texttt{pyseobnr} interface gives access to the quaternion $q_{J2P}(t)$ describing the time-dependent rotation from the J-frame to the P-frame, and the quaternion $q_{I2J}$ describing the fixed rotation from the I-frame to the J-frame, but does not return $q_{P2I}(t)$ directly \cite{Mihaylov:2023bkc}. We will instead build surrogate models for $q_{J2P}(t)$ and $q_{I2J}$ separately, from which we can obtain $q_{P2I} = q_{J2P}(t)^*\cdot q_{I2J}^*$ 
Knowledge of $q_{J2P}(t)$ also allows us to evaluate the J-frame modes and strain using Eqs.~\eqref{eq:P2I_mode_rotation} and \eqref{eq:Iframe_strain_opt} [with the substitution I $\rightarrow$ J] if required. Throughout this work we represent quaternions $q$ by their Cartesian components $(q_w, q_x, q_y, q_z) \in \mathbb{R}^4$ with $q = q_w + q_x i + q_yj + q_zk$ and $q^* = q_w - q_x i - q_yj  -q_zk$. Figure~\ref{fig:qJ2P_illustration} illustrates the components of $q_{J2P}(t)$ for the same intrinsic parameters as Fig.~\ref{fig:IPR_frames}.

Overall, therefore, we need to build surrogate models for 16 individual real-valued functions of time: the orbital phase, $\phi_{\rm orb}(t)$, the real and imaginary parts of $h_{\lm}^R(t)$ for all $6$ modes included in the P-frame [note that $\Im [h_{2,2}^R(t) ]\equiv 0$ by definition \eqref{eq:orb_phase_def}, leaving us with only 11 real functions to model], and finally the 4 Cartesian components of $q_{J2P}(t)$. In addition, we must provide a model for the (time-independent) quaternion $q_{I2J}$. We refer to these decomposed quantities as the waveform data pieces.

\subsection{Data compression: reduced bases and empirical interpolation}\label{sec:data_compression}
Consider the case that our data piece is a function of time, $f(t, \lamvec)$. Suppose we have calculated the values of $f(t, \lamvec)$ for times in some interval $I_t$, at discrete locations $\lamvec_i$ in our chosen region $\Lambda$ of parameter space. The set $\Lambda_N := \{\lamvec_i\}_{i=1}^N \subset \Lambda$ is known as the {\it training set}. 

The goal of this section is to obtain a compressed representation for $f$ on $\Lambda$. To be precise, we seek to express $f(t, \lamvec)$ for arbitrary $\lamvec \in \Lambda$ as a linear combination of $n \ll N$ basis functions $e_i(t)$ that approximately span $f(\Lambda)$,
\begin{align}
    f(t, \lamvec) \approx \mathcal{P}_n[f](t, \lamvec) :=\sum_{i=1}^n c_i(\lamvec)e_i(t),
\end{align}
where $\mathcal{P}_n[f]$ denotes the orthogonal projection of $f$ onto the span of the {\it reduced basis} $\{e_i(t)\}_{i=1}^n$.  The reduced basis is chosen to be orthonormal, such that the coefficients $c_i(\lamvec)$ are given by
\begin{align}
    c_i(\lamvec) = \int_{I_t} f(t, \lamvec)e_i(t)dt. \label{eq:time_inner_product}
\end{align}
The first element of our basis is chosen to be $e_1(t) = f(t, \lamvec_1)$. We then proceed iteratively, using a {\it greedy algorithm} \cite{Field:2013cfa, Field:2011mf} to add new elements to our basis in the following way. At each stage we compute the representation error with respect to our current basis $\{e_i(t)\}_{i=1}^m$,
\begin{align}
    \epsilon_j := \left\|f(t, \lamvec_j) - \mathcal{P}_m[f](t, \lamvec_j) \right\|_p
\end{align}
for each $\lamvec_j \in \Lambda_N$, where $\|\cdot\|_p$ is the $L^p$-norm on $I_t$ (the degree $p$ of which we are free to choose). We then select $j = \text{argmax}_j\>\epsilon_j$ and set $e_{m+1}(t) = \text{GS}_m[f](t, \lamvec_j)$, where $\text{GS}_m$ denotes orthonormalization with respect to $\{e_i(t)\}_{i=1}^m$ using a modified Gram-Schmidt procedure~\cite{Trefethen:1997}. The greedy algorithm terminates when either the maximum representation error across the training set falls below some pre-specified tolerance $\sigma$, i.e. $\max_j \epsilon_j \leq \sigma$, or if the selected element $f(t, \lamvec_j)$ is already a member of the basis. 

With the reduced basis in hand, we next construct an {\it empirical interpolant} ($\text{EI}$) \cite{Barrault2004, maday2009general}, which is the unique linear combination of the basis vectors such that
\begin{align}
    \text{EI}[f](T_j, \lamvec) = f(T_j, \lamvec)
\end{align}
for all $\lamvec \in \Lambda$ and some chosen {\it empirical time nodes} $\{T_j\}_{j=1}^n$. The empirical interpolant is given explicitly by
\begin{align}
    \text{EI}[f](t, \lamvec) := \sum_{j=0}^n B_j(t) f(T_j, \lamvec),\label{eq:empirical_interpolant}
\end{align}
where 
\begin{align}
    B_j(t) := \sum_{i=0}^n e_i(t)(V^{-1})_{ij}
\end{align}
and $V_{ij} := e_i(T_j)$ is the interpolation matrix. The values of the empirical time nodes $T_j$ are chosen using the {\it empirical interpolation method} \cite{Barrault2004, maday2009general}, another iterative greedy algorithm that proceeds as follows. The first time node is chosen to be $T_1 = \text{argmax}_t \>|e_1(t)|$, and subsequently in the $j$th step (with $j > 1$) we choose $T_j = \text{argmax}_t \>|e_j(t) - \text{EI}_{j-1}[e_j](t)|$, where $\text{EI}_{j-1}$ denotes the empirical interpolant constructed from the partial basis $\{e_i(t)\}_{i=1}^{j-1}$ with time nodes $(T_i)_{i=1}^{j-1}$. 

Finally, we note that in practical calculations, all continuous functions of time are stored as time series at $K$ discrete time nodes $\mathcal{T} := \{t_i\}_{i=0}^{K}$ that cover $I_t$ densely. In this case, Eq.~\eqref{eq:empirical_interpolant} is expressed in the form
\begin{align}
    \text{EI}[f](t_i, \lamvec) = \sum_{j=0}^n \mathcal{B}_{ij}f(T_j, \lamvec), \label{eq:empirical_interpolant_discrete}
\end{align}
where $\mathcal{B}_{ij} := B_j(t_i)$.

\subsection{Parameter space fits using artificial neural networks}\label{sec:ANN_review}
The empirical interpolant in Eq.~\eqref{eq:empirical_interpolant} provides a compact representation of the data piece $f(t, \lamvec)$ for any $\lamvec \in \Lambda$ on the time interval $t \in I_t$ in terms of its values at only the $n$ empirical time nodes, $T_j$. To complete our surrogate model we must thus use our knowledge of the values $f(T_j, \lamvec_i)$ from the training set $\lamvec_i \in \Lambda_N$ to create a scheme which can {\it predict} the values of $f(T_j, \lamvec)$ for generic $\lamvec \in \Lambda$. We must also devise a model to predict the value of the four components of the quaternion $q_{I2J}(\lamvec)$. The problem here is fundamentally one of interpolation, and has previously been tackled in a number of different ways, including conventional spline and polynomial fitting functions \cite{Field:2011mf, Blackman:2015pia, Varma:2019csw}, and machine learning techniques such as artificial neural networks \cite{Chua:2018woh, Khan:2020fso, Fragkouli:2022lpt, GramaxoFreitas:2024bpk, Thomas:2022rmc} and Gaussian process regression \cite{Moore:2015sza, Varma:2018aht, Williams:2019vub, Islam:2021mha}.

\begin{figure}[tb]
  \centering
  \includegraphics[width=\linewidth]{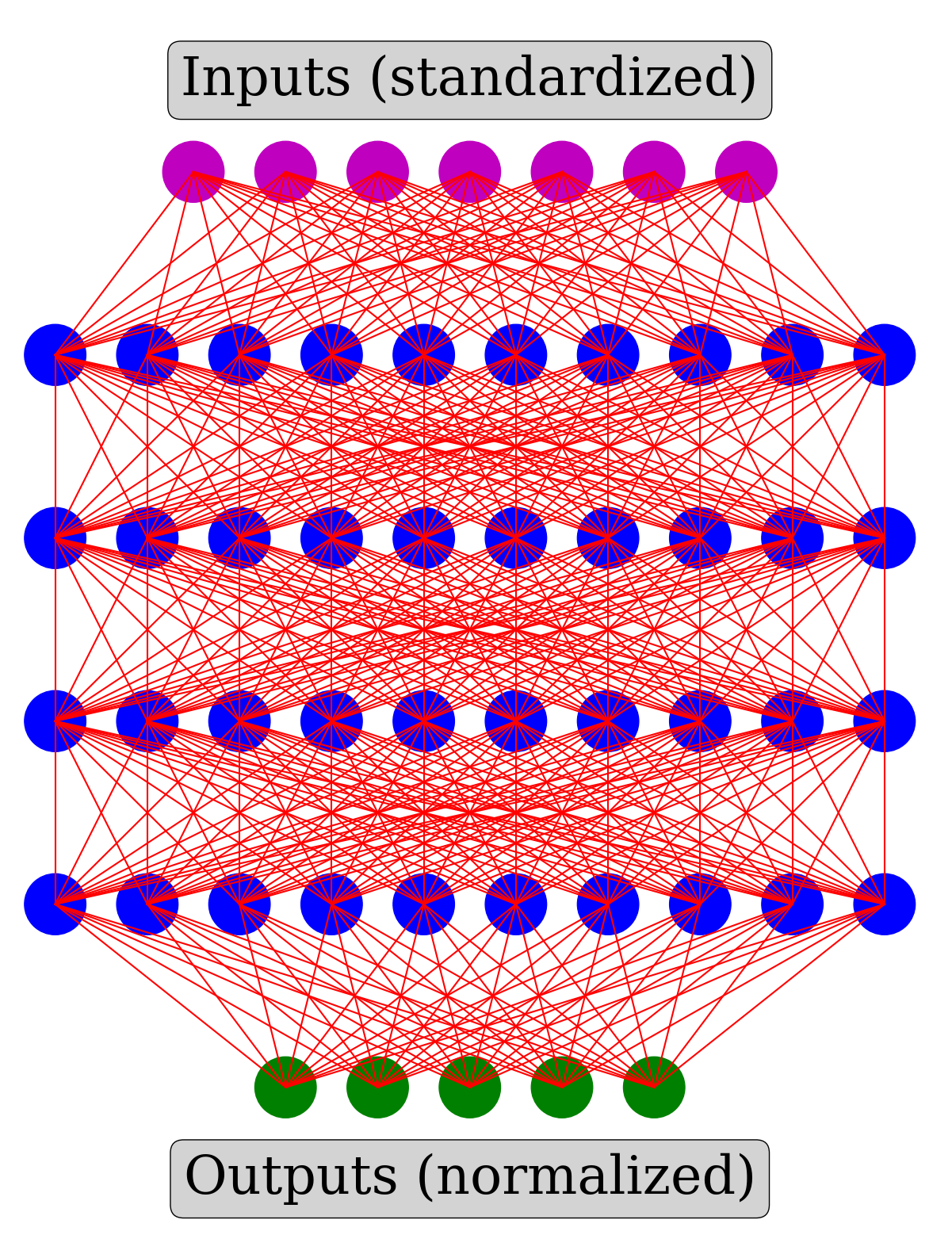}~\\~\\
  \caption{\label{fig:MLP_structure}An example multi-layer perceptron with 7 inputs (magenta) and 5 outputs, (green) with 4 fully-connected hidden layers (blue) containing 10 neurons each. Forward connections between neurons are indicated with red lines. All neural networks constructed in this work assume the inputs are standardized to have mean $0$ and standard deviation $1$, and outputs normalized to the range $[0,1]$, over the training set.}
\end{figure}

In this work, we will make use of artificial neural networks to perform the parameter space fits for the various data pieces. Specifically, we make use of fully-connected feed-forward networks (otherwise known as multi-layer perceptrons) of the form illustrated in Fig.~\ref{fig:MLP_structure}. We use one neural network per data piece. The first (input) layer of each network will contain 7 neurons, one for each of the intrinsic parameters $\lamvec$. The parameters $\lamvec$ are first standardized so that each input has mean $0$ and standard deviation $1$ over the training set. The number of neurons in the output layer varies between data pieces, with the outputs corresponding to the values $\{f(T_j, \lamvec)\}_{j=1}^n$ of the data piece at each of the empirical time nodes (for the time-series data pieces) or to the 4 Cartesian components of $q_{I2J}$. The raw outputs are normalized such that, for each output, the largest value in the training set corresponds to an output of $1$, and the smallest value to an output of $0$. When evaluating the network subsequently, the inverse transformation must be applied to map the normalized outputs back to the true range of the data piece. The intermediate layers of the network are known as the hidden layers; the illustration in Fig.~\ref{fig:MLP_structure} contains 4 hidden layers with 10 neurons each, but in practice the neural networks used in our surrogate model will be both deeper (more hidden layers) and wider (more neurons per layer). Nonlinearity is introduced into the network through the choice of activation function for the hidden layer neurons. Details of the final neural network architecture for each data piece can be found in Sec.~\ref{sec:model_construction}.

The values of the free network parameters (the weights of the neural connections and the biases of each neuron) are set during {\it training}, an iterative gradient-descent process in which the (randomly initialized) parameters are updated in such a way as to minimize the {\it training loss}, a measure of the error in the network's ability to reproduce the training data. In practice, updates are applied using small {\it mini-batches} of training data at a time, rather than using the entire training set at once. The size of the update to the parameters is controlled by the {\it learning rate}, with smaller learning rates producing smaller changes to the parameters. A complete pass through all the training data is called an {\it epoch}; training is carried out over a large number of epochs to allow for iterative improvements in model accuracy. The ultimate goal is to produce a neural network which {\it generalizes} well, i.e. which can also accurately predict the value of unseen data that were not included in the training set. We therefore also compute and monitor the {\it validation loss}, which is the network loss evaluated against a set of independent validation data that are not directly used to train the neural network.

Collectively, the network architecture (e.g. the number and size of the hidden layers and the choice of activation function) and the details of the training process (e.g. the choice of optimization algorithm, loss function, learning rate and mini-batch size) are known as the network hyperparameters. Careful selection of these hyperparameters is critical to obtain the lowest possible validation loss, and hence achieve optimal accuracy for the surrogate model \cite{Thomas:2022rmc, Thomas:2025rje}. We use an empirical approach to hyperparameter optimization, starting with short-duration training on reduced training sets to experiment with different configurations and then scaling
to the full training set using intuition, refining our candidate hyperparameters using further numerical experiments. Our experimentation primarily involved manual variation to find hyperparameter combinations with minimal validation loss, but on a small number of occasions we also made use of automated hyperparameter search routines (specifically, the stochastic Tree of Parzen Estimators approach \cite{TPE} implemented in \texttt{hyperopt} \cite{hyperopt}) to explore broad regions of hyperparameter space, and later to automatically refine parameter choices within small regions. We did not use these routines extensively or systematically. We assess that an automated search over the full hyperparameter space would have been computationally infeasible due to the large number of hyperparameter samples that would have to be drawn for a global search, and the costly need to train a neural network for each draw.


\subsection{Measures of faithfulness}\label{sec:mismatch_defs}
The purpose of our surrogate model is to faithfully predict the output of SEOBNRv5PHM given intrinsic parameters in the surrogate's training range. The primary quantities whose faithfulness shall be verified in Sec.~\ref{sec:model_performance} are the I- and P-frame modes $h_{\lm}^{I/P}(t)$ and the {\it real-valued strain}, defined as
\begin{align}
    h(t) := h^I_+(t) \cos(2\psi) + h^I_\times(t)\sin(2\psi), \label{eq:strain}
\end{align}
where $\psi$ is the {\it polarization angle}. For the remainder of this section, when we refer to a ``waveform", we mean either the modes or the real-valued strain. Our purpose now is to explain how we quantify the difference between the surrogate and SEOBNRv5PHM waveforms. 

As a first step, we introduce the overlap between two waveforms $h_1$ and $h_2$, defined as
\begin{align}
    \mathcal{O}[h_1, h_2] : = \frac{\langle h_1, h_2\rangle}{\sqrt{\langle h_1, h_1 \rangle\langle h_2, h_2\rangle}},
\end{align}
where the noise-weighted inner product is given by
\begin{align}
    \langle h_1, h_2\rangle := 4 \Re\displaystyle\int_{f_{\rm low}}^{f_{\rm high}} \frac{\tilde h_1(f) \tilde h_2^*(f)}{S_n(f)} df,
\end{align}
with $\tilde h_i(f)$ denoting the Fourier transform of $h_i(t)$ and $\tilde h^*_i$ the complex conjugate of $\tilde h_i$. The function $S_n(f)$ is the one-sided noise power spectral density (PSD) of the detector under consideration. For the mismatch calculations in this work, we consider two analytical PSDs available through \texttt{LALSimulation} \cite{lalsuite}: \texttt{aLIGOLateHighSensitivityP1200087} \cite{KAGRA:2013rdx} and \texttt{EinsteinTelescopeP1600143} \cite{LIGOScientific:2016wof}, hereafter known as the aLIGO and ET PSDs respectively. As the names suggest, the aLIGO PSD corresponds to the Advanced LIGO detectors, while the ET PSD corresponds to the Einstein Telescope. The {\it signal-to-noise ratio} (SNR) of a signal $h$ with a given detector PSD is given by
\begin{align}
    \rho[h] := \sqrt{\langle h,h\rangle}.
\end{align}

The {\it match} between two waveforms is defined to be their overlap, maximized over relative shifts in the time and phase, e.g.~\cite{Owen:1998dk}
\begin{align}
    \mathcal{M}_f[h_1, h_2] := \max_{\delta t_c, \delta\phi_0,...} \mathcal{O}[h_1, h_2], \label{eq:match}
\end{align}
where $t_c$ is the (geocenter) time of coalescence and $\phi_0$ is the reference phase. We compute our matches using the \texttt{pycbc.filters.matchedfilter.match} utility \cite{pycbc}, which maximizes the overlap over relative time and phase shifts between $h_1$ and $h_2$. When computing the strain mismatch, we also numerically maximize the time and phase optimized matches over small shifts $\delta\psi \in [-0.1, 0.1]$ in the $h_2$ polarization angle using the \texttt{L-BFGS-B} routine implemented in \texttt{scipy} \cite{scipy}. 

The difference between two similar waveforms $h_1$ and $h_2$ is best expressed in terms of the {\it mismatch}, 
\begin{align}
    \overline{\mathcal{M}}_f[h_1, h_2] := 1 - \mathcal{M}_f[h_1, h_2].
\end{align}
We will use the mismatch between the surrogate and SEOBNRv5PHM waveforms to measure the error in our surrogate model. We will also find it convenient to calculate orientation averaged strain mismatches, obtained by fixing the intrinsic parameters $\lamvec$ and $M$ and then computing the mismatch between the surrogate and SEOBNRv5PHM strain at different choices of $(\iota, \phi_0; \psi)$.
The averaged mismatch is defined by \cite{Harry:2016ijz}
\begin{align}
    \overline{\mathcal{M}}_{\rm SNR} := 1 - \left(\frac{\sum_i \mathcal{M}_i^3 \rho_i^3}{\sum_i \rho_i^3} \right)^{1/3}, \label{eq:snr_weighted_mm}
\end{align}
where $\mathcal{M}_i$ are the (time, phase and polarization shift-optimized) matches for the different choices of $(\iota, \phi_0; \psi)$, and $\rho_i$ is the corresponding (SEOBNRv5PHM) SNR at that orientation. Weighting by SNR in this manner accounts for the likelihood of detection at a given orientation and fixed luminosity distance.

The significance of the mismatch depends on the SNR of the signal under consideration, with higher SNRs placing stricter requirements on waveform accuracy. A commonly used criterion states that two waveforms $h_1$ and $h_2$ will be statistically indistinguishable for data analysis purposes if \cite{Lindblom:2008cm, McWilliams:2010eq, Hannam:2010ky, Baird:2012cu, Chatziioannou:2017tdw, Toubiana:2024car}
\begin{align}
    \mathcal{M}_f[h_1, h_2] < \frac{N}{2\rho^2}, \label{eq:indistinguishability_criterion}
\end{align}
where $N$ is the number of parameters we seek to measure and $\rho$ is the SNR of the signal under consideration. In this work, we take $N = 7$, which corresponds to the number of parameters $\lamvec$.\footnote{With the choice $N = 7$, Eq.~\eqref{eq:indistinguishability_criterion} gives the condition for indistinguishability at $1\sigma$ confidence for the joint 7-dimensional posterior distribution \cite{Thompson:2025hhc}. Requiring indistinguishability in all 1d marginal confidence intervals corresponds to the more stringent choice $N = 1$. Throughout this article we specify limiting SNRs assuming $N = 7$, but note that the limiting SNRs for different numbers of degrees of freedom can be obtained by applying a suitable scaling according to Eq.~\eqref{eq:limiting_SNR}.} Rearranging Eq.~\eqref{eq:indistinguishability_criterion}, the two signals are indistinguishable at any SNR 
\begin{align}
    \rho < \sqrt{\frac{N}{2\mathcal{M}_f}}. \label{eq:limiting_SNR}
\end{align}
This formula can be used to estimate the limiting SNR of our surrogate model at a given point in parameter space, beyond which we expect the surrogate may introduce parameter biases relative to SEOBNRv5PHM. We caution, however, that Eq.~\eqref{eq:limiting_SNR} is often found to be overly-conservative in practice \cite{Purrer:2019jcp, Thompson:2025hhc}, and the surrogate may remain unbiased even at SNRs well above this limit (see Ref.~\cite{Thompson:2025hhc} for a detailed explanation of why this can occur.) Nonetheless, we will use Eq.~\eqref{eq:limiting_SNR} in Sec.~\ref{sec:model_performance} to provide a convenient lower bound on the limiting SNR of our surrogate.

\section{Model construction}\label{sec:model_construction}
In this section we describe the construction of our surrogate model, starting with a description in Sec.~\ref{sec:data_generation} of how we calculate the decomposed waveform data pieces in practice, and how we choose the parameters to include in our training and validation datasets. We then describe the construction of the reduced bases and empirical interpolants in  Sec.~\ref{sec:rb_eim_construction}, and the training of our neural networks in Sec.~\ref{sec:ANN_training}. We conclude in Sec.~\ref{sec:surrogate_evaluation} by explaining the process of evaluating a surrogate waveform, including details of the practical implementation on both CPU and GPU architectures. 

\subsection{Data generation}\label{sec:data_generation}
For a single choice of intrinsic parameters $\lamvec$, we generate the P-frame modes $h_{\lm}^P(t; \lamvec)$ and the quaternions $q_{J2P}(t;\lamvec)$ and $q_{I2J}(\lamvec)$ using the \texttt{pyseobnr} \cite{Mihaylov:2023bkc} interface for the SEOBNRv5PHM model. The time-dependent quantities are returned on a time-grid $\mathcal{T}_{\rm SEOB}(\lamvec)$ which begins when the orbital frequency is equal to $\Omega_{\rm ref} = 0.007/M$. This starting frequency also acts at the reference frequency at which, recall, the $I$-frame is defined and the spin components $\chivec_i(t_{\rm ref})$ are specified.  

The quaternions $q_{\rm J2P}(t)$ are interpolated onto the {\it common time grid}, $\mathcal{T}_{\rm com}$, using the cubic \texttt{InterpolatedUnivariateSpline} from \texttt{scipy} \cite{scipy}. The common time grid is a fixed grid with spacing $\Delta t = 0.5M$, starting at $\tstart = -10^4M$ and ending at $\tend = 100M$ (where SEOBNRv5PHM waveforms are time-aligned such that the peak of the frame-invariant amplitude $\sum_{\lm}|h_{\lm}^P(t)|
^2$ occurs at $t = 0$ \cite{Ramos-Buades:2023ehm, Schmidt:2017btt}). Noting that the quaternions $q$ and $-q$ represent the same rotation, we initially condition the quaternions such that $q_w^{J2P}(\tstart) > 0$ and $q_w^{I2J} > 0$.

The orbital phase $\phiorb(t)$ is extracted from the values of $h_{2,2}^{\rm P}(t)$ sampled at the times $t \in \mathcal{T}_{\rm SEOB}$, and the $2\pi$ jumps caused by phase wrapping are removed using the \texttt{unwrap} utility from \texttt{numpy} \cite{numpy}. We note that Eq.~\eqref{eq:orb_phase_def} only defines the orbital phase up to the addition of multiples of $\pi$; we exploit this freedom by adding the necessary multiple of $\pi$ to ensure $\Re[h_{2,1}^R(\tstart)]>0$ and $\phiorb(\tref) \in (-\pi, \pi]$. The unwrapped, shifted phase is then interpolated onto $\mathcal{T}_{\rm com}$. We find that the order of the extraction and interpolation steps is very important to ensure appropriate data conditioning. When we instead interpolate $h_{2,2}^P(t)$ onto $\mathcal{T}_{\rm com}$ and then extract $\phiorb(t)$ from the interpolated modes, the resulting phases vary rapidly and apparently randomly across the parameter space, making it practically impossible to train a neural network to predict the values of $\phiorb(t)$ at the empirical time nodes. This behaviour is understood to occur because of the large number of orbital cycles between $\Omega = \Omega_{\rm ref}$, the instance at which the initial phases are consistently defined in the SEOBNRv5PHM model, and the start time of our surrogate model, $\tstart$: the growth of $\phiorb(t)$ between the reference frequency and the start time pseudo-randomizes the value of the phase modulo $2\pi$ at $\tstart$. Finally, after extracting $\phiorb(t)$, all the P-frame modes are interpolated from $\mathcal{T}_{\rm SEOB}(\lamvec)$ to $\mathcal{T}_{\rm com}$, and the R-frame modes on $\mathcal{T}_{\rm com}$ are then constructed using Eq.~\eqref{eq:Rmode_def}. 

We build our training and validation sets to cover the range $q \in [1, 10]$ with arbitrary spin magnitudes $|\chivec_i| \in [0, 1]$ and directions. We choose to parameterize the spin vectors of the black holes at the reference frequency in terms of their magnitudes $|\chivec_i|$ and the spherical polar angles $\theta_i$ and $\phi_i$ describing the unit direction vector of each spin in the I-frame. We use several different strategies to sample the intrinsic parameters $\lamvec = (q, |\chivec_1|, \theta_1, \phi_1, |\chivec_2|, \theta_2, \phi_2) $ when building our training and validation sets. One approach is Latin hypercube sampling \cite{Eglajs1977, McKay1979}, for which we use the \texttt{LatinHypercube} algorithm from \texttt{scipy} \cite{scipy}. A Latin hypercube sample of size $N$ has the property that each dimension in parameter space is divided into $N$ equal probability bins, and each bin contains exactly one sample. We generate two different Latin hypercube samples, hereafter referred to as $\Lambda_{\rm 200k}$ and $\Lambda_{\rm 1M}$, which contain $2 \times 10^5$ and $10^6$ samples respectively. We use $\Lambda_{\rm 200k}$ and $\Lambda_{\rm 1M}$ to build our reduced bases and train our neural networks. For validation data during neural network training we use another sample, $\Lambda_{\rm val}$, which contains $10^5$ waveforms whose parameters were chosen randomly and uniformly, i.e. were drawn randomly from the distributions
\begin{align}
    q \sim U[1,10], \quad |\chivec_i| \sim U[0, 1], \nonumber \\\theta_i \sim U[0, \pi], \text{ and  } \phi_i \sim U[0, 2\pi]. \label{eq:random_polar_sample}
\end{align}
In Sec.~\ref{sec:model_performance} we make use of additional independent validation datasets, which will be detailed in that section.

\subsection{Reduced bases and empirical interpolants}\label{sec:rb_eim_construction}

\begin{table*}[t]
\begin{ruledtabular}
\begin{tabular}{cccccc}
Data piece & Part/component & RB norm & RB greedy tolerance & Training set & RB size \\
\colrule
$\phiorb(t)$ \rule{0pt}{3ex} & --- & $L^{\infty}$ & $5 \times 10^{-3}$ & $\Lambda_{200k}$ & 22 \\[1ex]
\colrule

$h_{2,2}^R(t)$ \rule{0pt}{3ex} & Real & $L^2$ & $10^{-6}$ & $\Lambda_{\rm old}$ & 18 \\[1ex]
\colrule

\multirow{2}{*}{$h_{2,1}^R(t)$} & Real & \multirow{2}{*}{$L^2$} & \multirow{2}{*}{$10^{-6}$} & \multirow{2}{*}{$\Lambda_{\rm old}$} & $23$ \\
 & Imaginary &  &  &  & $17$ \\
\colrule

\multirow{2}{*}{$h_{3,3}^R(t)$} & Real & \multirow{2}{*}{$L^2$} & \multirow{2}{*}{$10^{-6}$} & \multirow{2}{*}{$\Lambda_{200k}$} & $17$ \\
 & Imaginary &  &  &  & $14$ \\
\colrule

\multirow{2}{*}{$h_{3,2}^R(t)$} & Real & \multirow{2}{*}{$L^2$} & \multirow{2}{*}{$10^{-6}$} & \multirow{2}{*}{$\Lambda_{\rm old}$} & $26$ \\
 & Imaginary &  &  &  & $19$ \\
\colrule

\multirow{2}{*}{$h_{4,4}^R(t)$} & Real & \multirow{2}{*}{$L^2$} & \multirow{2}{*}{$10^{-6}$} & \multirow{2}{*}{$\Lambda_{\rm old}$} & $14$ \\
 & Imaginary &  &  &  & $13$ \\
\colrule

\multirow{2}{*}{$h_{4,3}^R(t)$} & Real & \multirow{2}{*}{$L^2$} & \multirow{2}{*}{$10^{-6}$} & \multirow{2}{*}{$\Lambda_{200k}$} & $24$ \\
 & Imaginary &  &  &  & $21$ \\
\colrule
\multirow{4}{*}{$q_{J2P}(t)$} & $w$ & \multirow{4}{*}{$L^2$} & \multirow{4}{*}{$10^{-4}$} & \multirow{4}{*}{$\Lambda_{200k}$} & $28$ \\
 & $x$ &  &  &  & $51$ \\
 & $y$ &  &  &  & $57$ \\
 & $z$ &  &  &  & $32$ \\
\end{tabular}
\end{ruledtabular}
\caption{\label{tab:rb_details} Details of the norm, greedy tolerance and training set used for the greedy algorithm for each reduced basis used in our final surrogate model, along with the size of the resulting bases.}
\end{table*}

We build all of our reduced bases and empirical interpolants using the greedy algorithms described in Sec.~\ref{sec:data_compression}. For the orbital phase $\phiorb(t)$, we use the $L^{\infty}$ norm to compute the basis representation errors, with a greedy error tolerance of $\sigma =  5 \times 10^{-4}$, and the necessary training data provided by the $\Lambda_{200k}$ data set. We also tried using the $L^2$ norm with $\sigma = 10^{-6}$, but found that the resulting basis caused greater mismatches in the P-frame modes. Our final reduced basis has size $22$, which is listed alongside the other greedy algorithm parameters in Table~\ref{tab:rb_details}. 

For the real and imaginary parts of the R-frame modes, $h_{\lm}^R(t)$, we use the $L^2$ norm to compute the basis representation errors, with a greedy tolerance of $\sigma = 10^{-6}$. In each case, two sets of reduced bases/empirical interpolants were constructed, one using $\Lambda_{200k}$ as training data, and the other using $\Lambda_{\rm old}$, an older data set that was generated for preliminary investigations. $\Lambda_{\rm old}$ also has size $2 \times 10^5$, but with the mass ratio and polar spin components chosen randomly and uniformly according to Eq.~\eqref{eq:random_polar_sample}. While experimenting with training the neural networks, it was noted that for some modes the accuracy of the resulting neural network surrogate was sensitive to which data set was used to generate the reduced basis, with neither data set producing uniformly more accurate surrogate models across all modes. Although this effect was not large (only the sub-dominant $(4,4)$ and $(4,3)$ modes showed significant sensitivity), it was decided to proceed with neural network training using both sets of reduced bases for all modes, and we subsequently chose the optimal reduced basis for each mode based on the accuracy of the full R-frame surrogate. The methodology used to train the neural networks and choose the final R-frame surrogate will be discussed in Sec.~\ref{sec:ANN_training}. For now, we note only that the data sets used to create the optimal reduced bases used by the final R-frame surrogates are given for each mode in Table~\ref{tab:rb_details}, along with the size of each of the chosen bases and the details of the greedy algorithm parameters used.

For each component of $q_{J2P}(t)$ we use the $L^2$ norm with $\sigma = 10^{-6}$ and $\Lambda_{200k}$ for training data, and the size of each reduced basis is given in Table~\ref{tab:rb_details}. We note that the reduced bases for $q_x^{J2P}$ and $q_y^{J2P}$ are almost twice as long as those of $q_w^{J2P}$ and $q_z^{J2P}$, which can be traced to the greater degree of oscillation in the former components noted in Fig.~\ref{fig:qJ2P_illustration}. Having obtained the empirical interpolation matrices $\mathcal{B}_{ij}$ defined in Eq.~\eqref{eq:empirical_interpolant_discrete} for the common time grid $\mathcal{T}_{\rm com}$, we also construct {\it downsampled} interpolation matrices $\mathcal{B}_{ij}^{\rm ds}$, which give the values of the quaternions on the downsampled time grid, $\mathcal{T}_{\rm ds}$, defined to have piecwise uniform spacing,
\begin{equation}
\Delta t = 
\begin{cases}
    \begin{aligned}
        &7.5M  && \text{for} \quad -10^4M \leq t < -5000M, \\
        &5M    && \text{for} \quad -5000M \leq t < -1000M, \\
        &2M    && \text{for} \quad -1000M \leq t < -100M,  \\
        &0.5M  && \text{for} \quad -100M \leq t < 100M.
    \end{aligned}
\end{cases}
\end{equation}
Because $\mathcal{T}_{\rm ds} \subset \mathcal{T}_{\rm com}$ by construction, we can simply obtain $\mathcal{B}_{ij}^{\rm ds}$ for each component of the quaternion by deleting from $\mathcal{B}_{ij}$ those rows corresponding to times $t_i \in \mathcal{T}_{\rm com}\setminus\mathcal{T}_{\rm ds}$. The downsampling procedure is motivated by noting that Eqs.~\eqref{eq:P2I_mode_rotation} and \eqref{eq:Iframe_strain_opt} naively require us to evaluate $\WigD{m}{m'}{\ell}(q(t_i))$ at all times $t_i \in \mathcal{T}_{\rm com}$ in order to construct the inertial frame modes or strain from the P-frame modes. The density of $\mathcal{T}_{\rm com}$ (and hence its total number of points) is set by the need to resolve the merger-ringdown of the signal, whereas we only need to sample the quaternions at a rate inversely proportional to the (longer and time-varying) precession timescale. In practice the common time grid grossly oversamples the quaternions for most of the waveform, and we can safely downsample the quaternions to our non-uniform $\mathcal{T}_{\rm ds}$ --- minimizing the number of relatively costly $\WigD{m}{m'}{\ell}$ evaluations we need to perform --- and then interpolate the matrices back onto $\mathcal{T}_{\rm com}$. 
As we will see in Sec.~\ref{sec:model_performance}, this downsampling procedure can save significant amounts of time in some circumstances when the $P \rightarrow I$ rotation acts as the computational bottleneck in waveform generation, at negligible cost in accuracy. We save both the original and downsampled interpolation matrices. 

\subsection{Neural network training}\label{sec:ANN_training}
\begin{table*}[t]
\begin{ruledtabular}
\begin{tabular}{lcccc}
Parameter &  $\phiorb(t)$ & $h_{\lm}^R(t)$ & $q_{J2P}(t)$ & $q_{I2J}$ \\[1ex]
\colrule
Hidden layers \rule{0pt}{2ex} & 15 & 10 & 10 & 10 \\
Hidden layer width & 384 & 256 & 256 & 64 \\
Hidden activation & elu & elu & elu & elu \\
Output activation & Linear & Sigmoid & Linear & Linear \\
\colrule
Training set \rule{0pt}{2ex} & $\Lambda_{200k} \cup \Lambda_{1M} \cup \Lambda_{1M}^*$ & $\Lambda_{200k} \cup \Lambda_{1M}$ & $\Lambda_{200k} \cup \Lambda_{1M}$ & $\Lambda_{200k} \cup \Lambda_{1M}$ \\
Training set size & $2.2 \times 10^6$ & $1.2 \times 10^6$ & $1.2 \times 10^6$ & $1.2 \times 10^6$ \\
Validation set  & $\Lambda_{\rm val}$ & $\Lambda_{\rm val}$ & $\Lambda_{\rm val}$ & $\Lambda_{\rm val}$\\
Optimizer & adamax & adamax & adamax & adamax\\
Loss function & MSE & MSE & MSE & MSE \\
Minibatch size & 128 & 256 & 1024 & 1024 \\
\colrule
LR scheduler \rule{0pt}{2ex} & PlateauAGM & PlateauAGM & PlateauAGM & PlateauAGM\\
Initial LR & $2.0 \times 10^{-3}$ & $10^{-3}$ & $7.5 \times 10^{-4}$ & $10^{-3}$\\
LR patience & 75 & 75 & 100 & 75 \\
LR cooldown & 5 & 5 & 5 & 5 \\
LR decay factor & 0.8 & 0.8 & 0.8 & 0.8\\
\colrule
Early stopping \rule{0pt}{2ex} & EarlyStopping & DualES/EarlyStopping & EarlyStopping & EarlyStopping \\
ES patience & 250 & 300 & 300 & 300 \\
ES start epoch & 1250 & 150 & 500 & 500\\
\end{tabular}
\end{ruledtabular}
\caption{\label{tab:network_hyperparams_specs} Architectural and training hyperparameters for the different neural networks used in the final surrogate model. The neural network training for the R-frame modes used the \texttt{EarlyStopping} callback for the modes $(\ell, m) \in \{(3,3), (4,3)\}$ and \texttt{DualES} for $(\ell, m) \in \{(2,2), (2,1), (3,2), (4,4)\}$. Separate networks (with the same hyperparameters) are used for the real and imaginary part of each mode R-frame mode, and likewise for each Cartesian component of $q_{J2P}$.}
\end{table*}

\subsubsection{Common features}
Before describing in detail the neural network architecture and training process for each waveform data piece, we start by highlighting several common features. Our neural networks are implemented as \texttt{Sequential} models with fully-connected \texttt{Dense} layers using \texttt{TensorFlow} \cite{TensorFlow} and \texttt{Keras} \cite{Keras}. We experimented with the use of \texttt{Dropout} layers, but found no advantage to final model accuracy. The training data inputs and outputs are scaled using the \texttt{StandardScaler} and \texttt{MinMaxScaler} utilities from \texttt{scikit-learn} \cite{scikit-learn} respectively. Consistently throughout this work we use the mean squared error (MSE) loss function, defined by
\begin{align}
    {\rm{MSE}} = \frac{1}{N}\sum_{i=1}^N |\vec{y}_i^{\>\rm exact} - \vec{y}_i^{\>\rm pred}|^2,
\end{align}
where $\vec{y}_i^{\>\rm pred}$ is the prediction (output) of the neural network, $\vec{y}_i^{\>\rm exact}$ is the corresponding true value derived from SEOBNRv5PHM, and the sum is taken over all samples in the training minibatch. Training was always performed using the \texttt{adamax} optimizer \cite{Kingma:2014vow} with all parameters equal to their \texttt{TensorFlow} defaults, with the exception of the learning rate (LR), which we chose differently for different data pieces. We considered the use of other popular optimizers such as \texttt{Adam}  \cite{Kingma:2014vow}, \texttt{Nadam} \cite{dozat2016nadam} and \texttt{SGD} \cite{BottouSGD} in our initial work, but none showed any significant advantage over \texttt{adamax}, so we chose to hold the optimizer fixed in order to simplify the network hyperparameter optimization.

For each network we specify an initial LR (given for each example in Table~\ref{tab:network_hyperparams_specs}), and then gradually reduce it using a {\it learning rate scheduler}. An example LR scheduler is provided by the \texttt{ReduceLRonPlateau} callback in \texttt{keras} \cite{Keras}, which reduces the LR by a constant factor if a monitored quantity (typically the validation loss) has not reduced by a sufficiently large amount ($\Delta_{\rm min}$) in the last \texttt{patience} epochs, pausing for \texttt{cooldown} epochs after each reduction before resuming operation. We found two significant limitations to \texttt{ReduceLRonPlateau} when training our networks. The need to specify a fixed value of $\Delta_{\rm min}$ complicated the training of some of our networks, where the validation loss reduces by several orders of magnitude over the course of training. Once the loss becomes comparable in magnitude to $\Delta_{\rm min}$, the LR reduction will be activated every \texttt{patience} + \texttt{cooldown} epochs even if the validation loss is decreasing, causing the LR to decay exponentially and the training to stall rapidly. On the other hand, setting  $\Delta_{\rm min}$ to be very small (comparable to or smaller than the expected final validation loss) can make the scheduler insufficiently aggressive in the early stages of training when the loss is larger, slowing progress or stalling the training altogether at an early stage. We also find that the presence of noise in the validation loss can interfere with the operation of \texttt{ReduceLRonPlateau}, with noise fluctuations sometimes leading to inappropriate or delayed activation of the reduction mechanism. To mitigate these issues, we implemented our own custom callback, \texttt{PlateauAGM}, which mimics the logic of \texttt{ReduceLRonPlateau} with two changes. Firstly, rather than using a fixed absolute $\Delta_{\rm min}$ to decide whether a reduction in the loss is significant, we use a relative tolerance instead. To be precise, at each epoch we update $\Delta_{\rm min} = \max\left(0.01\cdot\texttt{val\_loss},\>10^{-9}\right)$, where \texttt{val\_loss} is the current value of the validation loss. Furthermore, rather than using the raw validation loss to decide upon LR reductions, we instead monitor the geometric mean of the loss calculated over the previous \texttt{patience} epochs. Together, we find that these changes produce a more robust learning rate scheduler for our purposes. The values of the LR decay factor and cooldown and patience periods used for each data piece are given in Table~\ref{tab:network_hyperparams_specs}. 

We set a conservative maximum of $5000$ training epochs for each of our networks, but the training and validation losses invariably plateaued long before this limit was reached. To prevent us from wasting computational resources performing unnecessary epochs of training for no gain in model accuracy, we used an {\it early stopping} callback to terminate training once the validation loss had stopped improving, and to then restore the network weights and biases to the point in training with the least validation loss. We primarily achieve this using the \texttt{keras} \texttt{EarlyStopping} callback,  which terminates training if there is no improvement in the validation loss over the previous \texttt{patience} epochs, starting to operate only after a minimum number of epochs (\texttt{start\_epoch}) have passed. We also experimented with a custom variation of this callback, \texttt{DualES}, which uses the rolling geometric mean of the validation loss over the past \texttt{patience} epochs to decide upon termination, but which still restores the model weights that gave the smallest non-averaged validation loss. We found that the much reduced level of noise in the validation loss in the terminal stages of training meant that \texttt{DualES} did not confer any advantage over \texttt{EarlyStopping}, but it was still used for a small number of the final $h_{\lm}^R(t)$ networks where the results with the standard callback were unavailable and it was felt that there would be no advantage in retraining. For details of the early stopping parameters used for each data piece, see Table~\ref{tab:network_hyperparams_specs}. 

\subsubsection{Orbital phase}

A list of all architectural and training hyperparameters for the neural network used to predict the values of the orbital phase at its empirical time nodes can be found in the first column of Table~\ref{tab:network_hyperparams_specs}. When choosing network dimensions, we considered a variety of different sizes, but restricted ourselves to ``rectangular" networks in which each hidden layer has the same width. The deepest network we considered had $20$ hidden layers with up to $512$ neurons each, while the shallowest had only $4$ hidden layers but up to $640$ neurons per layer. Our final network consists of $15$ hidden layers with $384$ neurons each, the largest network used for the final surrogate model out of all the waveform data pieces. We use the \texttt{elu} activation function for hidden layer neurons \cite{Clevert:2015qvd},
\begin{align}
    \mathrm{elu}(x) =
        \begin{cases}
        x, & \text{if } x > 0, \\
        \alpha \left( e^{x} - 1 \right), & \text{if } x \le 0,
        \end{cases}\label{eq:elu_def}
\end{align}
with the default value $\alpha=1.0$. The other options considered were \texttt{ReLU} \cite{Fukushima1969, Nair2010, Glorot2011}, \texttt{tanh}, \texttt{Softplus} \cite{Dugas2000, Glorot2011} and \texttt{Sigmoid}. 

The network was trained using data from the two Latin Hypercube samples $\Lambda_{200k}$ and $\Lambda_{1M}$ as well as data from an additional sample, $\Lambda_{1M}^*$ generated specifically for the orbital phase model. The parameters in $\Lambda_{1M}^*$ were drawn using the \texttt{ExpandLHS} \cite{Boschini:2025ymu} software package, chosen such that $\Lambda_{1M} \cup \Lambda_{1M}^*$ is an approximate Latin Hypercube sample of length $2 \times 10^6$. This additional set of training data was used only for the orbital phase network, which tests revealed to be the most significant source of error in the final waveform model at most places in the parameter space (see Sec.~\ref{sec:model_performance} for further details). A minibatch size of $128$ was used when training the final network, with the validation loss calculated against the random validation set, $\Lambda_{\rm val}$.

\begin{figure}[thb]
  \centering
  \includegraphics[width=\linewidth]{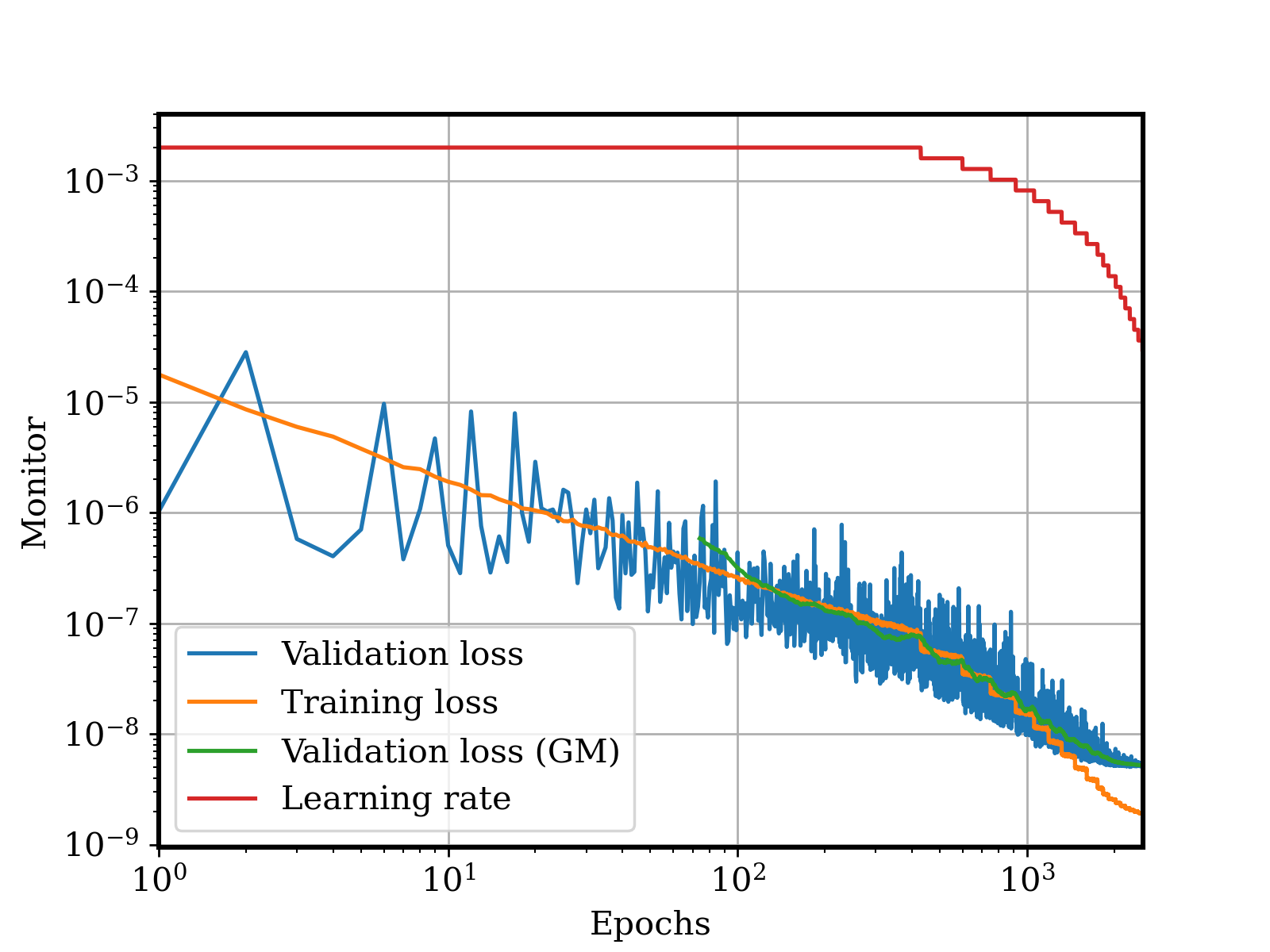}~\\~\\
  \caption{\label{fig:orb_phase_training_history}Training history for the orbital phase network, showing the evolution of the learning rate (red), training loss (orange), and the validation loss (blue) and its rolling geometric mean (green). }
\end{figure}

Figure~\ref{fig:orb_phase_training_history} illustrates the training history for the final orbital phase network, showing the training loss, the validation loss and its rolling geometric mean, and the learning rate as a function of training epoch. The validation loss shows the noise described previously, but the rolling geometric mean effectively produces a smooth quantity for the LR scheduler to monitor. The training and validation losses closely track each other until after the $1000$th epoch, with the validation loss plateauing a factor of approximately $2.7\times$ higher than the training loss. The stepwise reduction of the LR by the \texttt{PlateauAGM} scheduler can also be seen, with the associated downward steps in both training and averaged validation loss also visible, demonstrating that the loss decreases more quickly after LR reduction. Training was terminated by the \texttt{EarlyStopping} callback after $2518$ epochs, taking approximately $31.5$hrs wall time on a single Nvidia T4 GPU. The network weights were automatically restored to their values at the end of epoch $2268$, when the validation loss was at its least.

\subsubsection{R-frame modes}\label{sec:Rframe_networks}
To simplify the task of hyperparameter optimization, we used the same hyperparameters for each R-frame mode network --- tuned to improve the accuracy of the $\Re [h_{2,2}^R]$ network --- with the expectation that the subdominant modes can be slightly less accurate without impacting the accuracy of the final waveform model. Furthermore, after initially trying to optimize over the activation function, we ultimately decided to use the same \texttt{elu} function used for the orbital phase. These simplifying assumptions will be justified by the results of Sec.~\ref{sec:model_performance}, which demonstrate that the R-frame model accuracy is less significant than other sources of error. 

The complete hyperparameters for the final R-frame networks are given in the second column of Table~\ref{tab:network_hyperparams_specs}. Interestingly, our experiments found a weak preference for the use of the \texttt{Sigmoid} activation function in the output layer, the only instance in this work for which we tested a nonlinear activation for the outputs. The R-frame networks were always trained using the $\Lambda_{200k}$ and $\Lambda_{1M}$ data sets (with validation data provided by $\Lambda_{\rm val}$) but, as mentioned in Sec.~\ref{sec:rb_eim_construction}, we trained two sets of networks: one to predict the values at the empirical time nodes derived from the reduced basis generated from $\Lambda_{200k}$, and another for the empirical time nodes derived from $\Lambda_{\rm old}$. The network hyperparameters were identical for both sets in all respects except for the choice of early stopping callback, with those networks based on the $\Lambda_{\rm old}$ reduced bases using \texttt{DualES} and those based on the $\Lambda_{\rm 200k}$ reduced bases using the standard \texttt{EarlyStopping} callback. 

We decided which R-frame model to use for the final surrogate model on a mode-by-mode basis, aiming to minimize the mismatch in the corresponding P-frame surrogate. To be precise, for each mode we computed $h_{\lm}^P$ using the $h_{\lm}^R$ surrogates derived from (a) the $\Lambda_{\rm old}$ reduced bases and (b) the $\Lambda_{200k}$ reduced bases, with the exact orbital phase taken from SEOBNRv5PHM. We then calculated the frequency-domain mismatch in $\Re[h_{\lm}^P]$ against the exact modes from SEOBNRv5PHM, testing $10^4$ waveforms with parameters drawn on-the-fly from the uniform distribution in Eq.~\eqref{eq:random_polar_sample}. For each value of $(\ell, m)$ we selected the R-frame model with lowest median mismatch, which we found to be the model derived from the $\Lambda_{\rm old}$ reduced basis for the $(2,2), (2,1), (3,2)$ and $(4,4)$ modes, and the model derived from the $\Lambda_{200k}$ basis for the $(3,3)$ and $(4,3)$ modes. In our comparisons between the different reduced bases, we initially only considered the cases where the real and imaginary parts of each $(\ell, m)$ mode used the same reduced basis training set. Since we were ultimately satisfied with the accuracy of our final R-frame models (see Sec.~\ref{sec:model_performance}), we did not feel it was necessary to additionally test mixed basis combinations, or to retrain all of the neural networks to consistently use the same early stopping callback.

\subsubsection{Quaternions: $q_{J2P}(t)$}
We train separate networks to predict the values of each of the four Cartesian components of $q_{J2P}(t)$ evaluated at their respective empirical time nodes. As mentioned in Sec.~\ref{sec:data_generation}, the training and validation data were initially conditioned such that $q_w^{J2P}(\tstart) \geq  0$ for each sample. This choice was motivated by preliminary experiments with a reduced duration ($\approx 5000M$) model restricted to mass ratios $q \leq 2$, for which this choice of conditioning enabled us to train highly accurate predictive networks. Unfortunately, when we came to train the networks for the full $\approx 10^4 M$ duration, $q \leq 10$, model, we found that the networks would no longer train well, with the losses plateauing at a high level early in the training. After experimenting with different alternative strategies, it was decided to rescale all of the $q_{J2P}$ training and validation data according to
\begin{align}
    q_{J2P}(t; \lamvec) \rightarrow q_w^{J2P}(\tstart; \lamvec)\cdot q_{J2P}(t; \lamvec). \label{eq:qJ2P_recond}
\end{align}
We then trained neural networks to predict the rescaled quaternion components at their empirical time nodes, from which we can reconstruct the full (rescaled) time series $q_{J2P}(t)$ (on either $\mathcal{T}_{\rm com}$ or $\mathcal{T}_{\rm ds}$ as appropriate) by evaluating the corresponding empirical interpolants. Note that by the linearity of the empirical interpolant, there was no need to recompute the empirical interpolation matrices that were originally calculated using the non-rescaled $\Lambda_{200k}$ data set. Having computed all quaternion components, we can then re-normalize the quaternions such that $q_{J2P}(t)^*\cdot q_{J2P}(t)=1$ at each time step, inverting Eq.~\eqref{eq:qJ2P_recond} up to an insignificant sign. Unlike the original data conditioning, transformation \eqref{eq:qJ2P_recond} ensures that the rescaled $q_{J2P}(t; \lamvec)$ remain continuous in $\lamvec$ at points $\lamvec_0$ with $q_w^{J2P}(\tstart; \lamvec_0) = 0$, but the transformation is singular there. Overall this leads to significantly improved neural network performance throughout most of the parameter space, but the resulting neural networks cannot predict $q_{J2P}(t; \lamvec) = 0$ whenever $q_{J2P}(\tstart; \lamvec) = 0$, significantly imparing model accuracy in a small surrounding region of parameter space. The implications of this will be discussed in more detail in Sec.~\ref{sec:model_performance} and App.~\ref{app:mm_parameter_dependence}.

We focused our attention on optimizing the hyperparameters for the $q_x^{J2P}$ network, and later found that the same architecture and training parameters produce adequate results for all other components as well. Hyperparameters were tuned until we began to see diminishing returns and, based on interim tests, we were satisfied that the final surrogate errors due to the $q_{J2P}$ surrogate model were smaller than those caused by the orbital phase model in most circumstances (see Sec.~\ref{sec:model_performance} for a comparison of how the different data pieces contribute to the waveform error). The final network hyperparameters can be found in Table~\ref{tab:network_hyperparams_specs}. To maintain stability during training, we found it necessary to decrease the initial LR and significantly increase the minibatch size compared to the other data pieces. Without these changes, the loss tended to explode early in the training, never converging to an acceptable level of accuracy. 

\subsubsection{Quaternions: $q_{I2J}$}
Finally, we come to the task of interpolating the time-independent $I \rightarrow J$ quaternion across parameter space. We achieve this using our smallest network, consisting of the usual input layer with $7$ neurons for the standardized intrinsic parameters, $10$ hidden layers with $64$ neurons each, and an output layer with $4$ outputs, one for each of the (normalized) components of $q_{I2J}$. As in the case of $q_{J2P}$, the $q_{I2J}$ training and validation data were also reconditioned according to $q_{I2J} \rightarrow q_w^{I2J}\cdot q_{I2J}$. The complete architecture and training hyperparameters for this network are summarized in Table~\ref{tab:network_hyperparams_specs}.

\subsection{Surrogate evaluation}\label{sec:surrogate_evaluation}

\begin{figure*}[thb]
  \centering
  \includegraphics[width=0.85\linewidth]{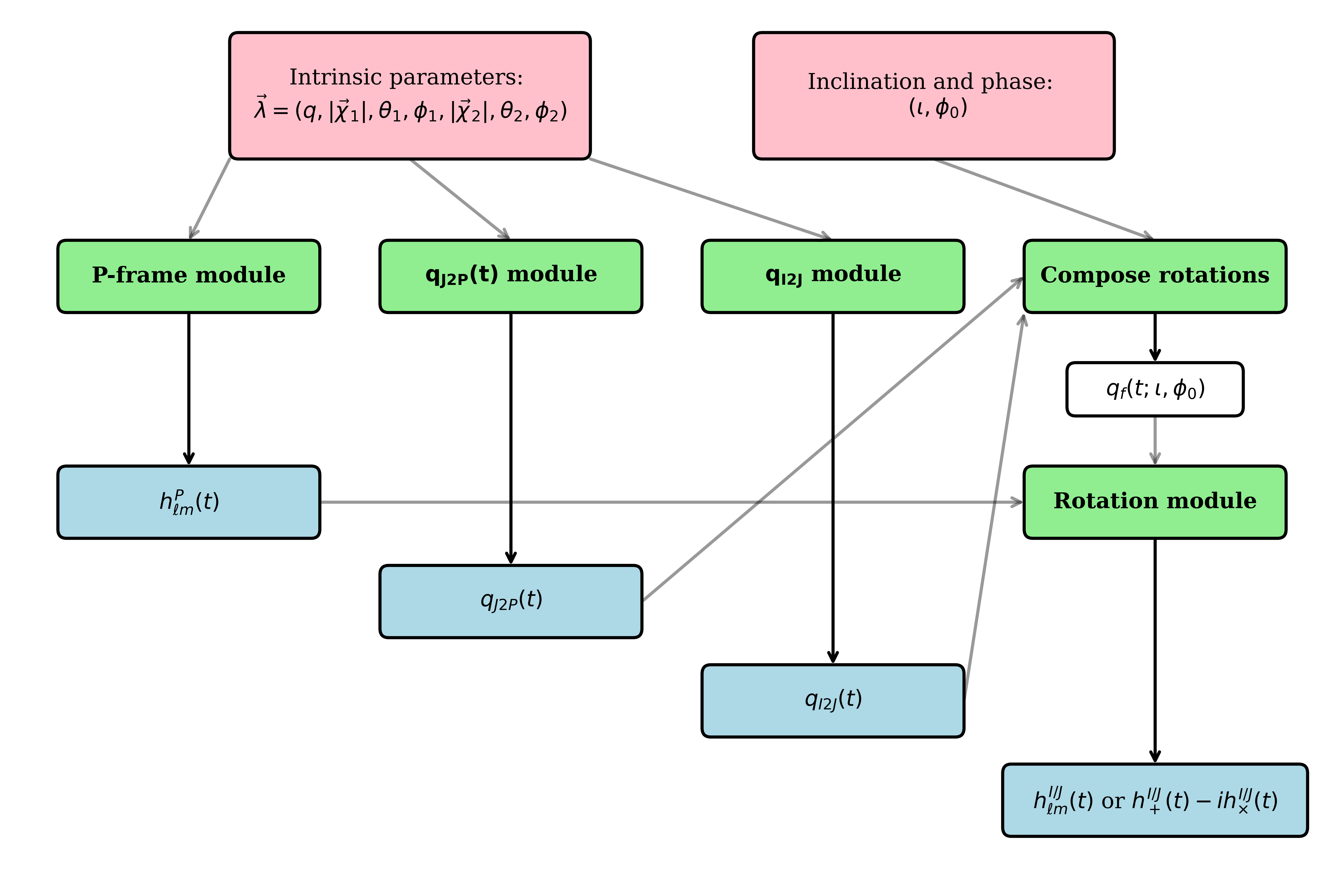}~\\~\\
  \caption{\label{fig:waveform_flowchart} Structure of \modelname, showing the main computational modules (green boxes) with their necessary inputs (gray arrows) and outputs (black arrows). Blue boxes denote possible model outputs. The main waveform generator computes the separate waveform data pieces sequentially and then combines, returning either the I- or J-frame modes $h_{\lm}^{I/J}(t)$, or the polarizations $h_{+,\times}^{I/J}(t; \iota, \phi_0)$ as desired. The P-frame modes $h_{\lm}^P(t)$, $q_{J2P}(t)$ and $q_{I2J}$ can also be computed and returned individually. The raw inputs to the model (pink) are the intrinsic parameters $\lamvec = (q, |\chivec_1|, \theta_1, \phi_1, |\chivec_2|, \theta_2, \phi_2)$, with the inclination and reference phase $(\iota, \phi_0$) also necessary when evaluating the polarizations.}
\end{figure*}

The final surrogate model is implemented entirely in \texttt{Python}, using library-agnostic \texttt{ndarray} operations for the most part to support waveform evaluation on the CPU (via \texttt{numpy} \cite{numpy}) and GPU (via \texttt{CuPy} \cite{cupy}), with some minimal architecture-dependent optimizations discussed below. Array broadcasting rules naturally enable the use of {\it batched} waveform calls, in which the polarizations for multiple sets of intrinsic parameters $\lamvec$ and sky locations $(\iota, \phi_0)$ can be computed in parallel using a single call to the waveform generator. As we shall see in Sec.~\ref{sec:model_timings}, this can lead to a significant reduction in average waveform evaluation cost, especially when running on a GPU. We save additional time by evaluating the neural networks using single precision arithmetic, which we find does not significantly impair the accuracy of our final result.

Although we made use of \texttt{StandardScaler} and \texttt{MinMaxScaler} from \texttt{scikit-learn} \cite{scikit-learn} to condition our training and validation data during neural network training, this library does not natively support GPU execution, so we replaced these scalers with our own custom implementations when evaluating the model. We also find that the native \texttt{TensorFlow} evaluation of the neural networks is slow, and found much improved performance by instead extracting the network weights and biases from the \texttt{TensorFlow} model, and then using our own custom implementation to evaluate the networks. The piecewise \texttt{elu} activation function [recall Eq.~\eqref{eq:elu_def}] was implemented on the GPU using a custom \texttt{CuPy} \texttt{ElementwiseKernel}, which proved significantly faster than a naive application of the $\texttt{cupy.where}$ array utility for piecewise functions. Further computational time was saved by combining the addition of the neuron biases and the application of the activation function into a single CUDA kernel. The evaluation of the empirical interpolants \eqref{eq:empirical_interpolant_discrete} is completely library-agnostic, using the in-built dot product utility of whichever array library is being used.

I-frame modes and polarizations are computed using Eqs.~\eqref{eq:P2I_mode_rotation} and \eqref{eq:Iframe_strain_opt} respectively, with negative $m$ modes in the P-frame obtained using Eq.~\eqref{eq:Pmode_symmetry}. The Wigner D-matrices are evaluated from the unit quaternions using \cite{Boyle:2013nka, Wigner1959}
\begin{align}
    \WigD{m}{m'}{\ell}(q) = &\sqrt{\frac{(\ell+m)!(\ell-m)!}{(\ell+m')!(\ell-m')!}}|Q_a|^{2\ell-2m}Q_a^{m+m'}Q_b^{m-m'}\nonumber\\ &\times\sum_{k=0}^{\ell+m'}(-1)^k\binom{\ell+m'}{k}\binom{\ell-m'}{\ell-k-m}\left(\frac{|Q_b|}{|Q_a|}\right)^{2k} \label{eq:WigD_formula}
\end{align}
where $Q_a := q_w + iq_z$ and $Q_b := q_y + iq_x$. Evaluating the D-matrices at each time step along the orbit is an expensive process, so the polynomial is evaluated using Horner's method for maximum efficiency. Further time is saved by computing and storing each of the combinatorial coefficients in Eq.~\eqref{eq:WigD_formula} only once when the waveform generator is initialized, and by pre-computing and storing the necessary powers of $|Q_a|$, $Q_a$ and $Q_b$ appearing in the pre-factor to the sum (for each value of $q$ required), preventing duplicate evaluations where different combinations of $(\ell, m, m')$ give rise to the same power. When evaluating $q_{J2P}(t)$ on the downsampled time grid, the D-matrices are upsampled onto $\mathcal{T}_{\rm com}$ using linear interpolation, for which we created optimized versions that support batched evaluation for both CPU and GPU execution. This includes batching over multiple waveforms as well as batching over multiple D-matrices for a single waveform.

Figure~\ref{fig:waveform_flowchart} illustrates the waveform generation process. In the P-frame module, the individual surrogate models for the orbital phase and each R-frame mode are computed sequentially for the given intrinsic parameters $\lamvec$, and then combined to give each of the ($m \geq 0$) P-frame modes $h_{\lm}^P(t;\lamvec)$ on $\mathcal{T}_{\rm com}$. The $q_{J2P}(t)$ module likewise evaluates the surrogate model for each component of $q_{J2P}$ and returns these quantities on either $\mathcal{T}_{\rm com}$ or $\mathcal{T}_{\rm ds}$ as appropriate; the value of $q_{I2J}$ is obtained from a separate module. Each of the P-frame, $q_{J2P}$ and $q_{I2J}$ modules can be called externally to return the individual data pieces, or one may call the overall waveform generation functions which evaluate the individual modules sequentially. The $P \rightarrow J$ and $J \rightarrow I$ rotations are composed (along with the rotation onto the line of sight of the observer, if relevant) and the composite quaternion $q_f(t; \iota, \phi_0)$ is passed into the rotation module along with the P-frame modes. The final output is either the inertial frame modes, or the inertial frame polarizations at a given $(\iota, \phi_0)$, in both cases evaluated on the common time grid $\mathcal{T}_{\rm com}$. The output is by default in the I-frame, but the J-frame quantities can be explicitly requested (in which case the calculation of $q_{I2J}$ is omitted).

\section{Model benchmarking}\label{sec:model_performance}
\subsection{Domain of validity: total mass}\label{sec:domain_of_validity}

Our surrogate model returns the requested waveform in geometric units on a fixed geometric time grid stretching from $t = -10^4M$ before merger until $t = 100M$ after merger. A consequence of this is that the starting frequency of the waveform is not controlled directly, and will be different for different choices of intrinsic binary parameters. This is an important consideration for gravitational wave data analysis, for which analyses typically take place with a fixed lower frequency cut-off $f_{\rm low}$ dictated by the sensitivity of the interferometer. It is important that our waveforms cover the entire period for which the signal is in the detector's sensitive frequency band, without missing the early low-frequency content, which in practice imposes a lower mass limit for which our model is valid for any given $f_{\rm low}$. 

To be more precise, the starting frequency $f_{2,2}^{\rm start}$ of the $(2,2)$ P-frame mode for a given waveform (that is, the frequency at $t = -10^4M$) is inversely proportional to the total mass of the binary,
\begin{align}
    f_{2,2}^{\rm start}(M, q, \chivec_i) = \frac{\tilde f_{2,2}^{\rm start}(q, \chivec_i)}{M}, \label{eq:f22_nodim}
\end{align}
so that $f_{2,2}^{\rm start} < f_{\rm low}$ if, and only if, 
\begin{align}
    M > \tilde f_{2,2}^{\rm start}(q, \chivec_i)/f_{\rm low}. 
\end{align}
This limit depends on all parameters $\lamvec = (q, \chivec_1, \chivec_2)$, so we find it convenient to define a spin-maximized {\it threshold mass},
\begin{align}
    \Mthr(q_{\rm max}; f_{\rm low}) := \displaystyle\max_{q \leq q_{\rm max}, \> \chivec_i} \frac{\tilde f_{2,2}^{\rm start}(q, \chivec_i)}{f_{\rm low}}.\label{eq:threshold_mass}
\end{align}
This mass has the property that, for any $q \leq q_{\rm max}$ with $M > \Mthr(q_{\rm max}; f_{\rm low})$, the dominant $(2,2)$ P-frame mode of our surrogate waveform starts below the minimum frequency cutoff $f_{\rm low}$ regardless of the spin magnitudes or directions. On the other hand, if $M < \Mthr(q_{\rm max}; f_{\rm low})$ then there exist parameter combinations $\lamvec=(q \leq q_{\rm max}, \chivec_1, \chivec_2)$ for which $f_{2,2}^{\rm start} > f_{\rm low}$, and in that instance our surrogate model will be missing some of the low-frequency content of the signal. 

To compute Eq.~\eqref{eq:threshold_mass} in practice, we extract the starting orbital frequency $\tilde\Omega^{\rm start}(q, \chivec_i) =  \pi\tilde f_{2,2}^{\rm start}(q, \chivec_i)$ in geometric units 
for each waveform in the training set $\Lambda_{1M}$, using finite differences to obtain the orbital frequency from the orbital phase. We then compute the value of 
\begin{align}
    \tilde\Omega_{\rm max}^{\rm start}(q_{\rm max}) &:= \displaystyle\max_{q \leq q_{\rm max}, \> \chivec_i}\tilde\Omega^{\rm start}(q, \chivec_i)
\end{align}
 for $18$ different values of $q_{\rm max}$ uniformly spaced between $1.5$ and $10$. The maximum is computed by finding the $99.99$th percentile value across $\Lambda_{1M}$ (subject to the mass ratio constraint) rather than the true maximum, to mitigate the impact of outliers in the numerical data. To estimate the $q_{\rm max} \rightarrow 1$ limit, we generate $2.5 \times 10^4$ waveforms with $q = 1$ and spins drawn uniformly according to Eq.~\eqref{eq:random_polar_sample} and again take the $99.99$th percentile value of the starting orbital frequency across this set. We perform a cubic least squares fit to the data,
\begin{align}
    \tilde\Omega_{\rm max}^{\rm start}(q_{\rm max}) \approx \sum_{k=0}^3 \alpha_k\left(\log q_{\rm max}\right)^k \label{eq:Omega_max_least_squares}
\end{align}
finding values
\begin{align}
    \alpha_0 &= 1.27182 \times 10^{-2}, \\
    \alpha_1 &= -7.74970 \times 10^{-5},\\
    \alpha_2 &= 1.10422 \times 10^{-3},\\
    \alpha_3 &= 7.11050 \times 10^{-5},
\end{align}
in geometric units with $G = 1 = c$, with a mean square fitting error of $5.6 \times 10^{-10}$.

\begin{figure}[tb]
  \centering
  \includegraphics[width=\linewidth]{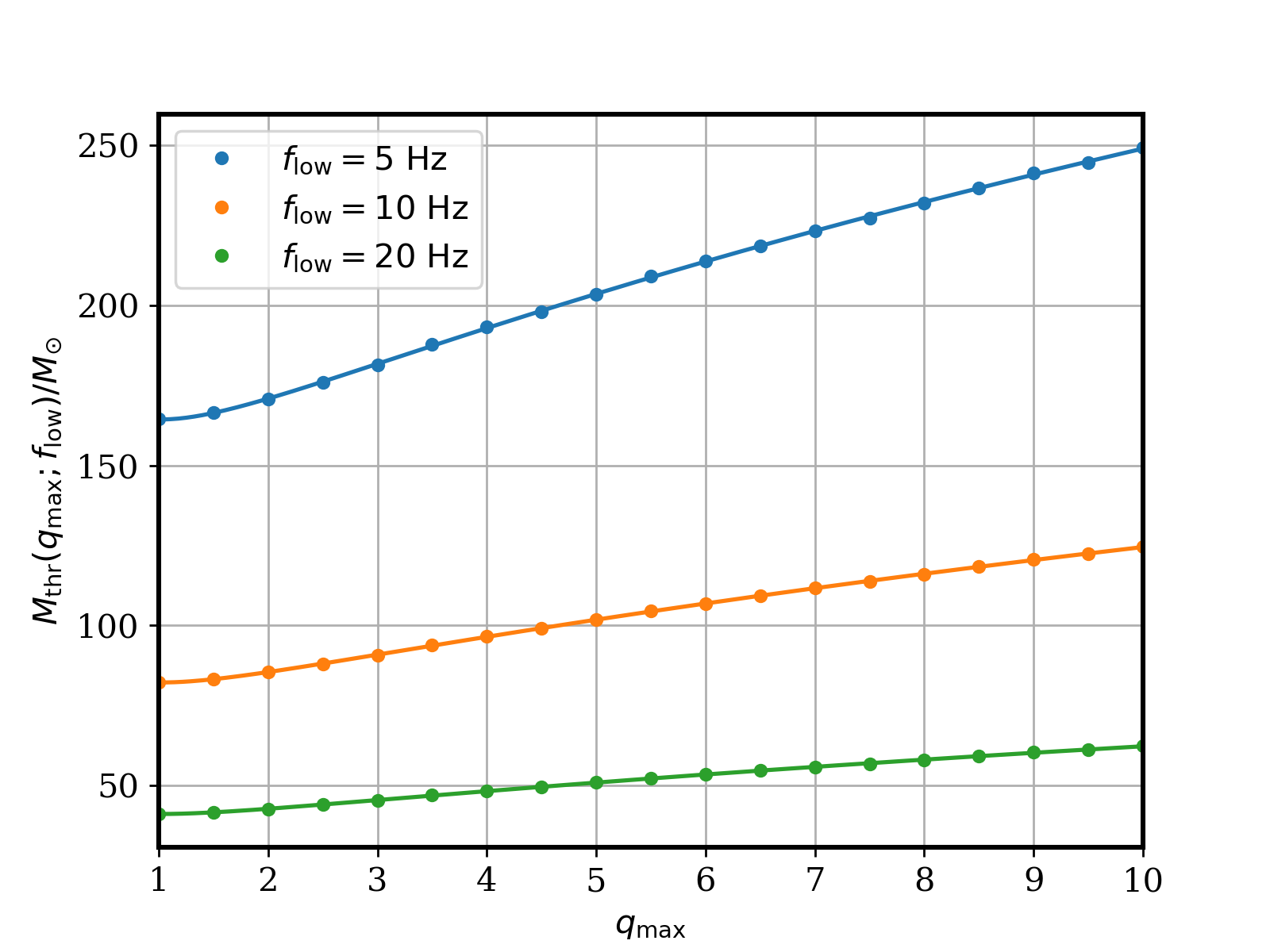}~\\~\\
  \caption{\label{fig:threshold_mass_vs_qmax} Detector frame threshold masses as a function of the maximum mass-ratio $q_{\rm max}$ under consideration, for three different choices of $f_{\rm low}$. Solid dots indicate the numerical values extracted directly from $\Lambda_{1M}$ or the dedicated $q = 1$ data set (see text), while solid lines are derived from the the least squares fit for $\tilde\Omega_{\rm max}^{\rm start}(q_{\rm max})$ given in Eq.~\eqref{eq:Omega_max_least_squares}.}
\end{figure}

Figure~\ref{fig:threshold_mass_vs_qmax} illustrates the spin-maximixed threshold mass as a function of $q_{\rm max}$ for three different low-frequency cut-offs: $f_{\rm low} = 20 \text{Hz}$, $10 \text{Hz}$ and $5 \text{Hz}$. $f_{\rm low} = 20 \text{ Hz}$ is commonly used as a lower-frequency cut-off for contemporary LVK analyses \cite{LIGOScientific:2025yae}, $10 \text{ Hz}$ is the LIGO design cut-off \cite{LIGOScientific:2014pky} and a possible cut-off for early runs of next-generation ground-based detectors \cite{Branchesi:2023mws, Evans:2023euw}, with $5 \text{ Hz}$ being an example design sensitivity target for those next-generation detectors \cite{Branchesi:2023mws, Evans:2023euw}. Solid curves denote fit values, while the dots indicate the ``exact" values obtained from the discrete numerical data for $\tilde\Omega_{\rm max}^{\rm start}$. The threshold mass grows monotonically with $q_{\rm max}$, rising from $41.1 M_{\odot}$ at $q_{\rm max} = 1$ to $62.3M_{\odot}$ at $q_{\rm max} = 10$ for $f_{\rm low} = 20 \text{ Hz}$, with threshold masses being proportionally higher for the lower values of $f_{\rm low}$ [recall Eq.~\eqref{eq:threshold_mass}]. For the planned space-based LISA observatory, assuming a low-frequency cut-off of $f_{\rm low} = 10^{-4} \text{ Hz}$ \cite{LISA:2024hlh}, the threshold mass with $q_{\rm max} = 10$ is $1.2 \times 10^7 M_{\odot}$. 

We emphasize that in the context of gravitational wave data analysis, the mass bounds illustrated in Fig.~\ref{fig:threshold_mass_vs_qmax} apply to the {\it detector frame} masses, the redshifted masses for which we evaluate the waveform templates for comparison to detector strain data. In the source frame of a binary at redshift $z$, the corresponding threshold masses are smaller by a factor of $1+z$,
\begin{align}
    \Mthr^{\rm source} = \frac{\Mthr^{\rm det}}{1+z}.
\end{align}
This effect can be significant, especially for next-generation detectors in which a significant fraction of events are expected to be observed at  redshifts $z > 1$ \cite{ET:2025xjr, Evans:2021gyd}. As an example, the spin-maximized detector frame threshold mass for an equal mass binary is $164.4 M_{\odot}$ for a hypothetical next-generation cut-off $f_{\rm low} = 5 \text{ Hz}$, corresponding to a source frame total mass of $56.7 M_{\odot}$ at the peak of star formation, $z \approx 1.9$ \cite{Madau:2014bja}.

Finally, we can compare the threshold masses of our model to the reported mass limits of NRSur7dq4 \cite{Varma:2019csw}, another fixed duration time-domain surrogate model which starts at time $\approx 4300M$ before merger and which is trained on precessing numerical relativity waveforms with mass ratios $1 \leq q \leq 4$ and spin magnitudes $|\chivec_i| < 0.8$. Recomputing our threshold masses with the added restriction $|\chivec_i| < 0.8$ in definition~\eqref{eq:threshold_mass}, we find $M_{\rm thr} = 47.5M_{\odot}$ at $q_{\rm max} = 4$ and $f_{\rm low} = 20 \text{ Hz}$\footnote{The equivalent threshold mass for our model without the restriction on spin magnitudes is $M_{\rm thr} = 48.3M_{\odot}$.}. This compares to a reported minimum mass of $66 M_{\odot}$ for NRSur7dq4 at this mass-ratio and cut-off \cite{Varma:2019csw}, showing the modest extension in total mass range provided by our longer pre-merger signal.

\subsection{Faithfulness to SEOBNRv5PHM} \label{sec:faithfulness}
Our purpose in this section is to validate the faithfulness of our surrogate model to the underlying base model, SEOBNRv5PHM. In Sec.~\ref{sec:ANN_training} we computed the validation loss against the $\Lambda_{\rm val}$ dataset when training our neural networks. Although the validation data was not used to train the neural networks directly, the validation loss was used to inform our selection of network hyperparameters, introducing a weak dependence of the final model on $\Lambda_{\rm val}$. To ensure fully independent verification of our surrogate's accuracy, we generate a new testing dataset containing $2.5 \times 10^4$ waveforms with intrinsic parameters $\Lambda_{\rm test}$ drawn uniformly in mass ratio and spin magnitudes,
\begin{align}
    q \sim U[1, 10], \quad |\chivec_i| \sim U[0,1], \label{eq:qa_uniform}
\end{align}
and {\it isotropically} in spin directions,
\begin{align}
    \cos \theta_i \sim U[-1, 1], \quad \phi_i \sim U[0,2\pi]. \label{eq:th_phi_isotropic}
\end{align}

We begin our tests by assessing the mismatch between the gravitational-wave strain \eqref{eq:strain} predicted by our surrogate and that of SEOBNRv5PHM. For each $\lamvec \in \Lambda_{\rm test}$, we compute the frequency-domain match \eqref{eq:match} for values of $(\iota, \phi_0, \psi)$ on a $17 \times 17 \times 9$ grid uniformly spaced in $\cos \iota \in [-1, 1]$, $\phi_0 \in [0, 2\pi]$ and $\psi \in [0, \pi/2]$, and then compute the orientation-averaged, SNR-weighted mismatch using Eq.~\eqref{eq:snr_weighted_mm}. We do this for both the aLIGO and ET PSDs, using lower frequency cut-offs $f_{\rm low} = 20 \text{ Hz}$ and $f_{\rm low} = 10 \text{ Hz}$ and fixed total masses $M = 65 M_{\odot}$ and $M = 125M_{\odot}$ respectively. These mass values are, in both cases, the threshold mass derived in Sec.~\ref{sec:domain_of_validity} at the respective values of $f_{\rm low}$ and $q_{\rm max} = 10$, rounded upward to the nearest multiple of $5 M_{\odot}$. 

\begin{figure}[tb]
  \centering
  \includegraphics[width=\linewidth]{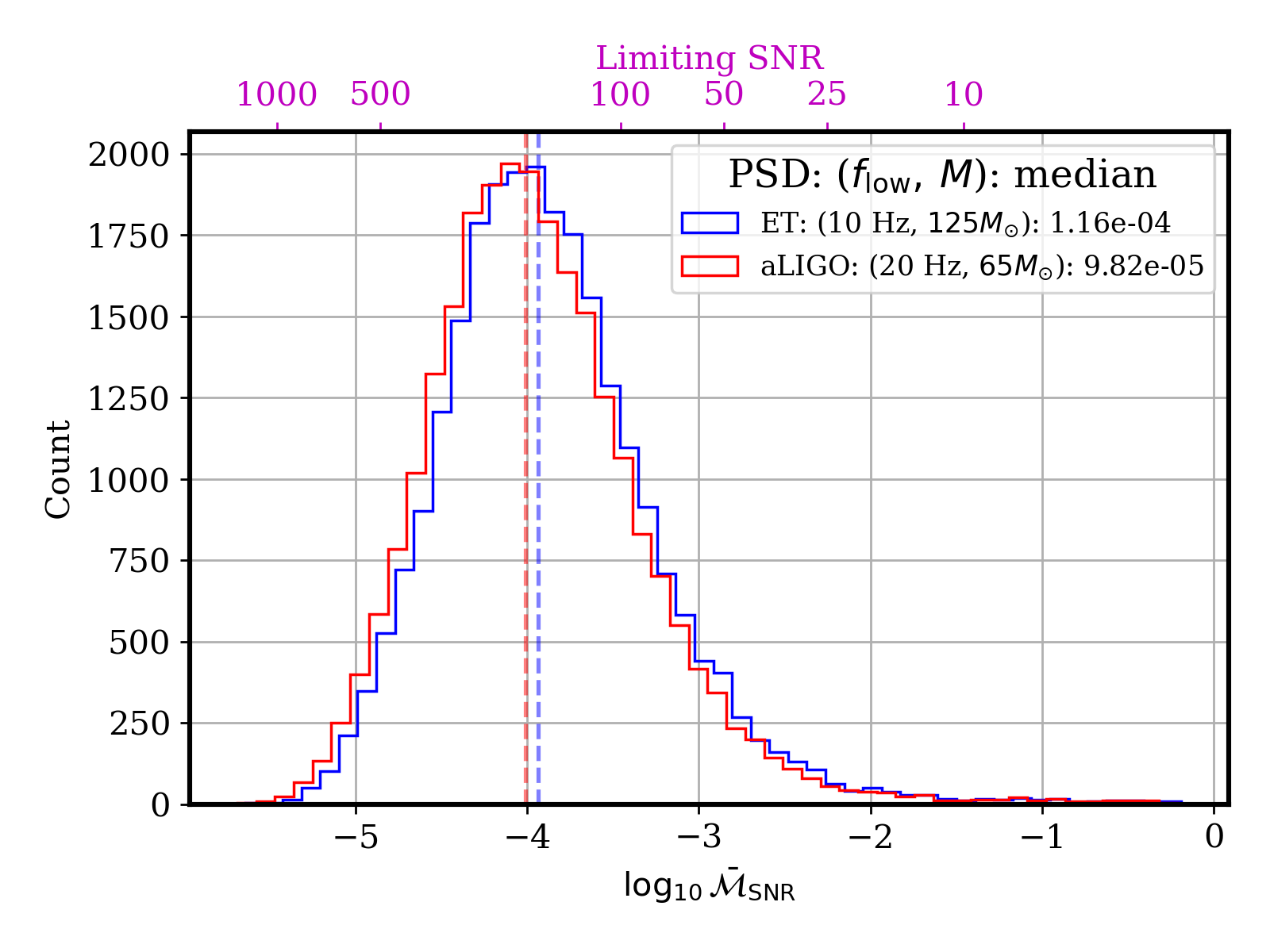}~\\~\\
  \caption{\label{fig:SNR_weighted_mismatches}SNR-weighted, $(\iota, \phi_0, \psi)$-averaged mismatches between the surrogate and SEOBNRv5PHM for the $I$-frame strain across $\Lambda_{\rm test}$ using aLIGO and ET PSDs, starting at $20 \text{ Hz}$ and $10 \text{ Hz}$ respectively.}
\end{figure}

The distribution of mismatches across the test set is displayed in Fig.~\ref{fig:SNR_weighted_mismatches}. The distribution is similar for both PSDs, with a median mismatch of $9.82 \times 10^{-5}$ for the aLIGO PSD versus $1.16 \times 10^{-4}$ for ET, with similar tails in both cases (including an extended low-probability tail that reaches mismatches greater than 0.1 in rare instances). Illustrated along the top axis of Fig.~\ref{fig:SNR_weighted_mismatches} is the limiting SNR estimate from Eq.~\eqref{eq:limiting_SNR}, assuming $N = 7$ degrees of freedom. This conservatively suggests that our surrogate will be effectively indistinguishable from SEOBNRv5PHM at SNRs greater than at least $100$ throughout the majority of parameter space for both Advanced LIGO and the Einstein Telescope, with a limiting SNR $> 18.6 \>\> (17.4)$ for $99\%$ of points in our test set using the aLIGO (ET) PSDs and these total masses. 

\begin{figure*}[tb]
  \centering
  \includegraphics[width=0.495\linewidth]{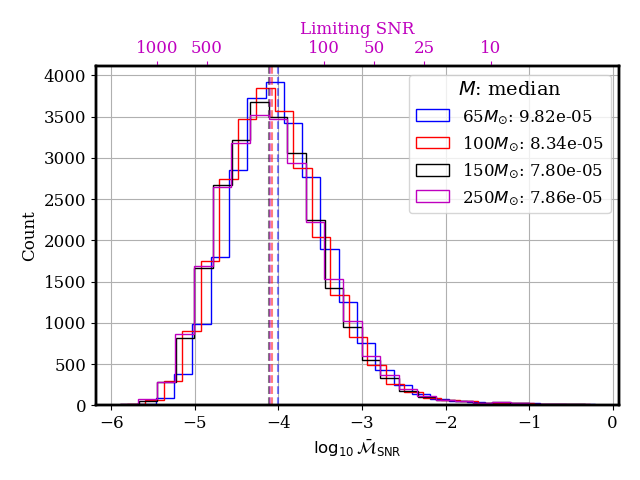}
  \includegraphics[width=0.495\linewidth]{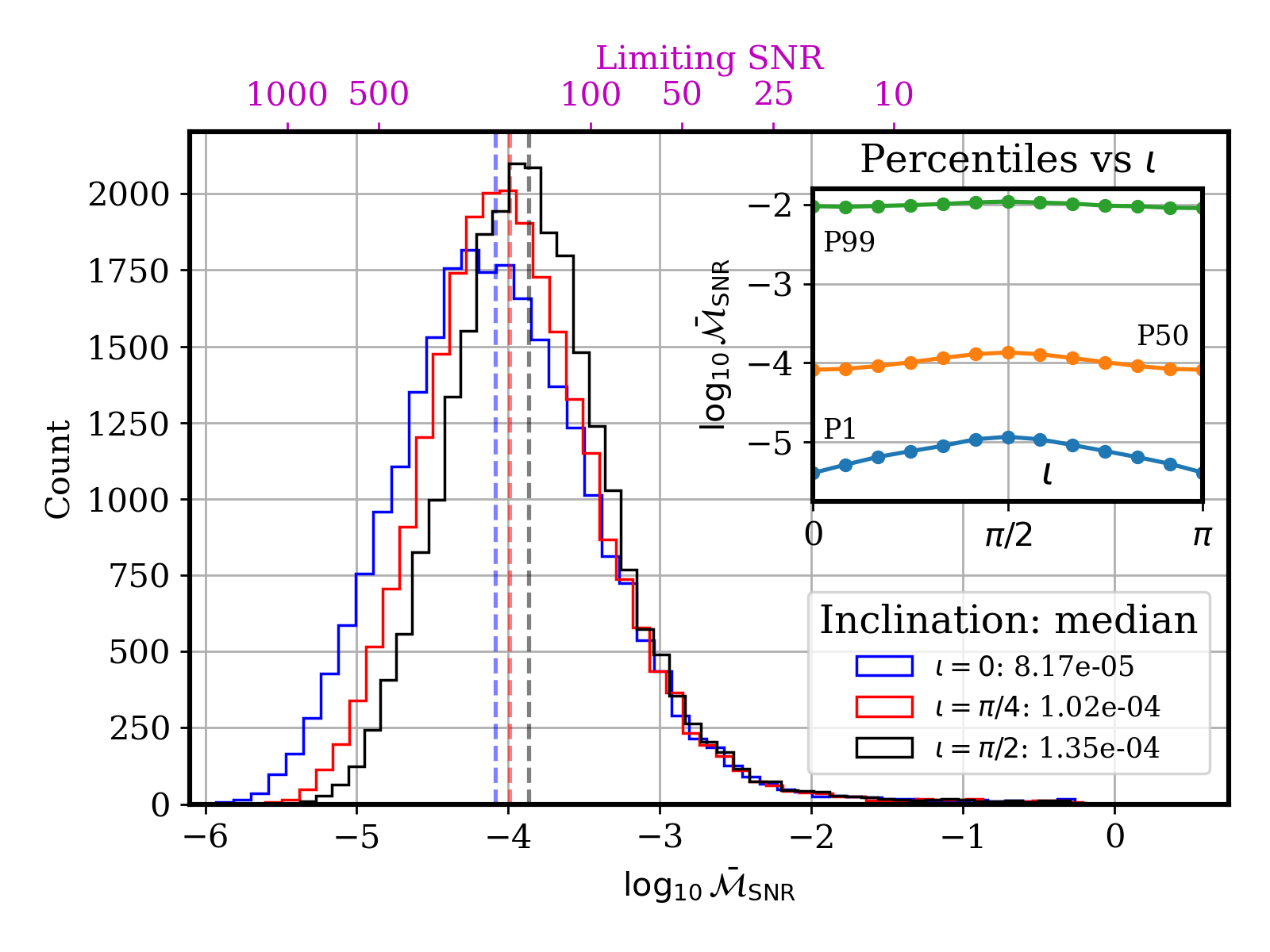}
  ~\\~\\
  \caption{\label{fig:avg_mm_mass_inc_tests}{\it Left panel:} SNR-weighted, $(\iota, \phi_0,\psi)$-averaged mismatches between the surrogate and SEOBNRv5PHM for the $I$-frame strain across $\Lambda_{\rm test}$ using the aLIGO PSD, starting at $20 \text{ Hz}$, for a range of different total masses. {\it Right panel:} SNR-weighted, $(\phi_0, \psi)$-averaged mismatches between the surrogate and SEOBNRv5PHM for the $I$-frame strain across $\Lambda_{\rm test}$ using the aLIGO PSD, starting at $20 \text{Hz}$ with $M = 65 M_{\odot}$, for a selection of different inclinations. {\it Inset:} Values of the 1st (P1), 50th (P50) and 99th (P99) percentiles of the SNR-weighted, $(\phi_0, \psi)$-averaged mismatch as a function of inclination.}
\end{figure*}

The left panel of Fig.~\ref{fig:avg_mm_mass_inc_tests} displays the distribution of $(\iota, \phi_0, \psi)$-averaged, SNR-weighted mismatches using the aLIGO PSD for different values of the total mass. The distribution is largely insensitive to the total mass, although the median shifts slightly from $9.82 \times 10^{-5}$ at $M = 65 M_{\odot}$ to a minimum of $7.80 \times 10^{-5}$ at $M = 150M_{\odot}$. The right panel  of Fig.~\ref{fig:avg_mm_mass_inc_tests}, meanwhile, displays the distribution of aLIGO mismatches across the test set for three different fixed inclinations ranging from face-on ($\iota = 0$) to edge-on ($\iota = \pi/2$). In each case, the mismatches are computed at a fixed $M = 65 M_{\odot}$, and for each $\iota$ and $\lamvec \in \Lambda_{\rm test}$ the mismatch is computed using the SNR-weighted average of the match over different values of $(\phi_0, \psi)$. The inset displays the 1st, 50th and 99th percentile values of the averaged mismatch over the test set as a function of inclination. Overall, moving from face-on to edge-on leads to an approximately $65\%$ increase in median mismatch (from $8.17 \times 10^{-5}$ to $1.35 \times 10^{-4}$), with a particularly large impact at the lowest mismatches ($185 \%$ increase in the 1st percentile value), but a much smaller impact on the large-mismatch tail ($14\%$ increase in the 99th percentile). 

We investigate the dependence of the strain mismatch on the intrinsic parameters in detail in App.~\ref{app:mm_parameter_dependence}, and note the general conclusions here. The mismatch between our surrogate and SEOBNRv5PHM is correlated with the spin magnitudes (especially that of the larger, primary black hole) and the mass ratio, with larger spins and mass ratios causing larger mismatches when averaged over all other parameters. The impact of the spin angles $\theta_i$ is less clear, although it appears that increasing misalignment between the spins and the orbital angular momentum ($|\cos\theta_i| < 1$) leads to increased mismatch at fixed (moderate-to-large) spin magnitude, particularly for the primary spin. There is no correlation between the azimuthal spin angles $\phi_i$ and the mismatch, as may be expected by symmetry.  We also investigate the parameter values $\lamvec \in \Lambda_{\rm test}$ that give the 1000 largest aLIGO mismatches in Fig.~\ref{fig:SNR_weighted_mismatches}, and note that the largest mismatches occur in regions of parameter space where either $q_w^{J2P}(\tstart)$ or $q_w^{I2J}$ is close to zero, suggesting that the singularity of the quaternion data conditioning noted in Sec.~\ref{sec:ANN_training} may be at least partially responsible for the large-mismatch tails in Fig.~\ref{fig:SNR_weighted_mismatches}.

\begin{figure}[tb]
  \centering
  \includegraphics[width=\linewidth]{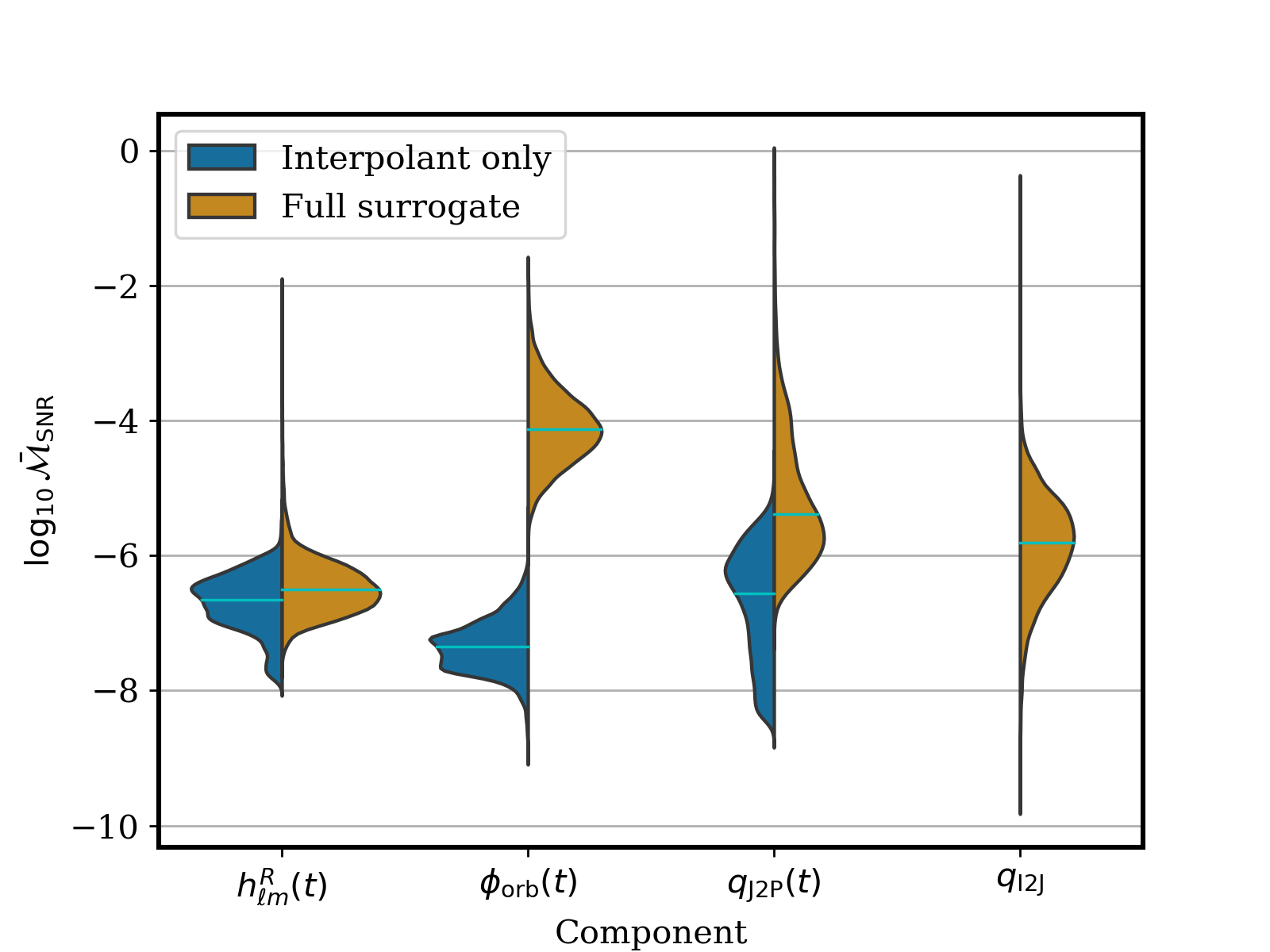}~\\~\\
  \caption{\label{fig:single_component_mismatches} SNR-weighted, $(\iota, \phi_0, \psi)$-averaged mismatches vs SEOBNRv5PHM for the I-frame strain when using the surrogate for only one component of the model at a time, with the other components obtained exactly from SEOBNRv5PHM. For the time-series components, the mismatch is computed for both the full surrogate (empirical interpolant plus predictive network) and the empirical interpolant alone with exact data at the empirical time nodes. All mismatches are computed using the aLIGO PSD, starting at $20 \text{ Hz}$ with fixed total mass $M = 65 M_{\odot}$. Median values are indicated by horizontal cyan lines.}
\end{figure}

To assess the contribution of different waveform data pieces to the final model error, we repeat the calculation of the $(\iota, \phi_0, \psi)$-averaged mismatches using the surrogate to provide only one component of the model at a time, using the exact values from SEOBNRv5PHM in place of the other components. In this context the model components are: the orbital phase surrogate, the R-frame mode surrogates (collectively), the $q_{J2P}$ surrogates and the $q_{I2J}$ surrogate. We do not consider the effects of individual R-frame mode surrogates (or individual Cartesian components of $q_{J2P}$), so we either use the surrogate for all modes or for none. Furthermore, for the time-series components we compute the mismatches twice, once using the full surrogate models (empirical interpolants with neural network predictions for the data values at empirical time nodes), and once using only the empirical interpolant (with the values at the empirical time nodes given by SEOBNRv5PHM). This allows us to disentangle the errors arising from the empirical interpolant/reduced basis from those introduced by the neural networks. The resulting mismatch distributions across the test set are illustrated in Fig.~\ref{fig:single_component_mismatches}. We see that the orbital phase surrogate produces the largest median error ($7.54 \times 10^{-5}$), although the maximum error is notably larger for the $q_{J2P}$ and $q_{I2J}$ surrogates, with a maximum mismatch of $0.618$ for the $q_{J2P}$ surrogate test. This further strengthens the link between the large-mismatch tails in Fig.~\ref{fig:SNR_weighted_mismatches} and the quaternion models.  We also note that for both the orbital phase and $q_{J2P}$, the mismatches using the empirical interpolant alone are significantly smaller than when using the full surrogate, indicating that the predictive neural network is the primary source of error. In particular, the largest mismatch when using the empirical interpolants for $q_{J2P}$ with exact data at the empirical time nodes was only $2.24 \times 10^{-5}$, indicating that the large mismatches are introduced by the neural network interpolants and subsequent re-normalization of the quaternions, and not by the reduced bases. The R-frame mode surrogates produce a significantly smaller mismatch than either the orbital phase or the different quaternion models (median $3.15 \times 10^{-7}$), and the distribution of mismatches is similar when using both the full surrogates and the empirical interpolants only. This suggests that the R-frame models are limited by the reduced bases/empirical interpolants, consistent with our observation in Sec.~\ref{sec:model_construction} that the accuracy of the final R-frame models are sensitive to which training set is used to generate the reduced bases. 

\begin{figure}[tb]
  \centering
  \includegraphics[width=\linewidth]{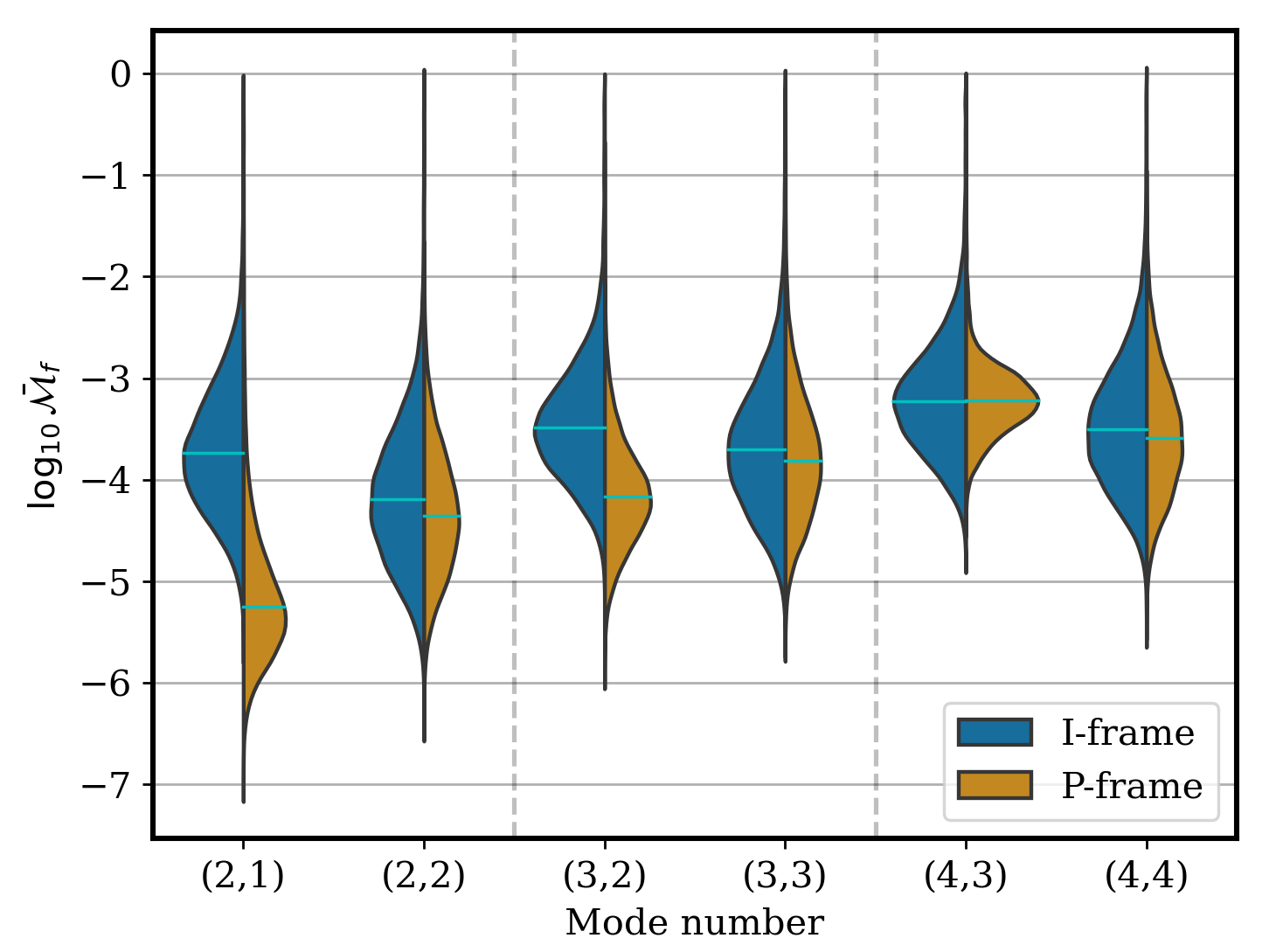}~\\~\\
  \caption{\label{fig:IPmode_mismatches}Mismatches between the surrogate and SEOBNRv5PHM for $\Re\left[h_{\lm}^X\right]$ in the $X = I$ (blue) and $P$ (green) frames across the test set $\Lambda_{\rm test}$. Mismatches computed using the aLIGO PSD starting at $20 \text{ Hz}$ with a total binary mass $M = 65 M_{\odot}$. The median value for each distribution is indicated by the horizontal cyan line.}
\end{figure}

Finally, we can also consider the mismatch in individual modes. Figure ~\ref{fig:IPmode_mismatches} displays the distribution of frequency-domain mismatches in $\Re[h_{\lm}^I]$ and $\Re[h_{\lm}^P]$ over the test set for each of the $(\ell, m)$ modes included in the P-frame, using the aLIGO PSD with $f_{\rm low} = 20 \text{ Hz}$ and $M = 65 M_{\odot}$. It should be noted that the $P \rightarrow I$ rotation given in Eq.~\eqref{eq:P2I_mode_rotation} mixes the $m$-modes for a given value of $\ell$, and in particular excites additional I-frame $m$-modes that we do not include in the P-frame, and which we do not include in the figure. Starting with the P-frame modes, the median mismatches vary from $5.67 \times 10^{-6}$ for the $(2,1)$ mode up to $6.03 \times 10^{-4}$ for the $(4,3)$ mode, with a median mismatch of $4.43 \times 10^{-5}$ for the quadrupolar $(2,2)$ P-frame mode. The mismatches are typically somewhat larger for the I-frame modes, with medians ranging from  $6.42 \times 10^{-5}$ for the $(2,2)$ mode to $5.95 \times 10^{-4}$ for the $(4,3)$ mode. Due to mode-mixing, it is not meaningful to directly compare the mismatch in I- and P-frame modes with the same values of $(\ell, m)$, but we can note that both of the $\ell = 2$ I-frame modes included in Fig.~\ref{fig:IPmode_mismatches} have a greater median mismatch than both of the $\ell = 2$ P-frame modes included in the model, indicating that the $P \rightarrow I$ frame rotation contributes significantly to the I-frame mode mismatch. The same is true for the $\ell = 3$ modes, but for $\ell = 4$ the mismatches in the I-frame and P-frame modes are broadly similar, suggesting that the error in the underlying P-frame modes is a much more significant source of error than the rotation in this case. 

We note that the tests in this section consistently used the CPU implementation of our surrogate, without the use of quaternion downsampling. We conclude in App.~\ref{app:mm_diff_configs} that the mismatches between different surrogate model configurations are several orders of magnitude smaller than the mismatches against SEOBNRv5PHM, so that the conclusions of this section do not depend on whether the surrogate waveforms are evaluated on the CPU or GPU, and whether or not downsampling is enabled.


\subsection{Waveform timing}\label{sec:model_timings}
Having validated the faithfulness of the surrogate against SEOBNRv5PHM, we now benchmark the computational saving we achieve when evaluating the polarizations $h_{+,\times}^I$. For this purpose, we introduce a set of $5000$ parameters $(\lamvec, \iota, \phi_0)$ -- collectively known as the timing set $\tilde\Lambda_{\rm time}$ -- with intrinsic parameters $\lamvec$ chosen according to Eq.~\eqref{eq:qa_uniform} and \eqref{eq:th_phi_isotropic}, and inclination and reference phase chosen isotropically,
\begin{align}
    \cos \iota \sim U[-1, 1], \quad \phi_0 \sim U[0,2\pi].
\end{align}
We focus specifically on the average wall time cost to generate a single surrogate waveform, and how this compares to an equivalent figure for SEOBNRv5PHM. The reported wall time costs are averaged over all sets of parameters in the timing set, as well as averaged over several repetitions of each waveform. When evaluating batches of waveforms, the cost per waveform is computed by normalizing the cost of the batch by its size. All GPU timings include the time to copy the final result from device (GPU) to host (accessible by the CPU) memory unless otherwise stated.  For further details of our timing methodology, see App.~\ref{app:timing_methodology}.

We find that it takes an average of approximately $65.57\text{ ms}$ to generate a single SEOBNRv5PHM waveform starting at $\tstart = -10^4M$ with grid spacing $\Delta t = 0.5M$ when running on a single thread of an Intel Core Ultra 7 155H laptop CPU (hereafter known simply as the laptop CPU). There is a small amount of uncertainty in this figure, due to the difficulty in generating SEOBNRv5PHM waveforms with fixed duration as outlined in App.~\ref{app:timing_methodology}, but we take this as our fiducial SEOBNRv5PHM cost regardless, noting that the conclusions below are not sensitive to small changes in the reference cost. 

{\setlength{\tabcolsep}{4pt}
\renewcommand{\arraystretch}{1.1}
\begin{table}[thb]
\centering
\begin{tabular}{ccccc}
\hline\hline
 &  & \multicolumn{2}{c}{Mean cost per wf [ms](speedup)}\\ 
 \cmidrule(lr){3-4}
Architecture& Batch size & With DS & Without DS\\
\hline
\multirow{2}{*}{\makecell{Intel Core \\ Ultra 7 155H}} & 1 & 12.52 (5.2x) & 15.47 (4.2x) \\
 & 250 & 7.30 (9.0x) & ---\\
\hline
\multirow{2}{*}{\makecell{Nvidia RTX \\ 1000 Ada}} & 1  & 13.47 (4.9x) & 13.01 (5.0x)  \\
     & 250  & 0.93 (70x) & 1.82 (36x)  \\
\hline
\multirow{3}{*}{\makecell{Nvidia \\ A100-40GB}}  & 1  & 25.09 (2.6x) & 24.35 (2.7x)  \\
       & 1500  & 0.16 (418x) & 0.25 (263x)  \\
 & 1500$^\dagger$ & 0.081 (813x) & --- \\
\hline\hline
\multicolumn{4}{l}{\footnotesize $^\dagger$Excluding time to copy the result from GPU to host memory.}

\end{tabular}
\caption{Average wall time cost per waveform with different computational architectures and batch sizes. Parenthetical figures give the speedup relative to the fiducial average cost of 65.57ms per SEOBNRv5PHM waveform on the Intel Core Ultra 7 155H CPU. 
} 
\label{tab:waveform_timings}
\end{table}}

We time the evaluation of the surrogate using the laptop CPU and two different GPU devices, with and without the use of quaternion downsampling, and using different waveform batch sizes. The resulting average wall time costs per waveform are given in Table~\ref{tab:waveform_timings}, along with the speedup relative to the fiducial SEOBNRv5PHM cost. Evaluating a single waveform using one thread of the same laptop CPU takes approximately 12.52ms on average with quaternion downsampling, a speedup of around $5.2\times$ compared to SEOBNRv5PHM, and faster than the 15.47ms taken to evaluate the surrogate without downsampling. Using an Nvidia RTX 1000 Ada Generation laptop GPU with 6GB VRAM, a single waveform takes 13.01ms without quaternion downsampling, slower than the downsampled CPU evaluation, but very slightly faster than GPU evaluation with downsampling (13.47ms). Although it may seem surprising that the GPU was slightly slower than the CPU in this instance, it should be noted that the advantage of the GPU lies in its ability to perform massively parallel calculations, but it is likely that our single waveform evaluation does not adequately utilize the GPU's full capacity, and hence does not achieve the maximum potential speedup. Likewise, it appears that the GPU was already able to compute all of the (non-downsampled) D-matrices so efficiently that the time saved by downsampling was minimal, and was outweighed by the cost of interpolation. It should also be noted that cross-architecture comparisons are inherently complicated by the use of two different devices, so that the fastest architecture may depend on the specific hardware in use, especially when the CPU and GPU costs are so similar.

\begin{figure}[tb]
  \centering
  \includegraphics[width=\linewidth]{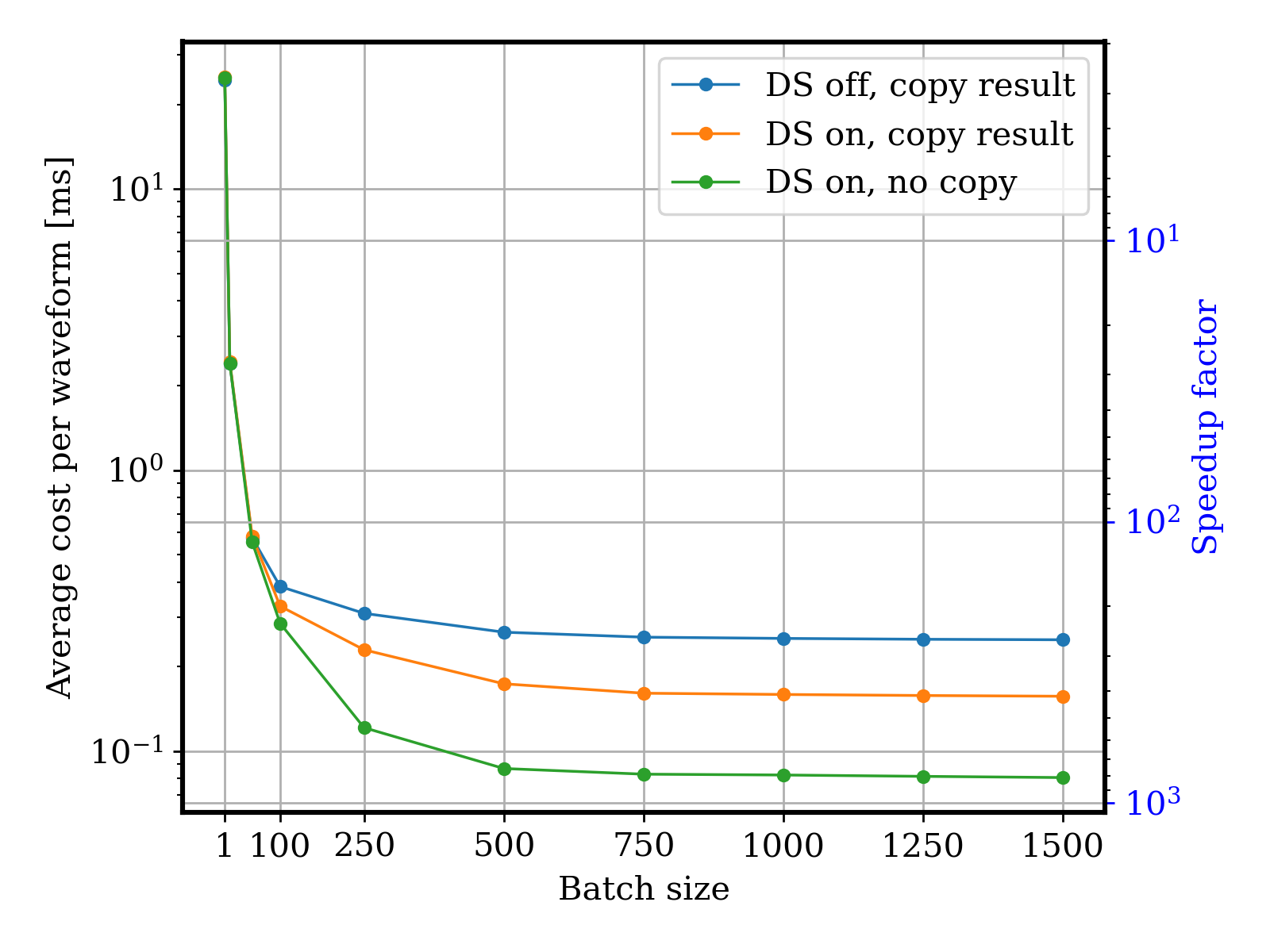}~\\~\\
  \caption{\label{fig:speedup_vs_batch_size} Average wall time per waveform as a function of batch size when evaluating waveform batches on an Nvidia A100-40GB GPU for three different configurations: downsampling off and copying the result to host; downsampling on and copying the result to host; downsampling on and leaving the result on the GPU. The right-hand axis indicates the speedup, measured against a fiducial average cost of $65.57 \text{ms}$ to generate a single SEOBNRv5PHM waveform on the laptop CPU.}
\end{figure}

The advantages of both GPU acceleration and quaternion downsampling become clear when evaluating multiple waveforms in a single batch. Averaged over a batch size of 250 waveforms and using quaternion downsampling, a single waveform takes an average of 7.30ms on the laptop CPU, compared to a mere 0.93ms on the RTX 1000 (speedups of $9\times$ and $70\times$ versus the fiducial SEOBNRv5PHM cost, respectively). Without downsampling, the average cost per waveform when evaluating a batch of 250 waveforms on the RTX 1000 is approximately double this at 1.82ms. We find even greater speedup when using a high-performance Nvidia A100 GPU with 40GB VRAM, achieving an average cost per waveform (with quaternion downsampling) of 0.16ms when amortized over a batch of size 1500, a speedup of approximately $418\times$ relative to SEOBNRv5PHM. We also highlight that until now we had been including the cost of copying the final result from the GPU back to host memory in all of our GPU timings, and we find this copy can be relatively expensive for large batches. Neglecting the cost of the final copy, the average surrogate cost is only 0.081ms per waveform when using quaternion downsampling and a batch size of 1500, a speedup of $813\times$ relative to SEOBNRv5PHM. We also include in Fig.~\ref{fig:speedup_vs_batch_size} an illustration of how the average cost per waveform on the Nvidia A100 depends on batch size, including three example configurations: downsampling disabled, downsampling enabled (both including the cost of the final copy), and downsampling enabled without performing the final copy to host. The difference between configurations is initially small, grows with batch size, and the cost rapidly plateaus in each case (at different levels) once the GPU becomes saturated. 

\begin{table}[h]
\centering
{\setlength{\tabcolsep}{8pt}
\renewcommand{\arraystretch}{1.3}
\begin{tabular}{lccc}
\hline\hline
 & CPU & GPU & GPU \\
Batch size & 1 & 1 & 250 \\
\hline
P-frame &		8.59 &	7.02 &	0.266\\
$q_{I2J}$ &		0.128 &	0.549 &	0.00225\\
$q_{J2P}(t)$ (no DS) &		3.19 &	2.07 &	0.164\\
$q_{J2P}(t)$ (DS) &		1.06 &	2.06 &	0.0216\\
Rotation (no DS) &		3.36 &	3.14 &	1.26\\
Rotation (DS) &		2.22 &	3.63 &	0.588\\
\hline\hline
\end{tabular}}
\caption{\label{tab:single_component_timings}Average wall time cost [ms] (normalized by batch size) of each major stage in the generation of the waveform polarizations: the P-frame module, the $q_{I2J}$ module, and the $q_{J2P}(t)$ and rotation modules with and without $q_{J2P}$ downsampling (DS). Columns include the times to evaluate a single parameter sample on the Intel Core Ultra 7 155H CPU, and to evaluate batches of $1$ and $250$ waveforms on the Nvidia RTX 1000 Ada GPU.}
\end{table}

Finally, in Table~\ref{tab:single_component_timings} we include the wall time cost (averaged over the timing set $\tilde\Lambda_{\rm time}$) to evaluate all of the major steps involved in generating the waveform polarizations: evaluating the orbital phase and R-frame surrogates and combining to obtain the P-frame modes $h_{\lm}^P(t)$; evaluating the models for $q_{I2J}$ and $q_{J2P}(t)$; and finally using the rotation module to construct the I-frame polarizations $h_{+,\times}^I(t)$ from the P-frame modes and quaternions. We time how long it takes to evaluate each of these steps for a single parameter value on the laptop CPU and the RTX 1000 Ada GPU, and for a batch of $250$ parameters on the GPU (as usual reporting the cost normalized by batch size). The P-frame module is the most expensive when evaluating a single waveform, on both the CPU and GPU. We also note how the use of downsampling does not significantly reduce the time to evaluate the $q_{J2P}(t)$ surrogates on the GPU when generating a single waveform, which we attribute to under-utilization of the GPU. For the batched waveform evaluation on the GPU, downsampling leads to a $7.6\times$ speedup in the evaluation of the $q_{J2P}(t)$ surrogates, and a $2.1\times$ speedup of the rotation module, although the rotation still remains the dominant cost.

\section{Applications to Bayesian inference}\label{sec:bayesian_inference}
In this section, we demonstrate the application of our surrogate model to Bayesian parameter inference using \texttt{bilby} \cite{bilby_paper}, starting with SEOBNRv5PHM waveforms injected into zero noise before analyzing several real-world gravitational wave events detected by the LVK collaboration. For all of our inference tests, we use the CPU implementation of the surrogate waveform, without quaternion downsampling. 

\subsection{EOB-injection recovery}
We performed two separate injection-recovery runs, in each case injecting an SEOBNRv5PHM waveform and recovering with our surrogate. We inject into 8s segments of data for the LIGO Hanford, LIGO Livingston and Virgo interferometers with zero noise, aligned such that merger occurs 2s before the end of the segment. We implemented our surrogate waveform as a custom time domain source model in \texttt{bilby}, and also wrote an equivalent wrapper for SEOBNRv5PHM (calling the waveform through \texttt{pyseobnr} with parameters specified at the surrogate's geometric reference orbital frequency $\Omega_{\rm ref} = 0.007/M$). These wrappers generate the time-domain polarizations on the surrogate's geometric time-grid, interpolate onto \texttt{bilby}'s requested GPS time array, and apply a Planck taper of width 0.1s to the start of the signal, and 0.01s to the end (which has already decayed significantly due to the quasinormal mode decay) to minimize artifacts when Fourier transforming into the frequency domain. We used the custom SEOBNRv5PHM wrapper to generate the injected waveform rather than the in-built \texttt{gwsignal} interface, to ensure that we were injecting and recovering like-for-like signals, with fully consistent parameter and frame conventions.  For both injection runs, the GPS merger time at the geocenter, $t_c$, was chosen to be $t_c^{\rm inj} = 1126259462.4$, and throughout this section we report the recovered discrepancy $\delta t_c = t_c - t_c^{\rm inj}$. We use the \texttt{dynesty} sampler \cite{dynesty_paper}, sampling over the (detector-frame) chirp mass $\mathcal{M}_c^{\rm det} := (m_1m_2)^{3/5}/(m_1+m_2)^{1/5}$ [where $m_1$ and $m_2$ are the detector-frame masses of each black hole], the inverse mass ratio $1/q = m_2/m_1 \leq 1$, the spin magnitudes $|\chivec_i|$, and angles $\theta_i$ and $\phi_i$, with all spin parameters specified at the geometric reference frequency $\Omega_{\rm ref} = 0.007/M$ of our surrogate model. We impose an appropriate mass constraint to ensure that our waveform starting frequency is below the lower-frequency cutoff used for the analysis, which is always taken to be $f_{\rm low} = 20 \text{ Hz}$. We also sample over the following extrinsic parameters: polarization angle $\psi$, right ascension $\alpha$, declination $\delta$, merger time at the geocenter, $t_c$, luminosity distance $d_L$, inclination $\iota$ and reference phase $\phi_0$. A full list of the priors used for each sampled parameter can be found for each run in Table~\ref{tab:pe_settings} of App.~\ref{app:PE_settings}, along with a list of the likelihood and sampler settings that were used.

{\setlength{\tabcolsep}{8pt}\begin{table}[h]
\renewcommand{\arraystretch}{1.3}
\centering
\begin{tabular}{ccc}
\hline\hline
 & Injected & Recovered \\
\hline
$\mathcal{M}_c^{\rm det}$ & 36.50 & $36.66_{-1.48}^{+1.59}$ \\
$1/q$ & 0.5 & $0.57_{-0.12}^{+0.16}$ \\
$|\chivec_1|$ & 0.7 & $0.63_{-0.28}^{+0.23}$ \\
$|\chivec_2|$ & 0 & $\leq 0.75$ \\
$\theta_1$ & $\frac{\pi}{3}\approx 1.05$ & $1.02_{-0.45}^{+0.35}$ \\
$\theta_2$ & 0 & $1.25_{-0.81}^{+1.00}$ \\
$\phi_1$ & 0 & $3.24_{-2.92}^{+2.71}$ \\
$\phi_2$ & 0 & $3.13_{-2.79}^{+2.85}$ \\
$\iota$ & $\frac{\pi}{4} \approx 0.79$ & $0.78_{-0.28}^{+0.27}$ \\
$\phi_0$ & 1.2 & $3.17_{-2.86}^{+2.82}$ \\
$\psi$ & 2.659 & $2.68_{-2.59}^{+0.39}$ \\
$\alpha$ & 1.375 & $1.374_{-0.031}^{+0.030}$  \\
$\delta$ & -1.210 & $-1.210_{-0.032}^{+0.032}$ \\
$d_L$ & 1500 & $1507_{-313}^{+242}$  \\
$\delta t_c$ & 0 & $(1.56_{-4.17}^{+3.75})\times10^{-4}$ \\
\hline\hline
\end{tabular}
\caption{\label{tab:inj1_params} Injection 1: injected and recovered parameters. Recovered parameters are given as median values with $90\%$ confidence interval. Chirp mass $\mathcal{M}_c^{\rm det}$ is measured in units of the solar mass $M_{\odot}$, luminosity distance $d_L$ in megaparsecs, geocenter time discrepancy $\delta t_c$ in seconds, and the remaining angular parameters are measured in radians. For the secondary spin magnitude, with injected value $0$, we reported the one-sided $90\%$ confidence bound.
}
\end{table}}

\begin{figure*}[tp]
  \centering
  \includegraphics[width=0.8\linewidth]{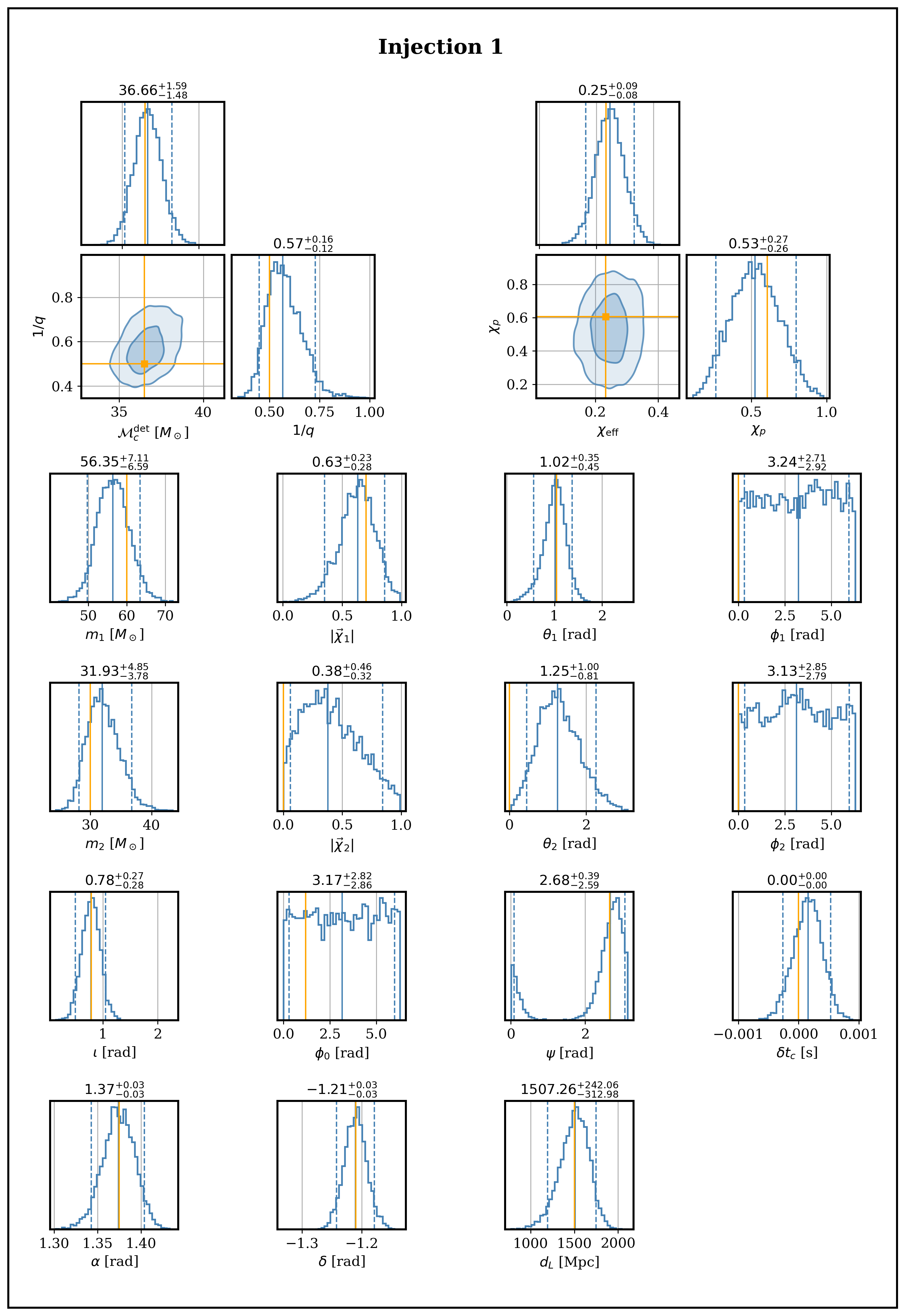}~\\~\\
  \caption{\label{fig:inj1_histograms} Recovered posterior distributions for all parameters for injection 1. All mass quantities are given in the detector frame, and time-evolving spin parameters are specified at the geometric reference frequency $\Omega_{\rm ref} = 0.007/M$. The injected parameters are highlighted with orange lines, dashed vertical lines indicate the $90\%$ confidence interval, and solid vertical lines indicate the median. The 50th and 90th percentile contours are drawn for the 2d posteriors. {\it Top row:} marginalized 1d and 2d posterior distributions on the mass parameters $(\mathcal{M}_c^{\rm det}, q)$ and effective spin parameters $(\chi_{\rm eff}, \chi_p)$. {\it Second and third row:} marginalized 1d posteriors for the intrinsic properties of the source black holes. {\it Fourth and fifth rows:} marginalized 1d posteriors on the extrinsic parameters.}
\end{figure*}

For the first injection, we choose a configuration in which a $60 M_{\odot}$ black hole with spin $|\chivec_1| = 0.7$ lying at angle $\theta_1 = \pi/3$ to the orbital angular momentum (at $\Omega_{\rm ref} = 0.007/M$) merges with a secondary non-spinning black hole of mass $30 M_{\odot}$. The injected luminosity distance is $d_L = 1500 \text{ Mpc}$, giving a network optimal SNR of 25.7. The full intrinsic and extrinsic parameters of the injected waveform are given in Table~\ref{tab:inj1_params}, along with the recovered median and $90\%$ confidence interval for each. Figure~\ref{fig:inj1_histograms} displays the marginalized 1d and 2d posteriors for the chirp mass and mass ratio, and for the effective inspiral spin $\chi_{\rm eff}$ \cite{Ajith:2009bn, Santamaria:2010yb} and effective precession spin $\chi_p$ \cite{Schmidt:2014iyl}, defined by
\begin{align}
    \chi_{\rm eff} &:= \frac{q|\chivec_1|\cos\theta_1 + |\chivec_2|\cos\theta_2}{1+q}, \label{eq:chi_eff}\\
    \chi_p &:= \max\left\{|\chivec_1|\sin\theta_1, \frac{(4+3q)}{(4q+3)q}|\chivec_2|\sin\theta_2 \right\}.\label{eq:chi_p}
\end{align}
Figure~\ref{fig:inj1_histograms} also includes full 1d marginalized posteriors for all component properties $(m_i, |\chivec_i|, \theta_i, \phi_i)$ for both black holes, plus the extrinsic parameters. We see that most parameters are well-measured, with narrow posteriors containing the injected value within the $90\%$ confidence region. The exceptions to this are the azimuthal spin angle $\phi_1$ and reference phase $\phi_0$, which have uninformative, flat posteriors, and the spin angles $\theta_2$ and $\phi_2$, which are not uniquely defined when $|\chivec_2|=0$. The secondary spin magnitude $|\chivec_2|$ is consistent with $0$ but only weakly constrained, $|\chivec_2| \leq 0.75 $ at $90\%$ confidence. 

\begin{figure*}[tp]
  \centering
  \includegraphics[width=0.8\linewidth]{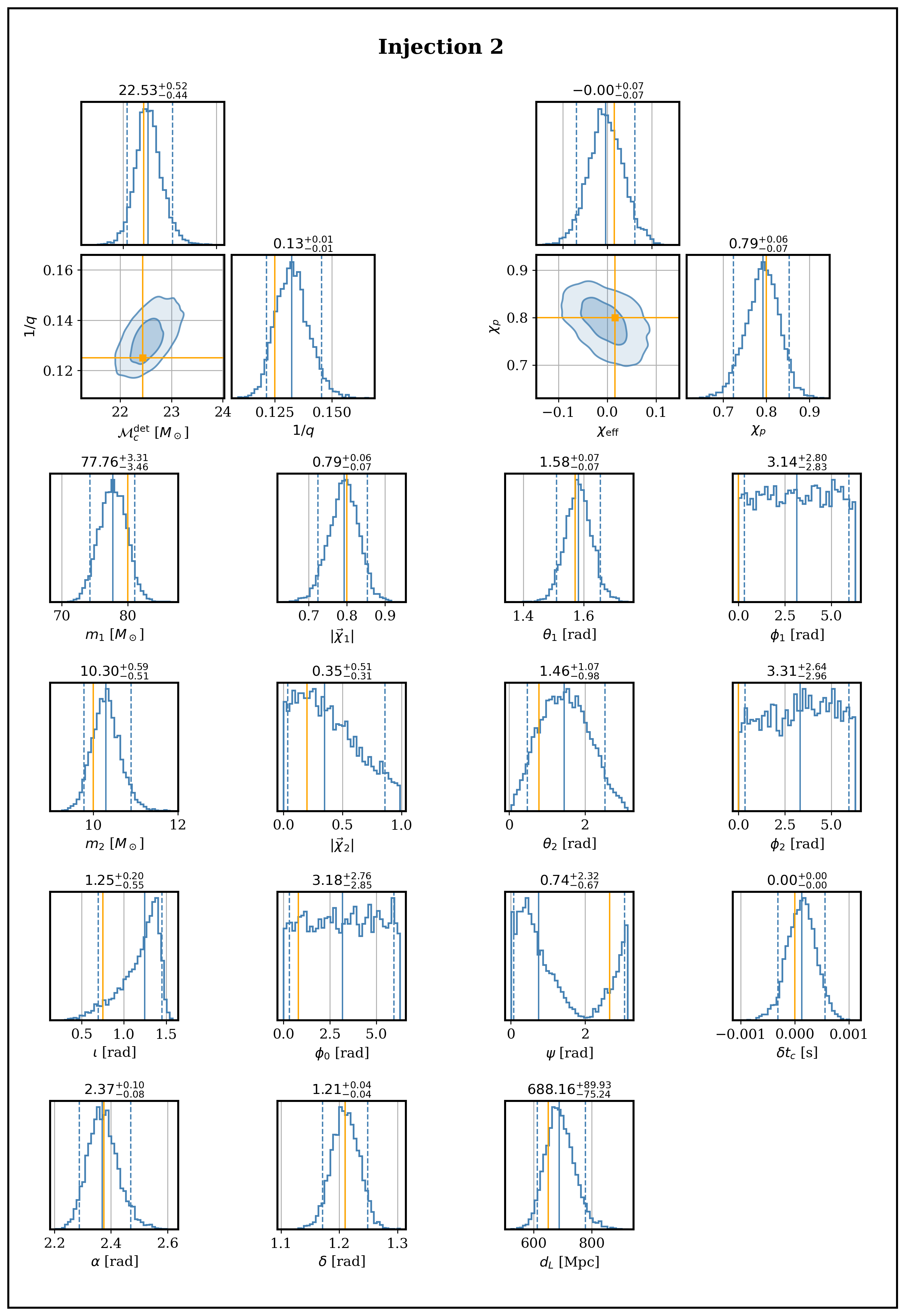}~\\~\\
  \caption{\label{fig:inj2_histograms} As Fig.~\ref{fig:inj1_histograms} but for Injection 2.}
\end{figure*}

For our second injection, we consider a significantly asymmetric, strongly-precessing binary, with a primary black hole of mass $m_1 = 80M_{\odot}$ and large spin $|\chivec_1| = 0.8$ lying entirely in the orbital plane, $\theta_1 = \pi/2$. The secondary is much less massive at $m_2 = 10 M_{\odot}$, with a low spin $|\chivec_2| = 0.2$ oriented at $\theta_2 = \pi/4$. The injected luminosity distance is $d_L = 650 \text{ Mpc}$, giving a network optimal SNR of 19.9. A complete list of the injected parameters can be found in Table~\ref{tab:inj2_params}, along with the median and $90\%$ confidence intervals of the recovered values. Figure~\ref{fig:inj2_histograms} again displays the marginalized 1d and 2d posteriors for $(\mathcal{M}_c^{\rm det},1/q)$ and $(\chi_{\rm eff}, \chi_p)$, as well as the marginalized 1d posteriors for all intrinsic and extrinsic parameters. Once again, most parameters are well-measured, and we draw particular attention to the tight measurements of the effective precessing spin, $\chi_p$ and primary spin magnitude $|\chivec_1|$ and tilt $\theta_1$. Measurements of the secondary spin magnitude and tilt are much less precise, as would be expected at this mass ratio, but consistent with the injected values. The posteriors on the azimuthal spin angles $\phi_1$ and $\phi_2$ and the reference phase $\phi_0$ remain uninformative, and the inclination posterior is notably shifted from the injected value (but still contains the injection at $90\%$ credibility). The injected mass ratio --- which also appeared slightly offset from the median value for injection 1 --- now appears more distinctly shifted from the median, but remains inside the $90\%$ confidence interval.

{\setlength{\tabcolsep}{8pt}\begin{table}[h]
\renewcommand{\arraystretch}{1.3}
\centering
\begin{tabular}{ccc}
\hline\hline
 & Injected & Recovered \\
\hline
$\mathcal{M}_c^{\rm det}$ & 22.44 &$22.53_{-0.44}^{+0.52}$ \\
$1/q$ & 0.125 & $0.132_{-0.011}^{+0.013}$ \\
$|\chivec_1|$ & 0.8 & $0.79_{-0.07}^{+0.06}$ \\
$|\chivec_2|$ & 0.2 &  $0.35_{-0.31}^{+0.51}$\\
$\theta_1$ & $\frac{\pi}{2} \approx 1.57$ & $1.582_{-0.072}^{+0.073}$ \\
$\theta_2$ & $\frac{\pi}{4} \approx 0.79$ & $1.46_{-0.98}^{+1.07}$ \\
$\phi_1$ & 0 & $3.14_{-2.83}^{+2.80}$ \\
$\phi_2$ & 0 & $3.31_{-2.96}^{+2.64}$ \\
$\iota$ & 0.75 & $1.25_{-0.55}^{+0.20}$ \\
$\phi_0$ & 0.8 & $3.18_{-2.85}^{+2.76}$ \\
$\psi$ & 2.659 & $0.74_{-0.67}^{+2.32}$ \\
$\alpha$ & 2.375 & $2.369_{-0.082}^{+0.100}$  \\
$\delta$ & -1.210 & $1.210_{-0.039}^{+0.039}$ \\
$d_L$ & 650.0 & $688_{-75}^{+90}$ \\
$\delta t_c$ & 0 & $(1.31_{-4.39}^{+4.29}) \times 10^{-4}$ \\
\hline\hline
\end{tabular}
\caption{\label{tab:inj2_params} As Table~\ref{tab:inj1_params} but for Injection 2 and including two-sided $90\%$ confidence interval for $|\chivec_2|$. 
}
\end{table}}

\subsection{Real gravitational wave events}
We now turn our attention to the analysis of three binary black hole mergers detected by the LVK collaboration: GW150914 \cite{LIGOScientific:2016aoc}, the first event ever detected; GW200129, an event that has been claimed \cite{Hannam:2021pit, Payne:2022spz} to show evidence of precession; and GW250114, the loudest binary black hole merger yet reported, with a network SNR $\sim 80$  \cite{LIGOScientific:2025rid}. We will analyze each of these events with three different waveform models: our surrogate, SEOBNRv5PHM (using the \texttt{gwsignal} interface) and IMRPhenomXPHM \cite{Pratten:2020ceb}, a fast phenomenological waveform model describing precessing binaries with higher-order modes, which shall act as our independent comparison point. 

When using the surrogate model to generate waveform templates, we continue to sample over the intrinsic parameters $(\mathcal{M}_c^{\rm det}, 1/q, |\chivec_i|, \theta_i, \phi_i)$ in addition to the extrinsic parameters $(\iota, \phi_0, \psi, t_c, \alpha, \delta, d_L)$, using the priors given in Table~\ref{tab:pe_settings} in App.~\ref{app:PE_settings}. For the SEOBNRv5PHM and IMRPhenomXPHM reference results, we sample over the angle between the line-of-sight and the total angular momentum, $\theta_{JN}$, instead of the inclination $\iota$, and also use the angles $\phi_{12} := \phi_2-\phi_1$ and $\phi_{JL}$ (the azimuthal angle of the orbital angular momentum $\vec L$ measured in the J-frame) in place of the azimuthal spin angles $\phi_1$ and $\phi_2$. The priors used for these parameters can likewise be found in Table~\ref{tab:pe_settings}. We highlight that for the SEOBNRv5PHM and IMRPhenomXPHM runs, the spins and other time-dependent quantities are specified at a constant physical waveform frequency $f_{\rm ref} = 20 \text{ Hz}$. This contrasts with our surrogate model, for which the parameters are specified at the fixed geometric orbital frequency, $\Omega_{\rm ref} = 0.007/M$, corresponding to a mass-dependent $(2,2)$-mode waveform frequency, $f_{\rm ref} = 0.007/(M\pi)$ that is approximately equal to $6.5\text{ Hz}$ 
when $M \approx 70M_{\odot}$. This reference frequency mismatch does not affect comparisons between the mass parameters $(\mathcal{M}_c^{\rm det}, 1/q)$, for example, nor for the sky location $(\alpha, \delta)$ or luminosity distance $d_L$, but it does complicate the comparison of the recovered spin parameters between the different models. To avoid the full complexity of evolving the spins to consistent reference frequencies, we will instead compare posteriors for $\chi_{\rm eff}$ and $\chi_p$, evaluated in each case at the respective waveform model's reference frequency. The effective inspiral spin $\chi_{\rm eff}$ is conserved at 2PN order under the orbit-averaged precession equations \cite{Racine:2008qv}, so that the evolution between $6.5 \text{ Hz}$ and $20 \text{ Hz}$ should be minimal, enabling meaningful comparisons despite the different reference frequency conventions. The effective precession spin $\chi_p$, on the other hand, is not conserved but exhibits oscillations on the precession timescale which typically have relatively small amplitude (see e.g. Fig.~1 of Ref.~\cite{Gerosa:2020aiw}, where Eq.~\eqref{eq:chi_p} corresponds to the blue ``Heuristic" curves). In particular, we expect that the surrogate and SEOBNRv5PHM should agree on whether $\chi_p$ is confidently greater than zero, with broadly similar median values of $\chi_p$ (although the difference between the two may be greater than for $\chi_{\rm eff}$). 

Regardless of the waveform model used, and in line with the standard LVK analyses \cite{LIGOScientific:2025yae}, our reported results are marginalized over the uncertainty in detector calibration. This is achieved by allowing frequency-dependent phase and amplitude shifts to the strain in each interferometer, described by spline models \cite{Farr2014Calib}, with the values at the spline nodes treated as sampling parameters. Using 10 spline nodes, this gives 20 extra sampling parameters per interferometer. For GW150914 we determine the priors for each calibration parameter from the in-situ calibration envelopes \cite{LIGOScientific:2018mvr} provided by the LVK collaboration, and use Gaussian priors for each node for GW200129 and GW250114 (see Table~\ref{tab:pe_settings} for further details). All of our inference runs make use of $8s$ data segments, starting $6s$ before the reported GPS time of merger, with data sourced from the Gravitational Wave Open Science Centre \cite{LIGOScientific:2025snk}. For GW150914 we use the published noise PSD for this event \cite{LIGOScientific:2018mvr}, and for GW200129 and GW250114 we estimate the PSD from the $256s$ stretch of interferometer data immediately before our analysis segment. For full details of all run settings, see Table~\ref{tab:pe_settings} in App.~\ref{app:PE_settings}. All tests in this section were performed on a cluster using Intel Xeon Platinum 8570 CPUs. 

{\setlength{\tabcolsep}{8pt}\begin{table}[h]
\renewcommand{\arraystretch}{1.3}
\centering
\begin{tabular}{cccc}
\hline\hline
 & Surrogate & SEOBNR & IMRPhenom \\
\hline
$\mathcal{M}_c^{\rm det}$ &	 $30.99^{+1.48}_{-1.40}$ &	 $31.01^{+1.56}_{-1.41}$ &	 $30.97^{+1.49}_{-1.49}$\\
$1/q$ &	 $0.88^{+0.10}_{-0.19}$ &	 $0.88^{+0.11}_{-0.17}$ &	 $0.88^{+0.11}_{-0.18}$\\
$\chi_{\rm eff}$ &	 $-0.04^{+0.11}_{-0.11}$ &	 $-0.04^{+0.11}_{-0.12}$ &	 $-0.03^{+0.11}_{-0.14}$\\
$\chi_p$ &	 $0.40^{+0.40}_{-0.31}$ &	 $0.42^{+0.41}_{-0.33}$ &	 $0.42^{+0.43}_{-0.34}$\\
$\alpha$ &	 $1.85^{+0.68}_{-0.82}$ &	 $1.59^{+0.93}_{-0.64}$ &	 $1.81^{+0.75}_{-0.80}$\\
$\delta$ &	 $-1.23^{+0.17}_{-0.06}$ &	 $-1.22^{+0.22}_{-0.06}$ &	 $-1.22^{+0.19}_{-0.06}$\\
$d_L$ &	 $504^{+124}_{-139}$ &	 $479^{+142}_{-138}$ &	 $499^{+134}_{-151}$\\
\hline
Runtime & 19hr 54m & 63hr 49m & 9hr 10m \\
\hline\hline
\end{tabular}
\caption{\label{tab:GW150914_params}Median values and symmetric $90\%$ confidence intervals for selected parameters of GW150914, recovered using the surrogate, SEOBNRv5PHM and IMRPhenomXPHM: detector-frame chirp mass $\mathcal{M}_c^{\rm det}$, inverse mass ratio $1/q$, effective inspiral and precession spins, $\chi_{\rm eff}$ and $\chi_p$, right-ascension and declination $(\alpha, \delta)$, and luminosity distance $d_L$. The final row gives the total walltime taken by each job.}
\end{table}}

\begin{figure*}[tb]
  \centering
  \includegraphics[width=0.85\linewidth]{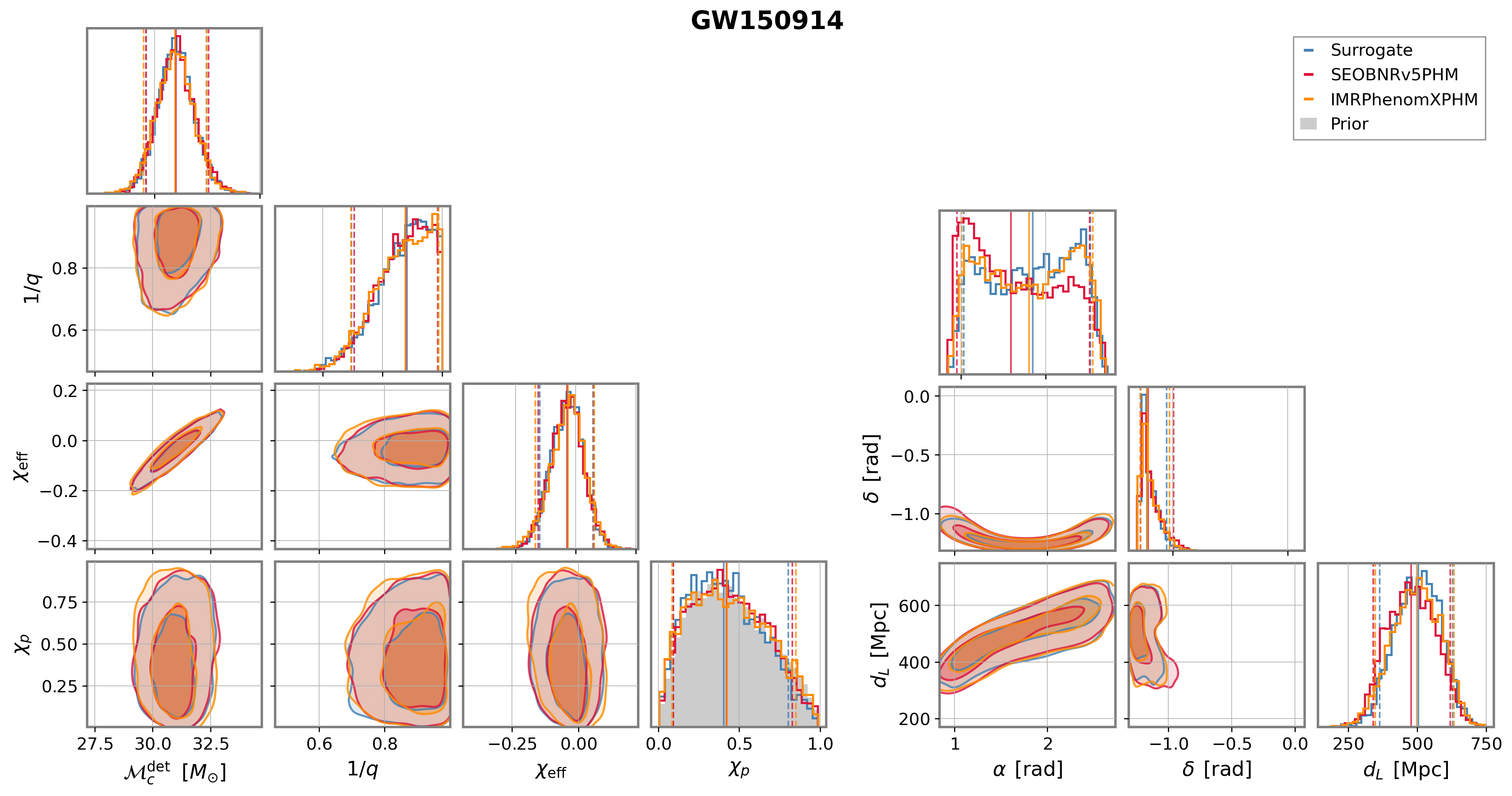}~\\~\\
  \caption{\label{fig:GW150914_corner} Selected intrinsic and extrinsic parameter corner plots for GW150914, comparing the surrogate to SEOBNRv5PHM and IMRPhenomXPHM. Dashed vertical lines in the 1d marginal posteriors ({\it diagonal}) indicate the $90\%$ confidence intervals, and solid vertical lines indicate the medians. The 50th and 90th percentile contours are drawn for the 2d marginal posteriors ({\it off-diagonal}). The prior for $\chi_p$ is shaded gray for illustrative purposes. }
\end{figure*}

\begin{figure}[tb]
  \centering
  \includegraphics[width=\linewidth]{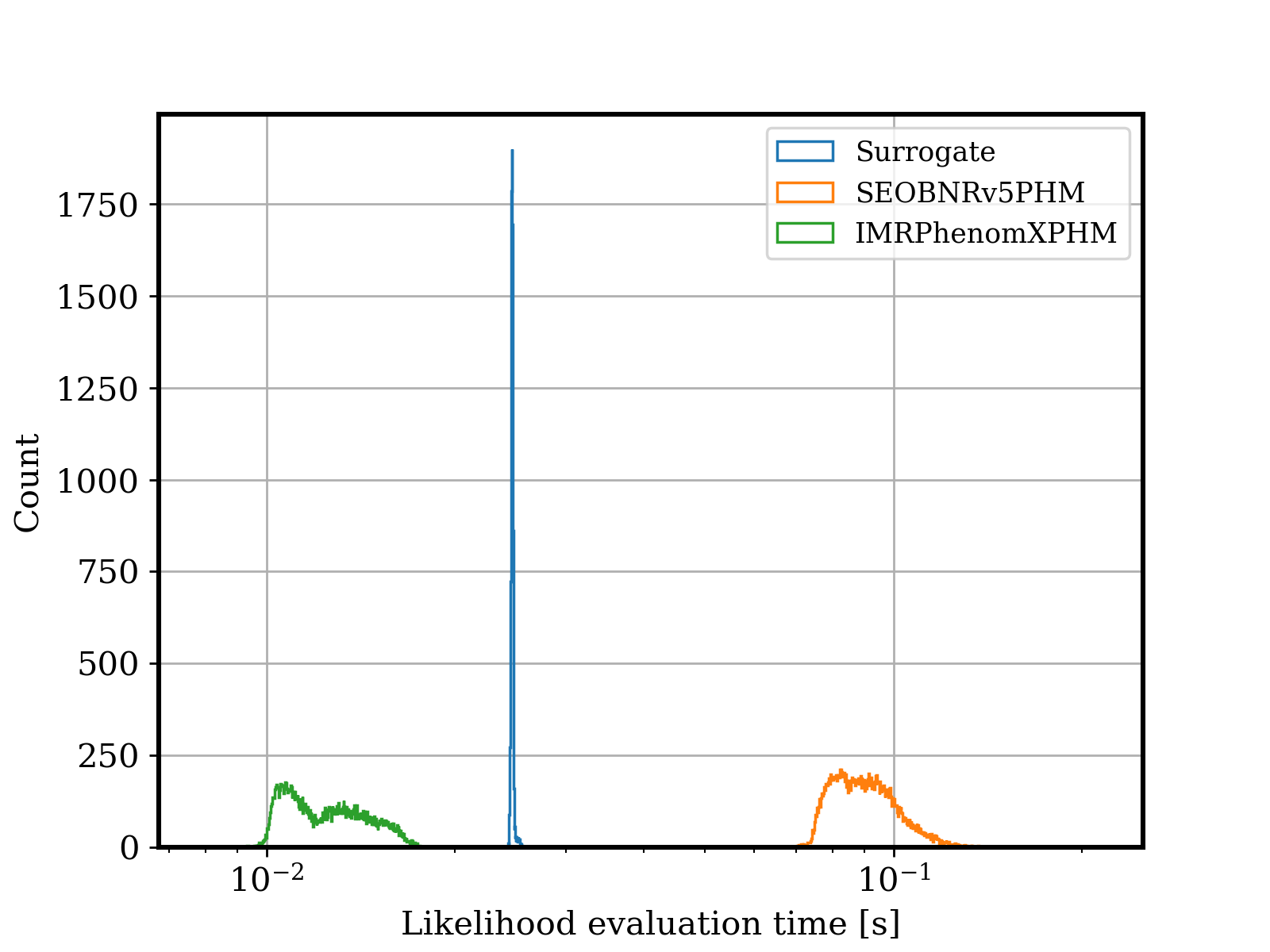}~\\~\\
  \caption{\label{fig:gw150914_likelihood_timing} Distribution of times taken to evaluate the likelihood for the different waveform models considered. Timings were performed for $2.5\times 10^4$ parameter samples drawn from the prior used in the GW150914 analysis, using the same Intel Xeon Platinum 8570 CPU used for the inference runs. }
\end{figure}

Table~\ref{tab:GW150914_params} contains the median and symmetric $90\%$ confidence intervals for the intrinsic parameters $(\mathcal{M}_c^{\rm det}, 1/q, \chi_{\rm eff}, \chi_p)$ and the extrinsic parameters $(\alpha, \delta, d_L)$ for GW150914, while Fig.~\ref{fig:GW150914_corner} shows the corner plots for the 1d and 2d marginalized posteriors for these sets. We see excellent agreement between all three waveform models, with almost identical posteriors for the intrinsic parameters in each case. Note that all three waveforms produce posteriors for $\chi_p$ that are extremely close to the prior (shaded in gray), indicating no strongly informative measurement. The confidence intervals for the extrinsic parameters are similar across waveforms, with only minor differences in posterior shape for the right ascension. The full wall time duration of the parameter estimation run was 19 hours, 54 minutes for the surrogate, $3.2\times$ faster than the run using the SEOBNRv5PHM base model (but around $2.2\times$ slower than IMRPhenomXPHM). As an additional test, we also draw $2.5\times 10^4$ sets of parameters from the prior for each waveform model, time the cost to evaluate the likelihood at those parameters, and show the resulting distribution in Fig.~\ref{fig:gw150914_likelihood_timing}. Evaluating the likelihood using the surrogate waveform model takes a mean of 24.65ms, compared to 91.79ms for SEOBNRv5PHM (around $3.7\times$ slower than the surrogate), and 12.80ms for IMRPhenomXPHM. We note also that the distribution of times is particularly narrow for the surrogate (standard deviation 0.17ms) compared to IMRPhenomXPHM (standard deviation 1.91ms), and especially SEOBNRv5PHM (standard deviation 11.28ms, maximum time 178ms). This is to be expected, since the number of operations required to evaluate the surrogate is independent of the parameters.

\begin{figure*}[tb]
  \centering
  \includegraphics[width=0.85\linewidth]{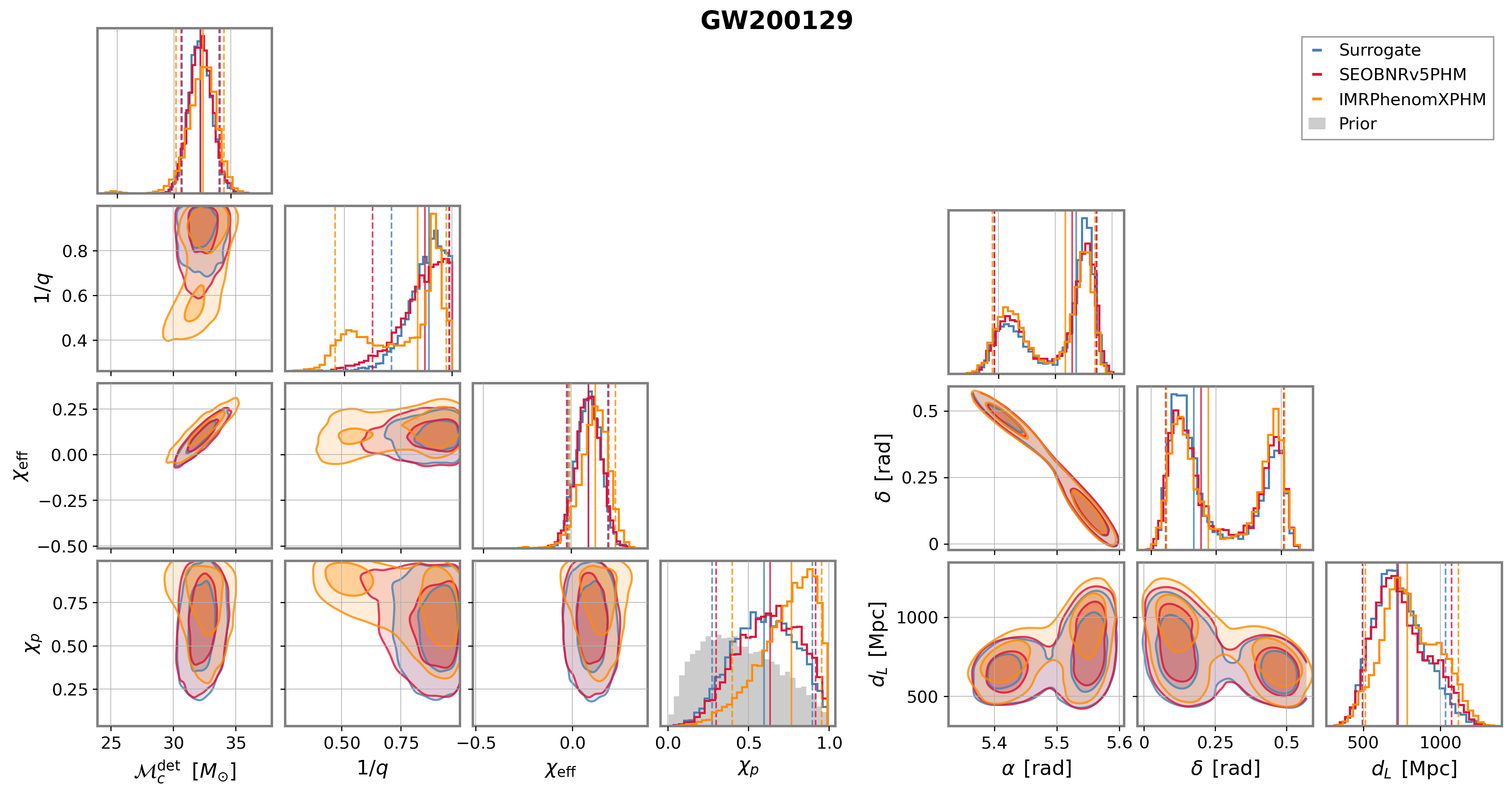}~\\~\\
  \caption{\label{fig:GW200129_corner} As Fig.~\ref{fig:GW150914_corner} but for GW200129.}
\end{figure*}

The second event of interest is GW200129, which has been claimed to show strong evidence of precession when analysed using the NRSur7dq4 \cite{Varma:2019csw} waveform model, although some studies \cite{Payne:2022spz} caution that the level of support for precession is sensitive to the approach taken to mitigate a glitch that occurred in the LIGO Livingston data stream around the time of GW200129. The original LVK analysis \cite{KAGRA:2021vkt} of this event concluded that there was evidence of strong precession and mass-asymmetry when the event was analyzed using IMRPhenomXPHM, but not when analyzed using SEOBNRv4PHM \cite{Ossokine:2020kjp} (the precursor of SEOBNRv5PHM). SEOBNRv5PHM itself has been shown to support greater values of $\chi_p$ than SEOBNRv4PHM, but still favours values smaller than IMRPhenomXPHM \cite{Ramos-Buades:2023ehm}.  Looking at the results of our runs --- corner plots in Fig.~\ref{fig:GW200129_corner} and confidence intervals in Table~\ref{tab:GW200129_params} --- we see that the surrogate and SEOBNRv5PHM are consistent for the mass ratio and $\chi_p$ (median $\approx 0.6$, distinctly shifted from the priors), but IMRPhenomXPHM shows an additional strong mode with high precession and more asymmetric mass ratio $1/q \approx 0.6$, consistent with the LVK analysis \cite{KAGRA:2021vkt}, and leading to support for greater marginal precession (median $\chi_p \approx 0.77$) and a bimodal mass ratio posterior. The chirp mass and $\chi_{\rm eff}$ posteriors are similar across all three waveform models, with differences much smaller than the widths of the confidence intervals, so that the different models can be considered consistent. In particular, the surrogate and SEOBNRv5PHM agree closely for these parameters. All three models produce essentially indistinguishable posteriors for the sky localization $(\alpha, \delta)$ and luminosity distance $d_L$. The analysis using the surrogate model took 42 hours, 37 minutes using 32 CPUs, around $2.6\times$ faster than SEOBNRv5PHM and $2.2\times$ slower than IMRPhenomXPHM. 

{\setlength{\tabcolsep}{8pt}\begin{table}[ht]
\renewcommand{\arraystretch}{1.3}
\centering
\begin{tabular}{cccc}
\hline\hline
 & Surrogate & SEOBNR & IMRPhenom \\
\hline
$\mathcal{M}_c^{\rm det}$ &	 $32.28^{+1.69}_{-1.60}$ &	 $32.35^{+1.68}_{-1.71}$ &	 $32.54^{+1.83}_{-2.38}$\\
$1/q$ &	 $0.89^{+0.09}_{-0.17}$ &	 $0.87^{+0.11}_{-0.24}$ &	 $0.84^{+0.13}_{-0.38}$\\
$\chi_{\rm eff}$ &	 $0.10^{+0.11}_{-0.12}$ &	 $0.10^{+0.11}_{-0.12}$ &	 $0.14^{+0.12}_{-0.15}$\\
$\chi_p$ &	 $0.60^{+0.30}_{-0.32}$ &	 $0.63^{+0.28}_{-0.33}$ &	 $0.77^{+0.19}_{-0.37}$\\
$\alpha$ &	 $5.54^{+0.04}_{-0.14}$ &	 $5.53^{+0.04}_{-0.14}$ &	 $5.52^{+0.05}_{-0.13}$\\
$\delta$ &	 $0.16^{+0.35}_{-0.11}$ &	 $0.19^{+0.32}_{-0.13}$ &	 $0.22^{+0.29}_{-0.16}$\\
$d_L$ &	 $717^{+315}_{-222}$ &	 $726^{+346}_{-230}$ &	 $785^{+331}_{-271}$\\
\hline
Runtime & 42hr 37m & 112hr 54m & 19hr 20m \\
\hline\hline
\end{tabular}
\caption{\label{tab:GW200129_params}As Table~\ref{tab:GW150914_params} but for GW200129.}
\end{table}}

{\setlength{\tabcolsep}{8pt}\begin{table}[h]
\renewcommand{\arraystretch}{1.3}
\centering
\begin{tabular}{cccc}
\hline\hline
 & Surrogate & SEOBNR & IMRPhenom \\
\hline
$\mathcal{M}_c^{\rm det}$ &	 $30.81^{+0.52}_{-0.48}$ &	 $30.87^{+0.61}_{-0.59}$ &	 $31.06^{+0.58}_{-0.64}$\\
$1/q$ &	 $0.96^{+0.04}_{-0.06}$ &	 $0.95^{+0.04}_{-0.07}$ &	 $0.96^{+0.04}_{-0.08}$\\
$\chi_{\rm eff}$ &	 $-0.06^{+0.05}_{-0.05}$ &	 $-0.06^{+0.05}_{-0.06}$ &	 $-0.03^{+0.05}_{-0.06}$\\
$\chi_p$ &	 $\leq 0.32$ &	 $\leq 0.39$ &	 $\leq 0.43$\\
$\alpha$ &	 $2.43^{+0.40}_{-0.13}$ &	 $2.69^{+0.35}_{-0.41}$ &	 $2.45^{+0.56}_{-0.17}$\\
$\delta$ &	 $0.34^{+0.35}_{-0.20}$ &	 $0.61^{+0.16}_{-0.51}$ &	 $0.38^{+0.39}_{-0.26}$\\
$d_L$ &	 $392^{+73}_{-59}$ &	 $399^{+75}_{-74}$ &	 $404^{+90}_{-89}$\\
\hline
Runtime$^\dagger$ & 33hr 39m & 70hr 56m & 12hr 45m \\
\hline\hline
\multicolumn{4}{l}{\footnotesize $^\dagger$Includes time to resume jobs after file system error.}
\end{tabular}
\caption{\label{tab:GW251014_params}As Table~\ref{tab:GW150914_params} but for GW250114. We report one-sided $90\%$ confidence bounds for the values of $\chi_p$, which is measured to be consistent with $0$ by all waveform models.}
\end{table}}

\begin{figure*}[tb]
  \centering
  \includegraphics[width=0.85\linewidth]{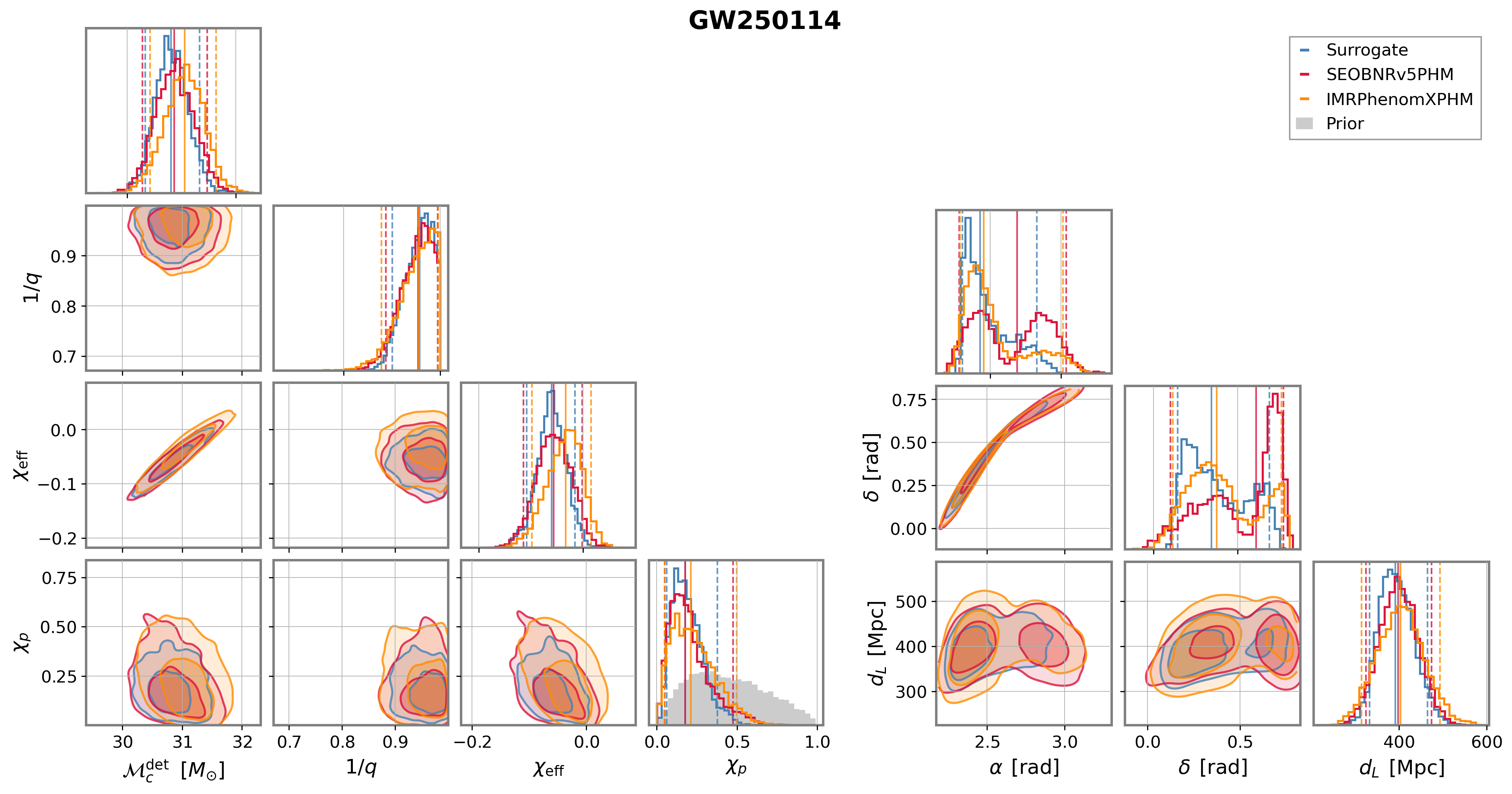}~\\~\\
  \caption{\label{fig:GW250114_corner} As Fig.~\ref{fig:GW150914_corner} but for GW250114.}
\end{figure*}

We finally consider GW250114, an exceptionally loud event with reported SNR $\approx 80$ \cite{LIGOScientific:2025rid} but properties otherwise much like those of GW150914 (SNR $\approx 24$ \cite{LIGOScientific:2016aoc}). Selected corner plots for this event are included in Fig.~\ref{fig:GW250114_corner}, with corresponding $90\%$ confidence intervals in Table~\ref{tab:GW251014_params}. There is a small but perceptible shift between the median recovered chirp mass for all three waveform models, but the shift is small compared to the size of the $90\%$ confidence intervals, and the surrogate and SEOBNRv5PHM agree more closely with each other than with IMRPhenomXPHM. There is little difference between the posteriors for inverse chirp mass. For all three waveform models the $\chi_p$ posteriors are noticeably shifted towards $0$, away from the prior (shaded gray in Fig.~\ref{fig:GW250114_corner}), indicating evidence against precession in the signal, with a bound $\chi_p \lesssim 0.32\>\>(0.39)$ at $90\%$ confidence using the surrogate model (SEOBNRv5PHM). Looking to the extrinsic parameters, the luminosity distance posteriors are similar for all three waveform models, with a greater degree of difference for the right ascension and declination. We note in particular that SEOBNRv5PHM is distinctly multimodal in $\alpha$, but the surrogate posterior does not place any great probability to the mode with $\alpha \approx 2.85$. IMRPhenomXPHM exhibits a weak second peak at this mode, but places greater weight to the $\alpha \approx 2.4$ mode detected by all three models. We do not assess in detail whether these differences are caused by genuine systematic differences between the waveform models or artifacts of our inference settings, and consider that our surrogate does not introduce significant biases for most parameters overall.  The analysis using the surrogate took 33 hours, 39 minutes of wall time using 56 CPUs,  approximately $2.1\times$ faster than the SEOBNRv5PHM run. \footnote{Note that all our GW250114 runs were interrupted by a transient file system error that caused the runs to fail when creating a progress checkpoint. The runs were resumed from the last successful checkpoint and completed without further issue, but the runtimes in Table~\ref{tab:GW251014_params} include the progress that was lost at the failed checkpoint (approximately 10 minutes of wall time) and the time taken for the run to resume (less than approximately 5 minutes).} 

\section{Conclusions}\label{sec:conclusions}
In this paper, we described the development and validation of \modelname, a reduced-order neural network surrogate of SEOBNRv5PHM trained on waveforms from binary black holes with generically oriented spins, arbitrary spin magnitudes, and mass ratios up to $q = 10$. The surrogate natively returns waveforms in the time-domain, starting at a fixed time $\tstart = -10^4M$ before merger, ensuring a $(2,2)$-mode starting frequency below $20 \text{ Hz}$ for any quasicircular binary with $ q \leq 10$ and total mass $M > 62.3 M_{\odot}$.

The full surrogate model comprises multiple individual surrogates describing the orbital phase of the binary, the mode amplitudes $h_{\lm}^R$ in the co-rotating frame, and the quaternions describing the rotations from the non-inertial co-precessing frame into the inertial source frame. The decomposition into waveform data pieces was described in Sec.~\ref{sec:wvform_decomp}, and the construction of each individual surrogate was detailed in Sec.~\ref{sec:model_construction}. We found in Sec.~\ref{sec:model_performance} that the orbital phase model contributes the largest median error out of all model components, although the quaternion models appear to cause a larger share of mismatches $\gtrsim 10^{-2}$, which we link in  App.~\ref{app:mm_parameter_dependence} to a singularity in the procedure used to condition our quaternion training data.

Overall, in Sec.~\ref{sec:faithfulness} we demonstrated that our surrogate has median orientation-averaged unfaithfulness $\approx 10^{-4}$ compared to SEOBNRv5PHM, consistent across tests using noise power spectral densities representing Advanced LIGO and the planned Einstein Telescope at  fixed total masses $M = 65 M_{\odot}$ and $M = 125M_{\odot}$ respectively, and tests using different values for the fixed total mass and inclination with the Advanced LIGO PSD. Using the conservative indistinguishability criterion given in Eq.~\eqref{eq:indistinguishability_criterion} and orientation-averaged mismatch data from Fig.~\ref{fig:SNR_weighted_mismatches}, we estimate that our surrogate is statistically indistinguishable from SEOBNRv5PHM for at least 98.6\% (81.6\%) of examples at an SNR of 25 (100) for the Advanced LIGO detectors and fixed total mass $M = 65 M_{\odot}$. We also validated the performance of our surrogate for Bayesian parameter inference in Sec.~\ref{sec:bayesian_inference}, correctly recovering the parameters of two precessing SEOBNRv5PHM waveform injections, and achieving results consistent with SEOBNRv5PHM for the analysis of three real-world gravitational-wave events: GW150914, GW200129 and GW250114. The consistency with SEOBNRv5PHM for GW250114, an event with exceptionally high $\text{SNR}\approx 80$, provides a particularly strong testament to the fidelity of our surrogate.

In Sec.~\ref{sec:model_timings} we found that a single surrogate waveform takes an average of approximately 12.52ms to generate using a representative CPU, a speedup of approximately $5\times$ compared to SEOBNRv5PHM. When amortizing the cost of generation across a batch of waveforms evaluated on a high-performance GPU, we find that the average cost-per-waveform can be as low as $81 \mu s$, a speedup of over $800\times$ compared to SEOBNRv5PHM. When analyzing real-word events in Sec.~\ref{sec:bayesian_inference}, the analysis was between $2.1\times$ and $3.2\times$ faster using our surrogate compared to equivalent runs using SEOBNRv5PHM. Notably, our inference tests did not make use of batched waveform evaluation or GPU acceleration, and we anticipate significantly greater speedup if using a sampler that supports these features.

Looking to future development, an improved rotation module that avoids the singular quaternion re-conditioning step would be helpful for reducing the tail of large mismatches against SEOBNRv5PHM. This might be achieved in practice using alternative conditioning choices for the quaternions or the use of different parameterizations for the rotation (such as the use of Euler angles). Improving the performance of the orbital phase network will also likely help reduce the bulk of the mismatches. Possible avenues for investigation include improved neural network architectures or the use of hierarchical neural networks \cite{Thomas:2022rmc, Fragkouli:2022lpt} or conditioning with post-Newtonian results to extract dominant trends in the data before training a network to recreate the more subtle residual features. The modular construction of the full surrogate model means that improved quaternion or orbital phase models can be inserted without significant changes to any other model components. With the recent addition of co-precessing frame mode asymmetries to SEOBNRv5PHM \cite{Estelles:2025zah}, it would also be natural to include this effect in our surrogate. Specifically, following the approach of Ref.~\cite{Varma:2019csw}, the primary difference would be the need to construct separate surrogates for the symmetric and anti-symmetric combinations
\begin{align}
   h_{\lm}^{\pm}(t) := \frac{h_{\ell,m}^R(t) \pm h_{\ell, -m}^R(t)}{2}, 
\end{align}
with the definition of the orbital phase adjusted to
\begin{align}
    \phi_{\rm orb}(t) := \frac{\arg\left[h_{2,2}^P(t)\right] - \arg\left[h_{2,-2}^P(t)\right]}{4}. 
\end{align}

Our surrogate is trained on waveforms with mass ratio $q \leq 10$ only, but this was chosen arbitrarily and does not represent a hard limit to our techniques. In principle, the same methodology can be used to construct surrogates that extend to higher mass ratios, but the amount of training data required will also increase. We also find that larger parameter spaces typically require larger reduced bases, which in turn require larger predictive neural networks that are harder to train and ultimately less accurate and slower to evaluate. One possible solution is to divide the parameter space into sub-regions and build smaller reduced bases/surrogates on each sub-region. This division can be chosen manually (as in Ref.~\cite{Gadre:2022sed}, for example) or using automated decomposition algorithms \cite{Cerino:2022dhr, Cerino:2023iqv}. 

As we saw in Sec.~\ref{sec:domain_of_validity}, the fixed duration of our surrogate model's time grid restricts the applicability of our model to binaries above a certain threshold in detector-frame total mass, a challenge that will only get worse with the development of next-generation ground-based detectors with improved low-frequency sensitivity. The development of long duration surrogate models is therefore highly motivated, but we find that the reduced basis size grows with signal duration, posing the same challenges noted above. It is possible that parameter space decomposition may also assist us here, but we consider that additional techniques are likely to be required. Significant breakthroughs have recently been achieved in the context of surrogate waveform modelling for {\it eccentric} non-spinning binaries, using alternative time parameterizations \cite{Maurya:2025shc, Nee:2025nmh} to build surrogates up to $\sim 10^6M$ duration in length in some cases \cite{Maurya:2025shc}. Similar re-parameterizations may also prove useful for the quasicircular precessing problem considered in this work. New time parameterizations, or non-uniform time grids such as the downsampled quaternion time grid used herein, will also be increasingly important to manage computational cost and memory requirements with long-duration waveforms. We also highlight efforts \cite{Theodoropoulos:2026wkj} to build full machine learning surrogate models that additionally utilize neural networks to compress the data pieces instead of the traditional reduced basis/empirical interpolant approach taken in this work, which may be a productive direction of study to efficiently compress long-duration signals. 

We have demonstrated that combining neural network interpolation with conventional reduced-order modelling techniques is a highly effective strategy for building gravitational waveform surrogates describing the merger of quasicircular, precessing binary black hole systems, with the faithfulness and computational efficiency needed to analyze real-world gravitational wave events. Future opportunities for development include the application to eccentric surrogate models, and possibly the development of surrogates for waveform models that include the effects of both eccentricity and spin-precession. Models incorporating both phenomena are crucial to reliably distinguish between the two effects in real-world observations \cite{Romero-Shaw:2022fbf, Morras:2025nlp, Morras:2025xfu}, but the presence of multiple orbital timescales in addition to the precession timescale and the excitation of many higher-order waveform modes inevitably makes such models slower to evaluate~\cite{Morras:2025nlp}. Fast surrogate models would therefore be of great benefit, but to develop them we must overcome the even greater parameter space dimension that comes with the inclusion of eccentric degrees of freedom. With ongoing development, we anticipate that machine learning techniques will prove their ability to handle problems with more than the 7 dimensions of our quasicircular precessing model, and consequently have the potential to play a major role in the development of the accurate, physically extensive, and fast waveform models needed to fulfill the requirements of gravitational wave detectors in the coming years.

\begin{acknowledgments}
We are grateful for helpful advice from Qian Hu on Bilby integration, Rahul Dhurkunde on PyCBC integration, and Duncan MacLeod on software development and packaging practices. We would like to thank all members of the broader Work Package 6 of the UK 3G Infrastructure Grant for their useful questions and feedback throughout the preparation of this work. 
We thank H\'{e}ctor Estell\'{e}s, Jacopo Tissino, and Prayush Kumar for useful comments on the manuscript.
C.W. and G.P. acknowledge support from the STFC via Grant No.~ST/Y00423X/1. G.P. is very grateful for support from a Royal Society University Research Fellowship URF{\textbackslash}R1{\textbackslash}221500 and RF{\textbackslash}ERE{\textbackslash}221015, and a UK Space Agency grant UKRI/ST/B000971/1.
The computations described in this paper were performed using the University of Birmingham's BlueBEAR HPC service, which provides a High Performance Computing service to the University's research community. 

This research has made use of data or software obtained from the Gravitational Wave Open Science Center (gwosc.org), a service of the LIGO Scientific Collaboration, the Virgo Collaboration, and KAGRA. This material is based upon work supported by NSF's LIGO Laboratory which is a major facility fully funded by the National Science Foundation, as well as the Science and Technology Facilities Council (STFC) of the United Kingdom, the Max-Planck-Society (MPS), and the State of Niedersachsen/Germany for support of the construction of Advanced LIGO and construction and operation of the GEO600 detector. Additional support for Advanced LIGO was provided by the Australian Research Council. Virgo is funded, through the European Gravitational Observatory (EGO), by the French Centre National de Recherche Scientifique (CNRS), the Italian Istituto Nazionale di Fisica Nucleare (INFN) and the Dutch Nikhef, with contributions by institutions from Belgium, Germany, Greece, Hungary, Ireland, Japan, Monaco, Poland, Portugal, Spain. KAGRA is supported by Ministry of Education, Culture, Sports, Science and Technology (MEXT), Japan Society for the Promotion of Science (JSPS) in Japan; National Research Foundation (NRF) and Ministry of Science and ICT (MSIT) in Korea; Academia Sinica (AS) and National Science and Technology Council (NSTC) in Taiwan.
\end{acknowledgments}

\appendix
\section{Additional faithfulness tests}
We collect additional tests concerning the mismatch between our surrogate and SEOBNRv5PHM in this appendix.
\label{app:extra_mm_tests}

\subsection{Dependence on intrinsic parameters}\label{app:mm_parameter_dependence}

\begin{figure*}[tb]
  \centering
  \includegraphics[width=0.75\linewidth]{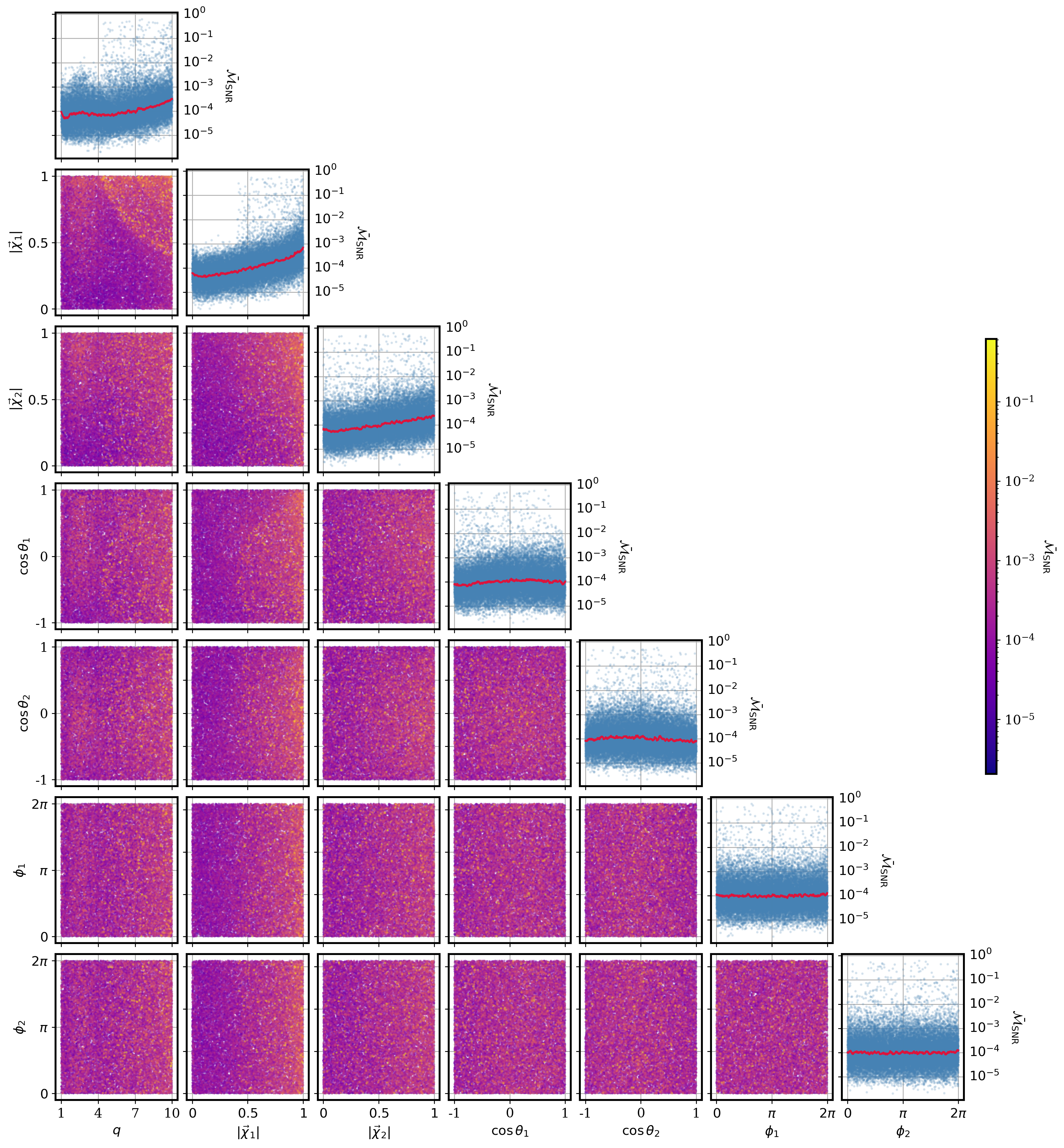}~\\~\\
  \caption{\label{fig:mm_corner} Plot of the SNR-weighted, $(\iota, \phi_0, \psi)$-averaged mismatch between the surrogate and SEOBNRv5PHM as a function of $q$, $|\chivec_i|$, $\cos\theta_i$ and $\phi_i$ ({\it diagonal}) and their unique pairwise combinations ({\it off-diagonal}). The red curves in the diagonal 1d plots show the median over all other parameters computed over a narrow rolling bin in the parameter of interest. For the off-diagonal 2d plots, the value of the mismatch is indicated by the color of the point. The mismatch data displayed in this figure is the same as the aLIGO mismatch data displayed in Fig.~\ref{fig:SNR_weighted_mismatches}, which uses $f_{\rm low} = 20 \text{ Hz}$ and $M = 65 M_{\odot}$. }
\end{figure*}

In this section we explore how the mismatch between our surrogate and SEOBNRv5PHM depends on the intrinsic parameters $\lamvec = (q, |\chivec_1|, \theta_1, \phi_1, |\chivec_2|, \theta_2, \phi_2)$, considering as a particular example the SNR-weighted, $(\iota, \phi_0, \psi)$-averaged mismatches using the aLIGO PSD illustrated in Fig.~\ref{fig:SNR_weighted_mismatches}. Figure~\ref{fig:mm_corner} displays the same mismatches as a function of the intrinsic parameters. The diagonal scatter plots show the mismatch versus each of the individual intrinsic parameters, with each point representing a single sample from the test set $\Lambda_{\rm test}$. The red line denotes the rolling median over all the other parameters, computed across a narrow bin in the parameter of interest. The off-diagonal plots display the location of the test set samples in the 2d plane of the given pair of parameters, with the value of $\overline{\mathcal{M}}_{SNR}$ for that sample denoted by the color of the point. We use the cosine of the spin angles $\theta_i$ [which, recalling Eq.~\eqref{eq:th_phi_isotropic}, are uniformly distributed across $\Lambda_{\rm test}$] instead of the angles themselves to simplify the interpretation of the figure. We notice that the median mismatch clearly grows with the spin magnitudes $|\chivec_i|$ and mass ratio $q$. This is visible from the 1d plots, but we can also clearly see the increased mismatches in the $q \rightarrow 10$, $|\chivec_1| \rightarrow 1$ corner of the $(q, |\chivec_1|$) plane, for example. The median mismatches as a function of the $\cos\theta_i$ appears broadly flat, although a close inspection suggests there may be slight peaks around $\cos \theta_i \approx 0$, corresponding to spins lying entirely in the orbital plane. Looking at the $(|\chivec_1|, \cos\theta_1)$ plane, we can see that the increase in mismatch with $|\chivec_1|$ is non-uniform, with a possible ``bulge" of increased mismatch reaching down to spins as low as $|\chivec_1| \approx 0.5$ at $\cos\theta_1 \approx 0$. This suggests that increasing misalignment between the primary spin and the orbital angular momentum leads to larger mismatches at fixed primary spin magnitude, at least when the spin magnitude is sufficiently large. A similar (but weaker) correlation may also exist for the secondary object. The azimuthal spin angles $\phi_i$ appear to be completely uncorrelated with the mismatch. 

\begin{figure}[tb]
  \centering
  \includegraphics[width=\linewidth]{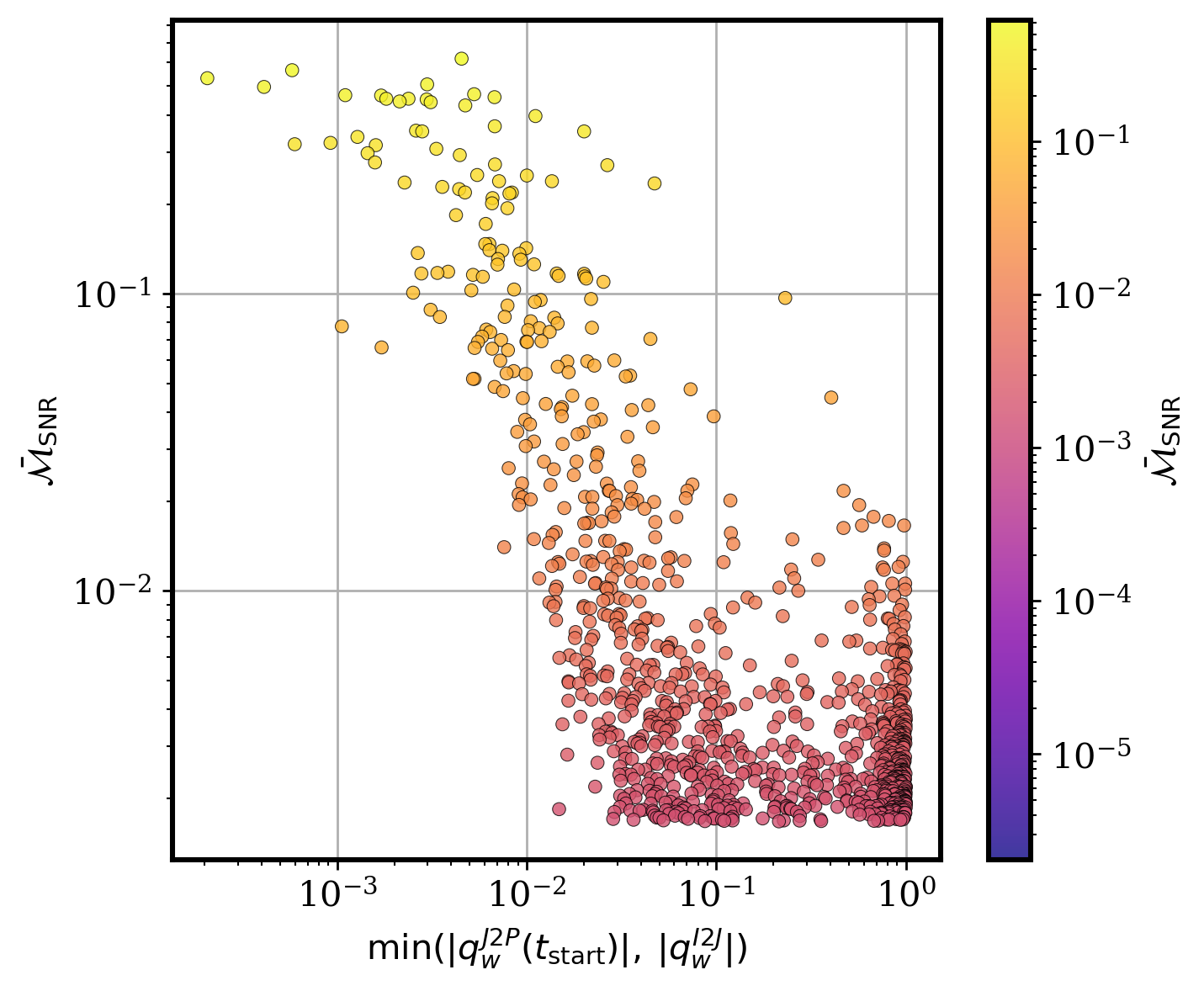}
  
  \caption{\label{fig:large_mismatches}Scatter plot of the values of the 1000 largest aLIGO mismatches from Fig.~\ref{fig:SNR_weighted_mismatches} against the corresponding values of $\min(|q_w^{J2P}(\tstart)|, |q_w^{I2J}|)$. The value of the mismatch is additionally highlighted by the color of the point.}
\end{figure}

We are also interested in understanding the origin of the tail of large mismatches observed in Fig.~\ref{fig:SNR_weighted_mismatches}. From Fig.~\ref{fig:mm_corner} we see that almost all points with mismatch $> 10^{-2}$ have $q > 4$ and $|\chivec_1| \gtrsim 0.5$, although it is not clear why these points specifically have such large mismatches when other, similarly highly spinning or asymmetric binaries result in much smaller mismatches. One possibility is the known flaw of the quaternion networks noted in Sec.~\ref{sec:ANN_training}. Recall that we reconditioned the training data for the $q_{J2P}$ and $q_{I2J}$ neural networks by rescaling
\begin{align}
    q_{J2P} \rightarrow q_w^{J2P}(\tstart)\cdot q_{J2P}(t), \quad q_{I2J} \rightarrow q_w^{I2J}\cdot q_{I2J}. \label{eq:quat_recond}
\end{align}
The neural networks are trained to predict these rescaled quaternions, with the unit quaternions recovered by re-normalizing the output of the networks. Although rescaling improved the global conditioning of the training data, allowing us to build a quaternion model that is highly accurate throughout most of the parameter space, it also introduced local difficulties in regions where the scale factors $q_w^{J2P}(\tstart)$ or $q_w^{I2J}$ become small, which suppresses the magnitude of the training data at these points (reducing their influence on training), and causes the inverse transformation to become singular. Even when the scale factors are small but non-zero, scaling by the large inverse scale factor magnifies the absolute errors in the quaternions, reducing the accuracy of the model. In Fig.~\ref{fig:large_mismatches}, we plot the value of $\min(|q_w^{J2P}(\tstart)|, |q_w^{I2J}|)$ against the mismatch for the largest $1000$ aLIGO mismatches in Fig.~\ref{fig:SNR_weighted_mismatches}.  There is a clear association between the smallest scale factor and the mismatch, and the vast majority of points with mismatch $> 10^{-2}$ have either $|q_w^{J2P}(\tstart)| < 0.1$ or $|q_w^{I2J}| < 0.1$.

\subsection{Equivalence of different model configurations}\label{app:mm_diff_configs}

\begin{figure}[t]
  \centering
  \includegraphics[width=\linewidth]{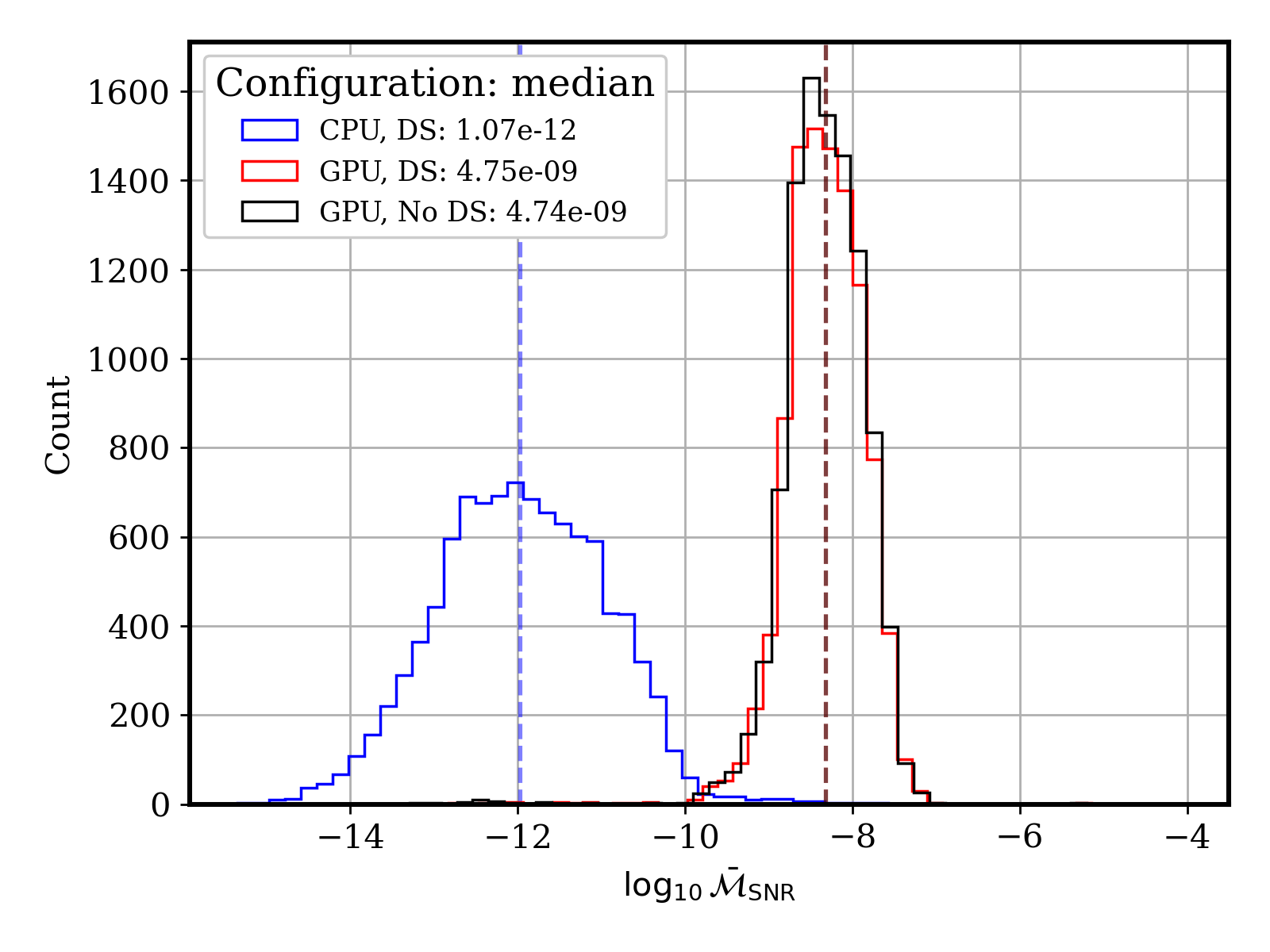}
  \caption{\label{fig:diff_config_mismatches}SNR-weighted, $(\iota, \phi_0, \psi)$-averaged mismatches between the reference configuration of the surrogate model (CPU evaluation without quaternion downsampling) and three other configurations of our model: CPU evaluation with downsampling (DS), GPU evaluation without downsampling, and GPU evaluation with downsampling.}
\end{figure}

In Sec.~\ref{sec:faithfulness} we evaluated the faithfulness of our surrogate model to SEOBNRv5PHM, using the CPU version of our model without quaternion downsampling as the reference configuration. In this appendix, we confirm that different configurations of our surrogate model produce consistent results with one another, investigating specifically the impact of quaternion downsampling and GPU evaluation.

Figure~\ref{fig:diff_config_mismatches} displays the SNR-weighted, $(\iota, \phi_0, \psi)$-averaged mismatch between the waveform generated using the reference configuration against three different configurations of our model: CPU evaluation with downsampling, GPU evaluation without downsampling, and GPU evaluation with downsampling, using the aLIGO PSD with fixed total mass $M = 65 M_{\odot}$. Compared to the results in Sec.~\ref{sec:faithfulness}, the average is taken over the more restricted sets $\iota \in \{0, \pi/3, \pi/2 \}$, $\phi_0 \in \{0, \pi/4, \pi/2\}$ and $\psi \in \{0, \pi/4, \pi/2\}$, and we compute the mismatch for only the first $10^4$ parameter samples in $\Lambda_{\rm test}$. We see from Fig.~\ref{fig:diff_config_mismatches} that downsampling produces a completely negligible mismatch against the reference configuration, with a median mismatch of only $1.07 \times 10^{-12}$ when the waveform is evaluated on the CPU with downsampling, and the mismatch is greater than $10^{-7}$ for only one parameter sample. When evaluating the waveform on the GPU with (without) downsampling, the median mismatch against the reference configuration is $4.75 \times 10^{-9}$ ($4.74\times 10^{-9}$), and only $9$ ($8$) parameter samples have a mismatch greater than $10^{-7}$. 

We note that the mismatches between the different surrogate configurations are in general much smaller than the mismatches between SEOBNRv5PHM and the surrogate using the reference configuration displayed in Fig.~\ref{fig:SNR_weighted_mismatches}, so that the results and conclusions of Fig.~\ref{sec:faithfulness} should be robust to changes in the surrogate model configuration. These tests also confirm that our CPU and GPU implementations agree with each other to an appropriate level (compared to the unfaithfulness against SEOBNRv5PHM), and prove that the use of quaternion downsampling does not introduce any significant error in the final waveform. Indeed, the exceptionally small mismatch between the model with and without downsampling suggests that we could have chosen $\mathcal{T}_{\rm ds}$ to be significantly more sparse without meaningfully altering the final waveform.

\section{Waveform timing methodology}
\label{app:timing_methodology}
In this appendix, we describe the methodology used to time the average waveform evaluation cost for both the surrogate and SEOBNRv5PHM in Sec.~\ref{sec:model_timings}. 

Achieving a like-for-like timing comparison between the surrogate and SEOBNRv5PHM is complicated by their different conventions. Recall that the surrogate model accepts as parameters the mass ratio and spins specified in the I-frame defined at a constant geometric reference frequency $\Omega_{\rm ref} = 0.007/M$, and returns the gravitational wave polarizations $h_{+,\times}^I(t; \lamvec)$ on a fixed geometric time grid stretching from $\tstart = -10^4M$ before merger through to $100M$ after merger, with spacing $\Delta t =0 .5M$. In practice, the orbital frequency at this starting time is typically much higher than the reference frequency. SEOBNRv5PHM, on the other hand, accepts the same parameters, but returns the polarizations on a (parameter-dependent) time grid covering the full evolution from $\Omega_{\rm ref}$ through to a few $100M$ after merger. For $\Omega_{\rm ref} = 0.007/M$, the SEOBNRv5PHM grid typically has much longer duration ($\sim 10^5M$) than the surrogate grid, so for a fair control we need a way to evaluate SEOBNRv5PHM waveforms with a fixed starting time. We do this in practice using a two-stage strategy: first, for a given set of parameters, we evaluate the SEOBNRv5PHM waveform starting at $\Omega_{\rm ref} = 0.007/M$, and extract the values of the orbital frequency, $\Omega_{\rm start}$, the spins, and the quaternion describing the $P \rightarrow I$ rotation at time $\tstart$. We then use the $P \rightarrow I$ rotation to transform the spins at time $\tstart$ into the P-frame at time $\tstart$ (which we treat as the new I-frame defined at reference frequency $\Omega_{\rm start}$), before calling SEOBNRv5PHM again with reference frequency $\Omega_{\rm start}$ and the evolved spin parameters. We found this gave us waveforms which start very close to $10^4M$, so we used these waveform calls for our timing tests. For simplicity we do not evolve the extrinsic parameters $\iota$ or $\phi_0$ because these do not significantly affect the waveform timing. We generate our SEOBNRv5PHM waveforms on a time grid of spacing $\Delta t = 0.5M$. Note that we do not control the amount of post-merger signal present in our SEOBNRv5PHM waveforms; in practice the excess signal is typically quite small ($\sim 100M$) compared to the length of the inspiral, and we do not expect a significant effect on the timing.

We begin all of our timing tests with 250 untimed waveform calls, to ensure that our subsequent timing runs accurately capture the steady-state evaluation cost, not including any fixed setup costs (such as just-in-time compilation) that contribute insignificantly to the overall runtime when evaluating large numbers of waveforms in real-world applications. When timing the cost to evaluate a single waveform (using either SEOBNRv5PHM or the surrogate) with given parameters $(\lamvec, \iota, \phi_0)$, we calculate the wall time cost to evaluate the waveform 5 times in a loop, and then compute the mean across these 5 runs. For the final mean cost per waveform reported in Table~\ref{tab:waveform_timings}, we report the mean of the run-averaged single-waveform costs over all parameters in the timing set $\tilde\Lambda_{\rm time}$. For the batched waveform tests running on a GPU, we randomly draw 500 parameter batches with the given batch size from among the parameters in the timing set, and for each batch compute the cost-per-batch across 5 looped runs. We then compute the mean cost-per-batch across each of the 500 test batches, and normalize by the batch size to get the mean cost-per-waveform, which we report in Table~\ref{tab:waveform_timings}. The methodology for the batched CPU evaluation is identical, except that we only draw 100 test batches due to the longer runtime on the CPU.

\section{Parameter estimation settings}
\label{app:PE_settings}
Table~\ref{tab:pe_settings} summarizes various settings that were used for the different parameter inference runs in Sec.~\ref{sec:bayesian_inference}.

\begin{table*}
\centering
{\setlength{\tabcolsep}{8pt}
\renewcommand{\arraystretch}{1.3}
\begin{tabular}{lccccc}
\hline\hline
 & Injection 1 & Injection 2 & GW150914 & GW200129 & GW250114 \\
\hline
\multicolumn{6}{l}{\textbf{Prior settings}} \\[2pt]
$\mathcal{M}_c^{\rm det}\>\>[M_{\odot}]$ & $U[20,\>50]$ & $U[15,\>30]$ & $U[10,\>50]$ & $U[15,\>44]$ & $U[28.5,\>33]$ \\
$1/q$ & $U[0.1,\>1]$ & $U[0.1,\>1]$ & $U[0.25,\>1.0]$ & $U[0.25,\>1.0]$ & $U[0.6,\>1]$ \\
$M \>\> [M_{\odot}]$ & $65 \leq M \leq 150$ & $65 \leq M \leq 120$ & $50 \leq M \leq 100$ & $55 \leq M \leq 100$ & $65 \leq M \leq 100$ \\
$|\chivec_i|\quad(i=1,2)$ & $U[0,\>0.99]$ & $U[0,\>0.99]$ & $U[0,\>0.99]$ & $U[0,\>0.99]$ & $U[0,\>0.99]$\\
$\theta_i \quad(i=1,2)$ [rad] & $\text{Sine}[0,\>\pi]$ & $\text{Sine}[0,\>\pi]$ & $\text{Sine}[0,\>\pi]$ & $\text{Sine}[0,\>\pi]$ & $\text{Sine}[0,\>\pi]$\\
$\phi_i \quad(i=1,2)$ [rad] & $U[0,\>2\pi]$ & $U[0,\>2\pi]$ & $U[0,\>2\pi]$ & $U[0,\>2\pi]$ & $U[0,\>2\pi]$\\
$\phi_{12}$  [rad] & Not used & Not used & $U[0,\>2\pi]$ & $U[0,\>2\pi]$ & $U[0,\>2\pi]$\\
$\phi_{JL}$ [rad] & Not used & Not used & $U[0,\>2\pi]$ & $U[0,\>2\pi]$  & $U[0,\>2\pi]$\\
$\iota$ [rad] & $\text{Sine}[0,\>\pi]$ & $\text{Sine}[0,\>\pi]$ & $\text{Sine}[0,\>\pi]$  & $\text{Sine}[0,\>\pi]$ & ---\\
$\theta_{JN}$ [rad] & Not used & Not used & $\text{Sine}[0,\>\pi]$ & $\text{Sine}[0,\>\pi]$ & $\text{Sine}[0,\>\pi]$\\
$\phi_0$ [rad] & $U[0,\>2\pi]$ & $U[0,\>2\pi]$ & $U[0,\>2\pi]$ & $U[0,\>2\pi]$ & $U[0,\>2\pi]$\\
$\psi$ [rad] & $U[0,\>\pi]$ & $U[0,\>\pi]$ & $U[0,\>\pi]$ & $U[0,\>\pi]$ & $U[0,\>\pi]$\\
$d_L$ [Mpc] & $U[0,\> 10^4]$ & $U[0,\> 10^4]$ & $U[0,\> 2000]$ & $U[0,\> 2000]$ & $U[0,\> 2000]$\\
$\alpha$ [rad] & $U[0,\>2\pi]$ & $U[0,\>2\pi]$ & $U[0,\>2\pi]$ & $U[0,\>2\pi]$ & $U[0,\>2\pi]$\\
$\delta$ [rad] & $U\left[-\frac{\pi}{2},\>\frac{\pi}{2}\right]$ & $U\left[-\frac{\pi}{2},\>\frac{\pi}{2}\right]$ & $\text{Cosine}\left[-\frac{\pi}{2},\>\frac{\pi}{2}\right]$ & $\text{Cosine}\left[-\frac{\pi}{2},\>\frac{\pi}{2}\right]$ & $\text{Cosine}\left[-\frac{\pi}{2},\>\frac{\pi}{2}\right]$\\
$\delta t_c$ [s]  & $U[-0.05,\>0.05]$ & $U[-0.05,\>0.05]$ & $U[-0.1\>0.1]$ & $U[-0.1\>0.1]$ & $U[-0.1\>0.1]$\\[6pt]

\multicolumn{6}{l}{\textbf{Calibration settings}} \\[2pt]
Calibration nodes & \multirow{3}{*}{\makecell{Not \\ used}} & \multirow{3}{*}{\makecell{Not \\used}} & 10 & 10 & 10 \\
Amplitude priors & & & \multirow{2}{*}{\makecell{From LVK \\ envelopes \cite{LIGOScientific:2018mvr}}} & $N[0,\>0.05]$ & $N[0,\>0.05]$ \\
Phase priors & & & & $N[0,\>0.05]$ & $N[0,\>0.05]$ \\[6pt]

\multicolumn{6}{l}{\textbf{Data settings}} \\[2pt]
Segment length [s] & 8.0 & 8.0 & 8.0 & 8.0 & 8.0\\
Sample rate [Hz] & 4096.0 & 4096.0 & 4096.0 & 4096.0 & 4096.0\\
PSD & Design & Design & LVK PSD \cite{LIGOScientific:2018mvr} & From 256s segment & From 256s segment \\
Interferometers used & HLV & HLV & HL & HLV & HL\\[6pt]

\multicolumn{6}{l}{\textbf{Likelihood settings}} \\[2pt]
$f_{\rm low}$ & 20.0 & 20.0 & 20.0 & 20.0 & 20.0\\
$f_{\rm high}$ & 2048.0 & 2048.0 & 1024.0 & 1024.0 & 1024.0 \\
Marginalization & Not used & Not used & Distance & Distance & Distance \\[6pt]

\multicolumn{6}{l}{\textbf{Sampler settings}} \\[2pt]
nlive & 1000 & 1000 & 1000 & 1000 & 1000\\
npool & 32 & 32 & 32 & 32 & 56 \\
sample & rwalk & acceptance-walk & acceptance-walk & acceptance-walk & acceptance-walk\\
Other arguments & walks=25 & naccept=60 & naccept=60 & naccept=60 & naccept=60\\
& nact=5 & & & & \\
\hline\hline
\end{tabular}}
\caption{\label{tab:pe_settings}PE settings for all analyses. $U[a,\>b]$ denotes the uniform distribution on the interval $[a,\>b]$,  $\text{Sine}[a,b]$ denotes the distribution on $x \in [a,\> b]$ uniform in $\sin(x)$, and $\text{Cosine}[a,\>b]$ denotes the distribution on $x \in [a,\>b]$ uniform in $\cos(x)$.
Interferometer names are abbreviated as follows: LIGO Hanford (H), LIGO Livingston (L) and Virgo (V). The parameter nlive denotes the number of live points used by the Dynesty nested sampler \cite{dynesty_paper} and npool is the number of CPUs used by the sampler. rwalk and acceptance-walk are different methods used by Dynesty to find new live points. These methods (and the other Dynesty parameters listed in the table) are explained in further detail in Ref.~\cite{bilby_dynesty_guide}}
\end{table*}

\clearpage

\bibliography{references}

@article{Ramos-Buades:2023ehm,
    author = "Ramos-Buades, Antoni and Buonanno, Alessandra and Estell{\'e}s, H{\'e}ctor and Khalil, Mohammed and Mihaylov, Deyan P. and Ossokine, Serguei and Pompili, Lorenzo and Shiferaw, Mahlet",
    title = "{Next generation of accurate and efficient multipolar precessing-spin effective-one-body waveforms for binary black holes}",
    eprint = "2303.18046",
    archivePrefix = "arXiv",
    primaryClass = "gr-qc",
    doi = "10.1103/PhysRevD.108.124037",
    journal = "Phys. Rev. D",
    volume = "108",
    number = "12",
    pages = "124037",
    year = "2023"
}

@article{Mihaylov:2023bkc,
    author = {Mihaylov, Deyan P. and Ossokine, Serguei and Buonanno, Alessandra and Estelles, Hector and Pompili, Lorenzo and P\"urrer, Michael and Ramos-Buades, Antoni},
    title = "{pySEOBNR: a software package for the next generation of effective-one-body multipolar waveform models}",
    eprint = "2303.18203",
    archivePrefix = "arXiv",
    primaryClass = "gr-qc",
    doi = "10.1016/j.softx.2025.102080",
    journal = "SoftwareX",
    volume = "30",
    pages = "102080",
    year = "2025"
}

@misc{Estelles:2025zah,
    author = "Estell{\'e}s, H{\'e}ctor and Buonanno, Alessandra and Enficiaud, Raffi and Foo, Cheng and Pompili, Lorenzo",
    title = "{Adding equatorial-asymmetric effects for spin-precessing binaries into the SEOBNRv5PHM waveform model}",
    eprint = "2506.19911",
    archivePrefix = "arXiv",
    primaryClass = "gr-qc",
    month = "6",
    year = "2025"
}

@misc{Boyle:2014ioa,
    author = "Boyle, Michael and Kidder, Lawrence E. and Ossokine, Serguei and Pfeiffer, Harald P.",
    title = "{Gravitational-wave modes from precessing black-hole binaries}",
    howpublished="arXiv:1409.4431 [gr-qc]"  
}

@article{Boyle:2011gg,
    author = "Boyle, Michael and Owen, Robert and Pfeiffer, Harald P.",
    title = "{A geometric approach to the precession of compact binaries}",
    eprint = "1110.2965",
    archivePrefix = "arXiv",
    primaryClass = "gr-qc",
    doi = "10.1103/PhysRevD.84.124011",
    journal = "Phys. Rev. D",
    volume = "84",
    pages = "124011",
    year = "2011"
}

@article{Apostolatos:1994mx,
    author = "Apostolatos, Theocharis A. and Cutler, Curt and Sussman, Gerald J. and Thorne, Kip S.",
    title = "{Spin induced orbital precession and its modulation of the gravitational wave forms from merging binaries}",
    reportNumber = "GRP-382",
    doi = "10.1103/PhysRevD.49.6274",
    journal = "Phys. Rev. D",
    volume = "49",
    pages = "6274--6297",
    year = "1994"
}

@article{Kidder:1995zr,
    author = "Kidder, Lawrence E.",
    title = "{Coalescing binary systems of compact objects to postNewtonian 5/2 order. 5. Spin effects}",
    eprint = "gr-qc/9506022",
    archivePrefix = "arXiv",
    reportNumber = "NU-GR-11, WUGRAV-94-6A",
    doi = "10.1103/PhysRevD.52.821",
    journal = "Phys. Rev. D",
    volume = "52",
    pages = "821--847",
    year = "1995"
}

@article{Schmidt:2010it,
    author = "Schmidt, Patricia and Hannam, Mark and Husa, Sascha and Ajith, P.",
    title = "{Tracking the precession of compact binaries from their gravitational-wave signal}",
    eprint = "1012.2879",
    archivePrefix = "arXiv",
    primaryClass = "gr-qc",
    doi = "10.1103/PhysRevD.84.024046",
    journal = "Phys. Rev. D",
    volume = "84",
    pages = "024046",
    year = "2011"
}

@article{Schmidt:2012rh,
    author = "Schmidt, Patricia and Hannam, Mark and Husa, Sascha",
    title = "{Towards models of gravitational waveforms from generic binaries: A simple approximate mapping between precessing and non-precessing inspiral signals}",
    eprint = "1207.3088",
    archivePrefix = "arXiv",
    primaryClass = "gr-qc",
    doi = "10.1103/PhysRevD.86.104063",
    journal = "Phys. Rev. D",
    volume = "86",
    pages = "104063",
    year = "2012"
}

@article{Schmidt:2014iyl,
    author = "Schmidt, Patricia and Ohme, Frank and Hannam, Mark",
    title = "{Towards models of gravitational waveforms from generic binaries II: Modelling precession effects with a single effective precession parameter}",
    eprint = "1408.1810",
    archivePrefix = "arXiv",
    primaryClass = "gr-qc",
    doi = "10.1103/PhysRevD.91.024043",
    journal = "Phys. Rev. D",
    volume = "91",
    number = "2",
    pages = "024043",
    year = "2015"
}

@article{OShaughnessy:2011pmr,
    author = "O'Shaughnessy, R. and Vaishnav, B. and Healy, J. and Meeks, Z. and Shoemaker, D.",
    title = "{Efficient asymptotic frame selection for binary black hole spacetimes using asymptotic radiation}",
    eprint = "1109.5224",
    archivePrefix = "arXiv",
    primaryClass = "gr-qc",
    reportNumber = "LIGO-DCC-P1100113",
    doi = "10.1103/PhysRevD.84.124002",
    journal = "Phys. Rev. D",
    volume = "84",
    pages = "124002",
    year = "2011"
}

@book{Wigner1959,
  author    = {E. P. Wigner},
  title     = {Group Theory and Its Application to the Quantum Mechanics of Atomic Spectra},
  publisher = {Academic Press},
  address   = {New York},
  year      = {1959}
}

@misc{Schmidt:2017btt,
    author = "Schmidt, Patricia and Harry, Ian W. and Pfeiffer, Harald P.",
    title = "{Numerical Relativity Injection Infrastructure}",
    eprint = "1703.01076",
    archivePrefix = "arXiv",
    primaryClass = "gr-qc",
    reportNumber = "LIGO-T1500606",
    month = "3",
    year = "2017"
}

@article{Ajith:2009bn,
    author = "Ajith, P. and others",
    title = "{Inspiral-merger-ringdown waveforms for black-hole binaries with non-precessing spins}",
    eprint = "0909.2867",
    archivePrefix = "arXiv",
    primaryClass = "gr-qc",
    doi = "10.1103/PhysRevLett.106.241101",
    journal = "Phys. Rev. Lett.",
    volume = "106",
    pages = "241101",
    year = "2011"
}

@article{Santamaria:2010yb,
    author = "Santamaria, L. and others",
    title = "{Matching post-Newtonian and numerical relativity waveforms: systematic errors and a new phenomenological model for non-precessing black hole binaries}",
    eprint = "1005.3306",
    archivePrefix = "arXiv",
    primaryClass = "gr-qc",
    reportNumber = "LIGO-P1000048, AEI-2010-122",
    doi = "10.1103/PhysRevD.82.064016",
    journal = "Phys. Rev. D",
    volume = "82",
    pages = "064016",
    year = "2010"
}

@article{Racine:2008qv,
    author = "Racine, Etienne",
    title = "{Analysis of spin precession in binary black hole systems including quadrupole-monopole interaction}",
    eprint = "0803.1820",
    archivePrefix = "arXiv",
    primaryClass = "gr-qc",
    doi = "10.1103/PhysRevD.78.044021",
    journal = "Phys. Rev. D",
    volume = "78",
    pages = "044021",
    year = "2008"
}

@article{Gerosa:2020aiw,
    author = "Gerosa, Davide and Mould, Matthew and Gangardt, Daria and Schmidt, Patricia and Pratten, Geraint and Thomas, Lucy M.",
    title = "{A generalized precession parameter $\chi_\mathrm{p}$ to interpret gravitational-wave data}",
    eprint = "2011.11948",
    archivePrefix = "arXiv",
    primaryClass = "gr-qc",
    doi = "10.1103/PhysRevD.103.064067",
    journal = "Phys. Rev. D",
    volume = "103",
    number = "6",
    pages = "064067",
    year = "2021"
}

@article{Ramos-Buades:2020noq,
    author = "Ramos-Buades, Antoni and Schmidt, Patricia and Pratten, Geraint and Husa, Sascha",
    title = "{Validity of common modeling approximations for precessing binary black holes with higher-order modes}",
    eprint = "2001.10936",
    archivePrefix = "arXiv",
    primaryClass = "gr-qc",
    doi = "10.1103/PhysRevD.101.103014",
    journal = "Phys. Rev. D",
    volume = "101",
    number = "10",
    pages = "103014",
    year = "2020"
}

@article{Field:2011mf,
    author = "Field, Scott E. and Galley, Chad R. and Herrmann, Frank and Hesthaven, Jan S. and Ochsner, Evan and Tiglio, Manuel",
    title = "{Reduced basis catalogs for gravitational wave templates}",
    eprint = "1101.3765",
    archivePrefix = "arXiv",
    primaryClass = "gr-qc",
    doi = "10.1103/PhysRevLett.106.221102",
    journal = "Phys. Rev. Lett.",
    volume = "106",
    pages = "221102",
    year = "2011"
}

@article{Field:2013cfa,
    author = "Field, Scott E. and Galley, Chad R. and Hesthaven, Jan S. and Kaye, Jason and Tiglio, Manuel",
    title = "{Fast prediction and evaluation of gravitational waveforms using surrogate models}",
    eprint = "1308.3565",
    archivePrefix = "arXiv",
    primaryClass = "gr-qc",
    doi = "10.1103/PhysRevX.4.031006",
    journal = "Phys. Rev. X",
    volume = "4",
    number = "3",
    pages = "031006",
    year = "2014"
}

@article{Purrer:2014fza,
    author = {P{\"u}rrer, Michael},
    title = "{Frequency domain reduced order models for gravitational waves from aligned-spin compact binaries}",
    eprint = "1402.4146",
    archivePrefix = "arXiv",
    primaryClass = "gr-qc",
    doi = "10.1088/0264-9381/31/19/195010",
    journal = "Class. Quant. Grav.",
    volume = "31",
    number = "19",
    pages = "195010",
    year = "2014"
}

@article{Blackman:2015pia,
    author = "Blackman, Jonathan and Field, Scott E. and Galley, Chad R. and Szil{\'a}gyi, B{\'e}la and Scheel, Mark A. and Tiglio, Manuel and Hemberger, Daniel A.",
    title = "{Fast and Accurate Prediction of Numerical Relativity Waveforms from Binary Black Hole Coalescences Using Surrogate Models}",
    eprint = "1502.07758",
    archivePrefix = "arXiv",
    primaryClass = "gr-qc",
    doi = "10.1103/PhysRevLett.115.121102",
    journal = "Phys. Rev. Lett.",
    volume = "115",
    number = "12",
    pages = "121102",
    year = "2015"
}

@article{Blackman:2017dfb,
    author = "Blackman, Jonathan and Field, Scott E. and Scheel, Mark A. and Galley, Chad R. and Hemberger, Daniel A. and Schmidt, Patricia and Smith, Rory",
    title = "{A Surrogate Model of Gravitational Waveforms from Numerical Relativity Simulations of Precessing Binary Black Hole Mergers}",
    eprint = "1701.00550",
    archivePrefix = "arXiv",
    primaryClass = "gr-qc",
    doi = "10.1103/PhysRevD.95.104023",
    journal = "Phys. Rev. D",
    volume = "95",
    number = "10",
    pages = "104023",
    year = "2017"
}

@misc{Galley:2016mvy,
    author = "Galley, Chad R. and Schmidt, Patricia",
    title = "{Fast and efficient evaluation of gravitational waveforms via reduced-order spline interpolation}",
    eprint = "1611.07529",
    archivePrefix = "arXiv",
    primaryClass = "gr-qc",
    reportNumber = "LIGO-P1600064",
    month = "11",
    year = "2016"
}

@article{Barrault2004,
title = {An ‘empirical interpolation’ method: application to efficient reduced-basis discretization of partial differential equations},
journal = {Comptes Rendus Mathematique},
volume = {339},
number = {9},
pages = {667-672},
year = {2004},
issn = {1631-073X},
doi = {https://doi.org/10.1016/j.crma.2004.08.006},
url = {https://www.sciencedirect.com/science/article/pii/S1631073X04004248},
author = {Maxime Barrault and Yvon Maday and Ngoc Cuong Nguyen and Anthony T. Patera},
}

@article{maday2009general,
  title={A general multipurpose interpolation procedure: the magic points},
  author={Maday, Yvon and Nguyen, Ngoc Cuong and Patera, Anthony T and Pau, SH},
  journal={Communications on Pure \&amp; Applied Analysis},
  volume={8},
  number={1},
  pages={383},
  year={2009},
  publisher={American Institute of Mathematical Sciences}
}

@article{Boyle:2013nka,
    author = "Boyle, Michael",
    title = "{Angular velocity of gravitational radiation from precessing binaries and the corotating frame}",
    eprint = "1302.2919",
    archivePrefix = "arXiv",
    primaryClass = "gr-qc",
    doi = "10.1103/PhysRevD.87.104006",
    journal = "Phys. Rev. D",
    volume = "87",
    number = "10",
    pages = "104006",
    year = "2013"
}

@article{Thomas:2025rje,
    author = "Thomas, Lucy M. and Chatziioannou, Katerina and Varma, Vijay and Field, Scott E.",
    title = "{Optimizing neural network surrogate models: Application to black hole merger remnants}",
    eprint = "2501.16462",
    archivePrefix = "arXiv",
    primaryClass = "gr-qc",
    reportNumber = "DCC:LIGO-P2400620",
    doi = "10.1103/PhysRevD.111.104029",
    journal = "Phys. Rev. D",
    volume = "111",
    number = "10",
    pages = "104029",
    year = "2025"
}

@article{Cerino:2022dhr,
    author = "Cerino, Franco and Diaz-Pace, J. Andr{\'e}s and Tiglio, Manuel",
    title = "{An automated parameter domain decomposition approach for gravitational wave surrogates using hp-greedy refinement}",
    eprint = "2212.08554",
    archivePrefix = "arXiv",
    primaryClass = "gr-qc",
    doi = "10.1088/1361-6382/acf4e7",
    journal = "Class. Quant. Grav.",
    volume = "40",
    number = "20",
    pages = "205003",
    year = "2023"
}

@article{Cerino:2023iqv,
    author = "Cerino, Franco and Diaz-Pace, J. Andr{\'e}s and Tassone, Emmanuel A. and Tiglio, Manuel and Villegas, Atuel",
    title = "{Hyperparameter Optimization of an hp-Greedy Reduced Basis for Gravitational Wave Surrogates}",
    eprint = "2310.15143",
    archivePrefix = "arXiv",
    primaryClass = "gr-qc",
    doi = "10.3390/universe10010006",
    journal = "Universe",
    volume = "10",
    number = "1",
    pages = "6",
    year = "2024"
}

@misc{Hamilton:2025xru,
    author = "Hamilton, Eleanor and others",
    title = "{PhenomXPNR: An improved gravitational wave model linking precessing inspirals and NR-calibrated merger-ringdown}",
    eprint = "2507.02604",
    archivePrefix = "arXiv",
    primaryClass = "gr-qc",
    month = "7",
    year = "2025"
}

@article{Pratten:2020ceb,
    author = "Pratten, Geraint and others",
    title = "{Computationally efficient models for the dominant and subdominant harmonic modes of precessing binary black holes}",
    eprint = "2004.06503",
    archivePrefix = "arXiv",
    primaryClass = "gr-qc",
    doi = "10.1103/PhysRevD.103.104056",
    journal = "Phys. Rev. D",
    volume = "103",
    number = "10",
    pages = "104056",
    year = "2021"
}

@article{Thompson:2023ase,
    author = "Thompson, Jonathan E. and Hamilton, Eleanor and London, Lionel and Ghosh, Shrobana and Kolitsidou, Panagiota and Hoy, Charlie and Hannam, Mark",
    title = "{PhenomXO4a: a phenomenological gravitational-wave model for precessing black-hole binaries with higher multipoles and asymmetries}",
    eprint = "2312.10025",
    archivePrefix = "arXiv",
    primaryClass = "gr-qc",
    reportNumber = "LIGO-P2300437",
    doi = "10.1103/PhysRevD.109.063012",
    journal = "Phys. Rev. D",
    volume = "109",
    number = "6",
    pages = "063012",
    year = "2024"
}

@misc{Cornish:2010kf,
    author = "Cornish, Neil J.",
    title = "{Fast Fisher Matrices and Lazy Likelihoods}",
    eprint = "1007.4820",
    archivePrefix = "arXiv",
    primaryClass = "gr-qc",
    month = "7",
    year = "2010"
}

@misc{Zackay:2018qdy,
    author = "Zackay, Barak and Dai, Liang and Venumadhav, Tejaswi",
    title = "{Relative Binning and Fast Likelihood Evaluation for Gravitational Wave Parameter Estimation}",
    eprint = "1806.08792",
    archivePrefix = "arXiv",
    primaryClass = "astro-ph.IM",
    month = "6",
    year = "2018"
}

@article{Leslie:2021ssu,
    author = "Leslie, Nathaniel and Dai, Liang and Pratten, Geraint",
    title = "{Mode-by-mode relative binning: Fast likelihood estimation for gravitational waveforms with spin-orbit precession and multiple harmonics}",
    eprint = "2109.09872",
    archivePrefix = "arXiv",
    primaryClass = "astro-ph.IM",
    doi = "10.1103/PhysRevD.104.123030",
    journal = "Phys. Rev. D",
    volume = "104",
    number = "12",
    pages = "123030",
    year = "2021"
}

@article{Vinciguerra:2017ngf,
    author = "Vinciguerra, Serena and Veitch, John and Mandel, Ilya",
    title = "{Accelerating gravitational wave parameter estimation with multi-band template interpolation}",
    eprint = "1703.02062",
    archivePrefix = "arXiv",
    primaryClass = "gr-qc",
    doi = "10.1088/1361-6382/aa6d44",
    journal = "Class. Quant. Grav.",
    volume = "34",
    number = "11",
    pages = "115006",
    year = "2017"
}

@article{Morisaki:2021ngj,
    author = "Morisaki, Soichiro",
    title = "{Accelerating parameter estimation of gravitational waves from compact binary coalescence using adaptive frequency resolutions}",
    eprint = "2104.07813",
    archivePrefix = "arXiv",
    primaryClass = "gr-qc",
    doi = "10.1103/PhysRevD.104.044062",
    journal = "Phys. Rev. D",
    volume = "104",
    number = "4",
    pages = "044062",
    year = "2021"
}

@article{Canizares:2013ywa,
    author = "Canizares, Priscilla and Field, Scott E. and Gair, Jonathan R. and Tiglio, Manuel",
    title = "{Gravitational wave parameter estimation with compressed likelihood evaluations}",
    eprint = "1304.0462",
    archivePrefix = "arXiv",
    primaryClass = "gr-qc",
    doi = "10.1103/PhysRevD.87.124005",
    journal = "Phys. Rev. D",
    volume = "87",
    number = "12",
    pages = "124005",
    year = "2013"
}

@article{Canizares:2014fya,
    author = "Canizares, Priscilla and Field, Scott E. and Gair, Jonathan and Raymond, Vivien and Smith, Rory and Tiglio, Manuel",
    title = "{Accelerated gravitational-wave parameter estimation with reduced order modeling}",
    eprint = "1404.6284",
    archivePrefix = "arXiv",
    primaryClass = "gr-qc",
    reportNumber = "LIGO-P1400038",
    doi = "10.1103/PhysRevLett.114.071104",
    journal = "Phys. Rev. Lett.",
    volume = "114",
    number = "7",
    pages = "071104",
    year = "2015"
}

@article{Smith:2016qas,
    author = {Smith, Rory and Field, Scott E. and Blackburn, Kent and Haster, Carl-Johan and P{\"u}rrer, Michael and Raymond, Vivien and Schmidt, Patricia},
    title = "{Fast and accurate inference on gravitational waves from precessing compact binaries}",
    eprint = "1604.08253",
    archivePrefix = "arXiv",
    primaryClass = "gr-qc",
    reportNumber = "LIGO-DOCUMENT-NUMBER-P1600096, LIGO-P1600096",
    doi = "10.1103/PhysRevD.94.044031",
    journal = "Phys. Rev. D",
    volume = "94",
    number = "4",
    pages = "044031",
    year = "2016"
}

@article{Chua:2019wwt,
    author = "Chua, Alvin J. K. and Vallisneri, Michele",
    title = "{Learning Bayesian posteriors with neural networks for gravitational-wave inference}",
    eprint = "1909.05966",
    archivePrefix = "arXiv",
    primaryClass = "gr-qc",
    doi = "10.1103/PhysRevLett.124.041102",
    journal = "Phys. Rev. Lett.",
    volume = "124",
    number = "4",
    pages = "041102",
    year = "2020"
}

@article{Dax:2021tsq,
    author = {Dax, Maximilian and Green, Stephen R. and Gair, Jonathan and Macke, Jakob H. and Buonanno, Alessandra and Sch{\"o}lkopf, Bernhard},
    title = "{Real-Time Gravitational Wave Science with Neural Posterior Estimation}",
    eprint = "2106.12594",
    archivePrefix = "arXiv",
    primaryClass = "gr-qc",
    reportNumber = "LIGO-P2100223",
    doi = "10.1103/PhysRevLett.127.241103",
    journal = "Phys. Rev. Lett.",
    volume = "127",
    number = "24",
    pages = "241103",
    year = "2021"
}

@article{Dax:2024mcn,
    author = {Dax, Maximilian and Green, Stephen R. and Gair, Jonathan and Gupte, Nihar and P{\"u}rrer, Michael and Raymond, Vivien and Wildberger, Jonas and Macke, Jakob H. and Buonanno, Alessandra and Sch{\"o}lkopf, Bernhard},
    title = "{Real-time inference for binary neutron star mergers using machine learning}",
    eprint = "2407.09602",
    archivePrefix = "arXiv",
    primaryClass = "gr-qc",
    reportNumber = "LIGO-P2400294",
    doi = "10.1038/s41586-025-08593-z",
    journal = "Nature",
    volume = "639",
    number = "8053",
    pages = "49--53",
    year = "2025"
}

@article{Hu:2024lrj,
    author = "Hu, Qian and Irwin, Jessica and Sun, Qi and Messenger, Christopher and Suleiman, Lami and Heng, Ik Siong and Veitch, John",
    title = "{Decoding Long-duration Gravitational Waves from Binary Neutron Stars with Machine Learning: Parameter Estimation and Equations of State}",
    eprint = "2412.03454",
    archivePrefix = "arXiv",
    primaryClass = "gr-qc",
    reportNumber = "LIGO-P2400567, ET-0666B-24",
    doi = "10.3847/2041-8213/ade42f",
    journal = "Astrophys. J. Lett.",
    volume = "987",
    pages = "L17",
    year = "2025"
}

@misc{Hu:2025vlp,
    author = "Hu, Qian",
    title = "{Hierarchical Subtraction with Neural Density Estimators as a General Solution to Overlapping Gravitational Wave Signals}",
    eprint = "2507.05209",
    archivePrefix = "arXiv",
    primaryClass = "gr-qc",
    reportNumber = "LIGO-P2500409, ET-0346A-25",
    month = "7",
    year = "2025"
}

@misc{TensorFlow,
title={ {TensorFlow}: Large-Scale Machine Learning on Heterogeneous Systems},
url={https://www.tensorflow.org/},
note={Software available from tensorflow.org},
author={
    Mart\'{i}n~Abadi and
    Ashish~Agarwal and
    Paul~Barham and
    Eugene~Brevdo and
    Zhifeng~Chen and
    Craig~Citro and
    Greg~S.~Corrado and
    Andy~Davis and
    Jeffrey~Dean and
    Matthieu~Devin and
    Sanjay~Ghemawat and
    Ian~Goodfellow and
    Andrew~Harp and
    Geoffrey~Irving and
    Michael~Isard and
    Yangqing Jia and
    Rafal~Jozefowicz and
    Lukasz~Kaiser and
    Manjunath~Kudlur and
    Josh~Levenberg and
    Dandelion~Man\'{e} and
    Rajat~Monga and
    Sherry~Moore and
    Derek~Murray and
    Chris~Olah and
    Mike~Schuster and
    Jonathon~Shlens and
    Benoit~Steiner and
    Ilya~Sutskever and
    Kunal~Talwar and
    Paul~Tucker and
    Vincent~Vanhoucke and
    Vijay~Vasudevan and
    Fernanda~Vi\'{e}gas and
    Oriol~Vinyals and
    Pete~Warden and
    Martin~Wattenberg and
    Martin~Wicke and
    Yuan~Yu and
    Xiaoqiang~Zheng},
  year={2015},
}

@misc{Keras,
  title={Keras},
  author={Chollet, Fran\c{c}ois and others},
  year={2015},
  howpublished={\url{https://keras.io}},
}

@inproceedings{hyperopt,
  title = 	 {Making a Science of Model Search: Hyperparameter Optimization in Hundreds of Dimensions for Vision Architectures},
  author = 	 {Bergstra, James and Yamins, Daniel and Cox, David},
  booktitle = 	 {Proceedings of the 30th International Conference on Machine Learning},
  pages = 	 {115--123},
  year = 	 {2013},
  volume = 	 {28},
  number =       {1},
  series = 	 {Proceedings of Machine Learning Research},
  month = 	 {17--19 Jun},
  publisher =    {PMLR},
  url = 	 {https://proceedings.mlr.press/v28/bergstra13.html}
}

@inproceedings{TPE,
 author = {Bergstra, James and Bardenet, R\'{e}mi and Bengio, Yoshua and K\'{e}gl, Bal\'{a}zs},
 booktitle = {Advances in Neural Information Processing Systems},
 editor = {J. Shawe-Taylor and R. Zemel and P. Bartlett and F. Pereira and K.Q. Weinberger},
 pages = {},
 publisher = {Curran Associates, Inc.},
 title = {Algorithms for Hyper-Parameter Optimization},
 url = {https://proceedings.neurips.cc/paper_files/paper/2011/file/86e8f7ab32cfd12577bc2619bc635690-Paper.pdf},
 volume = {24},
 year = {2011}
}

@article{scikit-learn,
  title={Scikit-learn: Machine Learning in {P}ython},
  author={Pedregosa, F. and Varoquaux, G. and Gramfort, A. and Michel, V.
          and Thirion, B. and Grisel, O. and Blondel, M. and Prettenhofer, P.
          and Weiss, R. and Dubourg, V. and Vanderplas, J. and Passos, A. and
          Cournapeau, D. and Brucher, M. and Perrot, M. and Duchesnay, E.},
  journal={Journal of Machine Learning Research},
  volume={12},
  pages={2825--2830},
  year={2011}
}

@article{numpy,
 title         = {Array programming with {NumPy}},
 author        = {Charles R. Harris and K. Jarrod Millman and St{\'{e}}fan J.
                 van der Walt and Ralf Gommers and Pauli Virtanen and David
                 Cournapeau and Eric Wieser and Julian Taylor and Sebastian
                 Berg and Nathaniel J. Smith and Robert Kern and Matti Picus
                 and Stephan Hoyer and Marten H. van Kerkwijk and Matthew
                 Brett and Allan Haldane and Jaime Fern{\'{a}}ndez del
                 R{\'{i}}o and Mark Wiebe and Pearu Peterson and Pierre
                 G{\'{e}}rard-Marchant and Kevin Sheppard and Tyler Reddy and
                 Warren Weckesser and Hameer Abbasi and Christoph Gohlke and
                 Travis E. Oliphant},
 year          = {2020},
 month         = sep,
 journal       = {Nature},
 volume        = {585},
 number        = {7825},
 pages         = {357--362},
 doi           = {10.1038/s41586-020-2649-2},
 publisher     = {Springer Science and Business Media {LLC}},
 url           = {https://doi.org/10.1038/s41586-020-2649-2}
}

@ARTICLE{scipy,
  author  = {Virtanen, Pauli and Gommers, Ralf and Oliphant, Travis E. and
            Haberland, Matt and Reddy, Tyler and Cournapeau, David and
            Burovski, Evgeni and Peterson, Pearu and Weckesser, Warren and
            Bright, Jonathan and {van der Walt}, St{\'e}fan J. and
            Brett, Matthew and Wilson, Joshua and Millman, K. Jarrod and
            Mayorov, Nikolay and Nelson, Andrew R. J. and Jones, Eric and
            Kern, Robert and Larson, Eric and Carey, C J and
            Polat, {\.I}lhan and Feng, Yu and Moore, Eric W. and
            {VanderPlas}, Jake and Laxalde, Denis and Perktold, Josef and
            Cimrman, Robert and Henriksen, Ian and Quintero, E. A. and
            Harris, Charles R. and Archibald, Anne M. and
            Ribeiro, Ant{\^o}nio H. and Pedregosa, Fabian and
            {van Mulbregt}, Paul and {SciPy 1.0 Contributors}},
  title   = {{{SciPy} 1.0: Fundamental Algorithms for Scientific
            Computing in Python}},
  journal = {Nature Methods},
  year    = {2020},
  volume  = {17},
  pages   = {261--272},
  adsurl  = {https://rdcu.be/b08Wh},
  doi     = {10.1038/s41592-019-0686-2},
}

@inproceedings{cupy,
  author       = "Okuta, Ryosuke and Unno, Yuya and Nishino, Daisuke and Hido, Shohei and Loomis, Crissman",
  title        = "{CuPy: A NumPy-Compatible Library for NVIDIA GPU Calculations}",
  booktitle    = "{Proceedings of Workshop on Machine Learning Systems (LearningSys) in The Thirty-first Annual Conference on Neural Information Processing Systems (NIPS)}",
  year         = "2017",
  url          = "http://learningsys.org/nips17/assets/papers/paper_16.pdf"
}

@misc{lalsuite,
       author         = "{LIGO Scientific Collaboration} and {Virgo Collaboration} and {KAGRA Collaboration}",
       title          = "{LVK} {A}lgorithm {L}ibrary - {LALS}uite",
       howpublished   = "Free software (GPL)",
       doi            = "10.7935/GT1W-FZ16",
       year           = "2018"
 }

@software{pycbc,
  author       = {Alex Nitz and
                  Ian Harry and
                  Duncan Brown and
                  Christopher M. Biwer and
                  Josh Willis and
                  Tito Dal Canton and
                  Collin Capano and
                  Thomas Dent and
                  Larne Pekowsky and
                  Gareth S Cabourn Davies and
                  Soumi De and
                  Miriam Cabero and
                  Shichao Wu and
                  Andrew R. Williamson and
                  Bernd Machenschalk and
                  Duncan Macleod and
                  Francesco Pannarale and
                  Prayush Kumar and
                  Steven Reyes and
                  dfinstad and
                  Sumit Kumar and
                  Márton Tápai and
                  Leo Singer and
                  Praveen Kumar and
                  veronica-villa and
                  maxtrevor and
                  Bhooshan Uday Varsha Gadre and
                  Sebastian Khan and
                  Stephen Fairhurst and
                  Arthur Tolley},
  title        = {gwastro/pycbc: v2.3.3 release of PyCBC},
  month        = jan,
  year         = 2024,
  publisher    = {Zenodo},
  version      = {v2.3.3},
  doi          = {10.5281/zenodo.10473621},
  url          = {https://doi.org/10.5281/zenodo.10473621},
}

@article{Clevert:2015qvd,
  title={{Fast and Accurate Deep Network Learning by Exponential Linear Units (ELUs)}},
  author={Clevert, Djork-Arné and Unterthiner, Thomas and Hochreiter, Sepp},
  journal={arXiv preprint arXiv:1511.07289},
  year={2015},
}

@article{Fukushima1969,
  author={Fukushima, Kunihiko},
  journal={{IEEE Transactions on Systems Science and Cybernetics}}, 
  title={{Visual Feature Extraction by a Multilayered Network of Analog Threshold Elements}}, 
  year={1969},
  volume={5},
  number={4},
  pages={322-333},
  doi={10.1109/TSSC.1969.300225}}

@inproceedings{Nair2010,
  title={Rectified Linear Units Improve Restricted Boltzmann Machines},
  author={Nair, Vinod and Hinton, Geoffrey E.},
  booktitle={Proceedings of the 27th International Conference on Machine Learning (ICML)},
  year={2010},
}

@InProceedings{Glorot2011,
  title = 	 {{Deep Sparse Rectifier Neural Networks}},
  author = 	 {Glorot, Xavier and Bordes, Antoine and Bengio, Yoshua},
  booktitle = 	 {{Proceedings of the Fourteenth International Conference on Artificial Intelligence and Statistics}},
  pages = 	 {315--323},
  year = 	 {2011},
  editor = 	 {Gordon, Geoffrey and Dunson, David and Dudík, Miroslav},
  volume = 	 {15},
  series = 	 {Proceedings of Machine Learning Research},
  month = 	 {11--13 Apr},
  publisher =    {PMLR},
  url = 	 {https://proceedings.mlr.press/v15/glorot11a.html},
}

@inproceedings{Dugas2000,
 author = {Dugas, Charles and Bengio, Yoshua and B\'{e}lisle, Fran\c{c}ois and Nadeau, Claude and Garcia, Ren\'{e}},
 booktitle = {{Advances in Neural Information Processing Systems}},
 editor = {T. Leen and T. Dietterich and V. Tresp},
 pages = {},
 publisher = {MIT Press},
 title = {{Incorporating Second-Order Functional Knowledge for Better Option Pricing}},
 url = {https://proceedings.neurips.cc/paper_files/paper/2000/file/44968aece94f667e4095002d140b5896-Paper.pdf},
 volume = {13},
 year = {2000}
}

@article{Kingma:2014vow,
    author = "Kingma, Diederik P. and Ba, Jimmy",
    title = "{Adam: A Method for Stochastic Optimization}",
    journal= {{arXiv preprint arXiv:1412.6980}},
    year = "2014"
}

@misc{dozat2016nadam,
  title={{Incorporating Nesterov Momentum into Adam}},
  author={Dozat, Timothy},
  year={2016},
  howpublished={Available online at \url{https://openreview.net/forum?id=OM0jvwB8jIp57ZJjtNEZ}},
  accessed={8th Feburary 2026}
}

@inproceedings{BottouSGD,
author="Bottou, L{\'e}on",
title="Large-Scale Machine Learning with Stochastic Gradient Descent",
booktitle="Proceedings of COMPSTAT'2010",
year="2010",
publisher="Physica-Verlag HD",
pages="177--186",
isbn="978-3-7908-2604-3"
}

@article{bilby_paper,
    author = "Ashton, Gregory and others",
    title = "{BILBY: A user-friendly Bayesian inference library for gravitational-wave astronomy}",
    eprint = "1811.02042",
    archivePrefix = "arXiv",
    primaryClass = "astro-ph.IM",
    doi = "10.3847/1538-4365/ab06fc",
    journal = "Astrophys. J. Suppl.",
    volume = "241",
    number = "2",
    pages = "27",
    year = "2019"
}

@article{dynesty_paper,
    author = "Speagle, Joshua S.",
    title = "{dynesty: a dynamic nested sampling package for estimating Bayesian posteriors and evidences}",
    eprint = "1904.02180",
    archivePrefix = "arXiv",
    primaryClass = "astro-ph.IM",
    doi = "10.1093/mnras/staa278",
    journal = "Mon. Not. Roy. Astron. Soc.",
    volume = "493",
    number = "3",
    pages = "3132--3158",
    year = "2020"
}

@misc{bilby_dynesty_guide,
    author       = {Ashton, Greg},
    title        = {Dynesty Guide --- \texttt{bilby} 2.8 Documentation},
    year         = {2019},
    howpublished = {\url{https://bilby-dev.github.io/bilby/dynesty-guide.html}},
    note         = {Accessed: 2025}
}

@article{Eglajs1977,
  author  = {Eglajs, V. and Audze, P.},
  title   = {New approach to the design of multifactor experiments},
  journal = {Problems of Dynamics and Strengths},
  volume  = {35},
  year    = {1977},
  pages   = {104--107},
  publisher = {Zinatne Publishing House},
  address = {Riga}
}

@article{McKay1979,
  author  = {McKay, M. D. and Beckman, R. J. and Conover, W. J.},
  title   = {A Comparison of Three Methods for Selecting Values of Input Variables in the Analysis of Output from a Computer Code},
  journal = {Technometrics},
  volume  = {21},
  number  = {2},
  pages   = {239--245},
  year    = {1979},
  doi     = {10.1080/00401706.1979.10489755}
}

@article{Boschini:2025ymu,
    author = "Boschini, Matteo and Gerosa, Davide and Crespi, Alessandro and Falcone, Matteo",
    title = "{LHS in LHS: A new expansion strategy for Latin hypercube sampling in simulation design}",
    eprint = "2509.00159",
    archivePrefix = "arXiv",
    primaryClass = "stat.ME",
    doi = "10.1016/j.softx.2025.102294",
    journal = "SoftwareX",
    volume = "31",
    pages = "102294",
    year = "2025"
}

@article{LIGOScientific:2014pky,
    author = "Aasi, J. and others",
    collaboration = "LIGO Scientific",
    title = "{Advanced LIGO}",
    eprint = "1411.4547",
    archivePrefix = "arXiv",
    primaryClass = "gr-qc",
    doi = "10.1088/0264-9381/32/7/074001",
    journal = "Class. Quant. Grav.",
    volume = "32",
    pages = "074001",
    year = "2015"
}

@article{LIGOScientific:2016aoc,
    author = "Abbott, B. P. and others",
    collaboration = "LIGO Scientific, Virgo",
    title = "{Observation of Gravitational Waves from a Binary Black Hole Merger}",
    eprint = "1602.03837",
    archivePrefix = "arXiv",
    primaryClass = "gr-qc",
    reportNumber = "LIGO-P150914",
    doi = "10.1103/PhysRevLett.116.061102",
    journal = "Phys. Rev. Lett.",
    volume = "116",
    number = "6",
    pages = "061102",
    year = "2016"
}

@misc{LIGOScientific:2025yae,
    author = "Abac, A. G. and others",
    collaboration = "LIGO Scientific, VIRGO, KAGRA",
    title = "{GWTC-4.0: Methods for Identifying and Characterizing Gravitational-wave Transients}",
    eprint = "2508.18081",
    archivePrefix = "arXiv",
    primaryClass = "gr-qc",
    reportNumber = "LIGO-P2400300",
    month = "8",
    year = "2025"
}

@misc{LIGOScientific:2025slb,
    author = "Abac, A. G. and others",
    collaboration = "LIGO Scientific, VIRGO, KAGRA",
    title = "{GWTC-4.0: Updating the Gravitational-Wave Transient Catalog with Observations from the First Part of the Fourth LIGO-Virgo-KAGRA Observing Run}",
    eprint = "2508.18082",
    archivePrefix = "arXiv",
    primaryClass = "gr-qc",
    reportNumber = "LIGO-P2400386",
    month = "8",
    year = "2025"
}

@misc{LVK:2026obs,
    author = "{LIGO Scientific Collaboration, Virgo Collaboration, and KAGRA Collaboration}",
    title = "{LIGO, Virgo and KAGRA Observing Run Plans}",
    year = "2026",
    month = feb,
    howpublished = "\url{https://observing.docs.ligo.org/plan/}",
    note = "Last updated 13 February 2026"
}

@article{LIGOScientific:2025rid,
    author = "Abac, A. G. and others",
    collaboration = "LIGO Scientific, Virgo, KAGRA",
    title = "{GW250114: Testing Hawking{\textquoteright}s Area Law and the Kerr Nature of Black Holes}",
    eprint = "2509.08054",
    archivePrefix = "arXiv",
    primaryClass = "gr-qc",
    reportNumber = "LIGO-P2500421",
    doi = "10.1103/kw5g-d732",
    journal = "Phys. Rev. Lett.",
    volume = "135",
    number = "11",
    pages = "111403",
    year = "2025"
}

@article{LIGOScientific:2018mvr,
    author = "Abbott, B. P. and others",
    collaboration = "LIGO Scientific, Virgo",
    title = "{GWTC-1: A Gravitational-Wave Transient Catalog of Compact Binary Mergers Observed by LIGO and Virgo during the First and Second Observing Runs}",
    eprint = "1811.12907",
    archivePrefix = "arXiv",
    primaryClass = "astro-ph.HE",
    reportNumber = "LIGO-P1800307",
    doi = "10.1103/PhysRevX.9.031040",
    journal = "Phys. Rev. X",
    volume = "9",
    number = "3",
    pages = "031040",
    year = "2019"
}

@misc{LIGOScientific:2025snk,
    author = "Abac, A. G. and others",
    collaboration = "LIGO Scientific, VIRGO, KAGRA",
    title = "{Open Data from LIGO, Virgo, and KAGRA through the First Part of the Fourth Observing Run}",
    eprint = "2508.18079",
    archivePrefix = "arXiv",
    primaryClass = "gr-qc",
    reportNumber = "LIGO-P2500167",
    month = "8",
    year = "2025"
}

@techreport{Farr2014Calib,
  author       = {Farr, Will and Farr, Benjamin and Littenberg, Tyson},
  title        = {Modelling Calibration Errors in {CBC} Waveforms},
  institution  = {LIGO Scientific Collaboration},
  year         = {2014},
  number       = {LIGO-T1400682-v1},
  url          = {https://dcc.ligo.org/LIGO-T1400682/public}
}

@article{KAGRA:2021vkt,
    author = "Abbott, R. and others",
    collaboration = "KAGRA, VIRGO, LIGO Scientific",
    title = "{GWTC-3: Compact Binary Coalescences Observed by LIGO and Virgo during the Second Part of the Third Observing Run}",
    eprint = "2111.03606",
    archivePrefix = "arXiv",
    primaryClass = "gr-qc",
    reportNumber = "LIGO-P2000318",
    doi = "10.1103/PhysRevX.13.041039",
    journal = "Phys. Rev. X",
    volume = "13",
    number = "4",
    pages = "041039",
    year = "2023"
}

@article{Hannam:2021pit,
    author = "Hannam, Mark and others",
    title = "{General-relativistic precession in a black-hole binary}",
    eprint = "2112.11300",
    archivePrefix = "arXiv",
    primaryClass = "gr-qc",
    reportNumber = "LIGO-P2100452",
    doi = "10.1038/s41586-022-05212-z",
    journal = "Nature",
    volume = "610",
    number = "7933",
    pages = "652--655",
    year = "2022"
}

@article{Payne:2022spz,
    author = "Payne, Ethan and Hourihane, Sophie and Golomb, Jacob and Udall, Rhiannon and Udall, Richard and Davis, Derek and Chatziioannou, Katerina",
    title = "{Curious case of GW200129: Interplay between spin-precession inference and data-quality issues}",
    eprint = "2206.11932",
    archivePrefix = "arXiv",
    primaryClass = "gr-qc",
    reportNumber = "LIGO DCC: P2200185",
    doi = "10.1103/PhysRevD.106.104017",
    journal = "Phys. Rev. D",
    volume = "106",
    number = "10",
    pages = "104017",
    year = "2022"
}

@article{Punturo:2010zz,
    author = "Punturo, M. and others",
    editor = "Ricci, Fulvio",
    title = "{The Einstein Telescope: A third-generation gravitational wave observatory}",
    doi = "10.1088/0264-9381/27/19/194002",
    journal = "Class. Quant. Grav.",
    volume = "27",
    pages = "194002",
    year = "2010"
}

@article{Branchesi:2023mws,
    author = "Branchesi, Marica and others",
    title = "{Science with the Einstein Telescope: a comparison of different designs}",
    eprint = "2303.15923",
    archivePrefix = "arXiv",
    primaryClass = "gr-qc",
    reportNumber = "ET-0084A-23",
    doi = "10.1088/1475-7516/2023/07/068",
    journal = "JCAP",
    volume = "07",
    pages = "068",
    year = "2023"
}

@misc{ET:2025xjr,
    author = "Abac, Adrian and others",
    collaboration = "ET",
    title = "{The Science of the Einstein Telescope}",
    eprint = "2503.12263",
    archivePrefix = "arXiv",
    primaryClass = "gr-qc",
    reportNumber = "ET-0036C-25",
    month = "3",
    year = "2025"
}

@article{Reitze:2019iox,
    author = "Reitze, David and others",
    title = "{Cosmic Explorer: The U.S. Contribution to Gravitational-Wave Astronomy beyond LIGO}",
    eprint = "1907.04833",
    archivePrefix = "arXiv",
    primaryClass = "astro-ph.IM",
    reportNumber = "LIGO-P1900316",
    journal = "Bull. Am. Astron. Soc.",
    volume = "51",
    number = "7",
    pages = "035",
    year = "2019"
}

@misc{Evans:2023euw,
    author = "Evans, Matthew and others",
    title = "{Cosmic Explorer: A Submission to the NSF MPSAC ngGW Subcommittee}",
    eprint = "2306.13745",
    archivePrefix = "arXiv",
    primaryClass = "astro-ph.IM",
    month = jun,
    year = "2023"
}

@misc{Evans:2021gyd,
    author = "Evans, Matthew and others",
    title = "{A Horizon Study for Cosmic Explorer: Science, Observatories, and Community}",
    eprint = "2109.09882",
    archivePrefix = "arXiv",
    primaryClass = "astro-ph.IM",
    reportNumber = "CE-P2100003-v7, Cosmic Explorer technical report CE-P2100003-v6",
    month = "9",
    year = "2021"
}

@misc{LISA:2024hlh,
    author = "Colpi, Monica and others",
    collaboration = "LISA",
    title = "{LISA Definition Study Report}",
    eprint = "2402.07571",
    archivePrefix = "arXiv",
    primaryClass = "astro-ph.CO",
    month = "2",
    year = "2024"
}

@article{Varma:2019csw,
    author = "Varma, Vijay and Field, Scott E. and Scheel, Mark A. and Blackman, Jonathan and Gerosa, Davide and Stein, Leo C. and Kidder, Lawrence E. and Pfeiffer, Harald P.",
    title = "{Surrogate models for precessing binary black hole simulations with unequal masses}",
    eprint = "1905.09300",
    archivePrefix = "arXiv",
    primaryClass = "gr-qc",
    doi = "10.1103/PhysRevResearch.1.033015",
    journal = "Phys. Rev. Research.",
    volume = "1",
    pages = "033015",
    year = "2019"
}

@article{Khan:2020fso,
    author = "Khan, Sebastian and Green, Rhys",
    title = "{Gravitational-wave surrogate models powered by artificial neural networks}",
    eprint = "2008.12932",
    archivePrefix = "arXiv",
    primaryClass = "gr-qc",
    doi = "10.1103/PhysRevD.103.064015",
    journal = "Phys. Rev. D",
    volume = "103",
    number = "6",
    pages = "064015",
    year = "2021"
}

@article{Fragkouli:2022lpt,
    author = "Fragkouli, Styliani-Christina and Nousi, Paraskevi and Passalis, Nikolaos and Iosif, Panagiotis and Stergioulas, Nikolaos and Tefas, Anastasios",
    title = "{Deep residual error and bag-of-tricks learning for gravitational wave surrogate modeling}",
    eprint = "2203.08434",
    archivePrefix = "arXiv",
    primaryClass = "astro-ph.IM",
    doi = "10.1016/j.asoc.2023.110746",
    journal = "Appl. Soft Comput.",
    volume = "147",
    pages = "110746",
    year = "2023"
}

@article{GramaxoFreitas:2024bpk,
    author = "Gramaxo Freitas, Osvaldo and Theodoropoulos, Anastasios and Villanueva, Nino and Fernandes, Tiago and Nunes, Solange and Font, Jos{\'e} A. and Onofre, Antonio and Torres-Forn{\'e}, Alejandro and Martin-Guerrero, Jos{\'e} D.",
    title = "{Deep learning powered numerical relativity surrogate for binary black hole waveforms}",
    eprint = "2412.06946",
    archivePrefix = "arXiv",
    primaryClass = "gr-qc",
    doi = "10.1103/7bkx-hs53",
    journal = "Phys. Rev. D",
    volume = "112",
    number = "4",
    pages = "043026",
    year = "2025"
}

@article{Thomas:2022rmc,
    author = "Thomas, Lucy M. and Pratten, Geraint and Schmidt, Patricia",
    title = "{Accelerating multimodal gravitational waveforms from precessing compact binaries with artificial neural networks}",
    eprint = "2205.14066",
    archivePrefix = "arXiv",
    primaryClass = "gr-qc",
    reportNumber = "LIGO{\_}DCC P2200161",
    doi = "10.1103/PhysRevD.106.104029",
    journal = "Phys. Rev. D",
    volume = "106",
    number = "10",
    pages = "104029",
    year = "2022"
}

@article{Chua:2018woh,
    author = "Chua, Alvin J. K. and Galley, Chad R. and Vallisneri, Michele",
    title = "{Reduced-order modeling with artificial neurons for gravitational-wave inference}",
    eprint = "1811.05491",
    archivePrefix = "arXiv",
    primaryClass = "astro-ph.IM",
    doi = "10.1103/PhysRevLett.122.211101",
    journal = "Phys. Rev. Lett.",
    volume = "122",
    number = "21",
    pages = "211101",
    year = "2019"
}

@article{Ossokine:2020kjp,
    author = "Ossokine, Serguei and others",
    title = "{Multipolar Effective-One-Body Waveforms for Precessing Binary Black Holes: Construction and Validation}",
    eprint = "2004.09442",
    archivePrefix = "arXiv",
    primaryClass = "gr-qc",
    doi = "10.1103/PhysRevD.102.044055",
    journal = "Phys. Rev. D",
    volume = "102",
    number = "4",
    pages = "044055",
    year = "2020"
}

@article{Williams:2019vub,
    author = "Williams, Daniel and Heng, Ik Siong and Gair, Jonathan and Clark, James A. and Khamesra, Bhavesh",
    title = "{Precessing numerical relativity waveform surrogate model for binary black holes: A Gaussian process regression approach}",
    eprint = "1903.09204",
    archivePrefix = "arXiv",
    primaryClass = "gr-qc",
    reportNumber = "DCC:ligo-p1800128",
    doi = "10.1103/PhysRevD.101.063011",
    journal = "Phys. Rev. D",
    volume = "101",
    number = "6",
    pages = "063011",
    year = "2020"
}

@article{Varma:2018aht,
    author = "Varma, Vijay and Gerosa, Davide and Stein, Leo C. and H{\'e}bert, Fran{\c{c}}ois and Zhang, Hao",
    title = "{High-accuracy mass, spin, and recoil predictions of generic black-hole merger remnants}",
    eprint = "1809.09125",
    archivePrefix = "arXiv",
    primaryClass = "gr-qc",
    doi = "10.1103/PhysRevLett.122.011101",
    journal = "Phys. Rev. Lett.",
    volume = "122",
    number = "1",
    pages = "011101",
    year = "2019"
}

@article{Islam:2021mha,
    author = "Islam, Tousif and Varma, Vijay and Lodman, Jackie and Field, Scott E. and Khanna, Gaurav and Scheel, Mark A. and Pfeiffer, Harald P. and Gerosa, Davide and Kidder, Lawrence E.",
    title = "{Eccentric binary black hole surrogate models for the gravitational waveform and remnant properties: comparable mass, nonspinning case}",
    eprint = "2101.11798",
    archivePrefix = "arXiv",
    primaryClass = "gr-qc",
    doi = "10.1103/PhysRevD.103.064022",
    journal = "Phys. Rev. D",
    volume = "103",
    number = "6",
    pages = "064022",
    year = "2021"
}

@article{Gadre:2022sed,
    author = {Gadre, Bhooshan and P{\"u}rrer, Michael and Field, Scott E. and Ossokine, Serguei and Varma, Vijay},
    title = "{Fully precessing higher-mode surrogate model of effective-one-body waveforms}",
    eprint = "2203.00381",
    archivePrefix = "arXiv",
    primaryClass = "gr-qc",
    reportNumber = "LIGO-P2200040",
    doi = "10.1103/PhysRevD.110.124038",
    journal = "Phys. Rev. D",
    volume = "110",
    number = "12",
    pages = "124038",
    year = "2024"
}

@article{Harry:2016ijz,
    author = "Harry, Ian and Privitera, Stephen and Boh{\'e}, Alejandro and Buonanno, Alessandra",
    title = "{Searching for Gravitational Waves from Compact Binaries with Precessing Spins}",
    eprint = "1603.02444",
    archivePrefix = "arXiv",
    primaryClass = "gr-qc",
    doi = "10.1103/PhysRevD.94.024012",
    journal = "Phys. Rev. D",
    volume = "94",
    number = "2",
    pages = "024012",
    year = "2016"
}

@article{Thompson:2025hhc,
    author = "Thompson, Jonathan E. and Hoy, Charlie and Fauchon-Jones, Edward and Hannam, Mark",
    title = "{Use and interpretation of signal-model indistinguishability measures for gravitational-wave astronomy}",
    eprint = "2506.10530",
    archivePrefix = "arXiv",
    primaryClass = "gr-qc",
    reportNumber = "LIGO-P2500361",
    doi = "10.1103/ddz7-x9zz",
    journal = "Phys. Rev. D",
    volume = "112",
    number = "6",
    pages = "064011",
    year = "2025"
}

@article{Lindblom:2008cm,
    author = "Lindblom, Lee and Owen, Benjamin J. and Brown, Duncan A.",
    title = "{Model Waveform Accuracy Standards for Gravitational Wave Data Analysis}",
    eprint = "0809.3844",
    archivePrefix = "arXiv",
    primaryClass = "gr-qc",
    doi = "10.1103/PhysRevD.78.124020",
    journal = "Phys. Rev. D",
    volume = "78",
    pages = "124020",
    year = "2008"
}

@article{McWilliams:2010eq,
    author = "McWilliams, Sean T. and Kelly, Bernard J. and Baker, John G.",
    title = "{Observing mergers of non-spinning black-hole binaries}",
    eprint = "1004.0961",
    archivePrefix = "arXiv",
    primaryClass = "gr-qc",
    doi = "10.1103/PhysRevD.82.024014",
    journal = "Phys. Rev. D",
    volume = "82",
    pages = "024014",
    year = "2010"
}

@article{Hannam:2010ky,
    author = "Hannam, Mark and Husa, Sascha and Ohme, Frank and Ajith, P.",
    title = "{Length requirements for numerical-relativity waveforms}",
    eprint = "1008.2961",
    archivePrefix = "arXiv",
    primaryClass = "gr-qc",
    doi = "10.1103/PhysRevD.82.124052",
    journal = "Phys. Rev. D",
    volume = "82",
    pages = "124052",
    year = "2010"
}

@article{Baird:2012cu,
    author = "Baird, Emily and Fairhurst, Stephen and Hannam, Mark and Murphy, Patricia",
    title = "{Degeneracy between mass and spin in black-hole-binary waveforms}",
    eprint = "1211.0546",
    archivePrefix = "arXiv",
    primaryClass = "gr-qc",
    doi = "10.1103/PhysRevD.87.024035",
    journal = "Phys. Rev. D",
    volume = "87",
    number = "2",
    pages = "024035",
    year = "2013"
}

@article{Chatziioannou:2017tdw,
    author = "Chatziioannou, Katerina and Klein, Antoine and Yunes, Nicol{\'a}s and Cornish, Neil",
    title = "{Constructing Gravitational Waves from Generic Spin-Precessing Compact Binary Inspirals}",
    eprint = "1703.03967",
    archivePrefix = "arXiv",
    primaryClass = "gr-qc",
    doi = "10.1103/PhysRevD.95.104004",
    journal = "Phys. Rev. D",
    volume = "95",
    number = "10",
    pages = "104004",
    year = "2017"
}

@misc{Toubiana:2024car,
    author = "Toubiana, Alexandre and Gair, Jonathan R.",
    title = "{Indistinguishability criterion and estimating the presence of biases}",
    eprint = "2401.06845",
    archivePrefix = "arXiv",
    primaryClass = "gr-qc",
    month = "1",
    year = "2024"
}

@article{Purrer:2019jcp,
    author = {P{\"u}rrer, Michael and Haster, Carl-Johan},
    title = "{Gravitational waveform accuracy requirements for future ground-based detectors}",
    eprint = "1912.10055",
    archivePrefix = "arXiv",
    primaryClass = "gr-qc",
    doi = "10.1103/PhysRevResearch.2.023151",
    journal = "Phys. Rev. Res.",
    volume = "2",
    number = "2",
    pages = "023151",
    year = "2020"
}

@article{Owen:1998dk,
    author = "Owen, Benjamin J. and Sathyaprakash, B. S.",
    title = "{Matched filtering of gravitational waves from inspiraling compact binaries: Computational cost and template placement}",
    eprint = "gr-qc/9808076",
    archivePrefix = "arXiv",
    reportNumber = "GRP-505",
    doi = "10.1103/PhysRevD.60.022002",
    journal = "Phys. Rev. D",
    volume = "60",
    pages = "022002",
    year = "1999"
}

@article{KAGRA:2013rdx,
    author = "Abbott, B. P. and others",
    collaboration = "KAGRA, LIGO Scientific, Virgo",
    title = "{Prospects for observing and localizing gravitational-wave transients with Advanced LIGO, Advanced Virgo and KAGRA}",
    eprint = "1304.0670",
    archivePrefix = "arXiv",
    primaryClass = "gr-qc",
    reportNumber = "LIGO-P1200087, VIR-0288A-12, JGW-P1808427",
    doi = "10.1007/s41114-020-00026-9",
    journal = "Living Rev. Rel.",
    volume = "19",
    pages = "1",
    year = "2016"
}

@article{LIGOScientific:2016wof,
    author = "Abbott, Benjamin P and others",
    collaboration = "LIGO Scientific",
    title = "{Exploring the Sensitivity of Next Generation Gravitational Wave Detectors}",
    eprint = "1607.08697",
    archivePrefix = "arXiv",
    primaryClass = "astro-ph.IM",
    reportNumber = "LIGO-P1600143",
    doi = "10.1088/1361-6382/aa51f4",
    journal = "Class. Quant. Grav.",
    volume = "34",
    number = "4",
    pages = "044001",
    year = "2017"
}

@article{Vecchio:2003tn,
    author = "Vecchio, Alberto",
    title = "{LISA observations of rapidly spinning massive black hole binary systems}",
    eprint = "astro-ph/0304051",
    archivePrefix = "arXiv",
    doi = "10.1103/PhysRevD.70.042001",
    journal = "Phys. Rev. D",
    volume = "70",
    pages = "042001",
    year = "2004"
}

@article{Chatziioannou:2014coa,
    author = "Chatziioannou, Katerina and Cornish, Neil and Klein, Antoine and Yunes, Nicolas",
    title = "{Spin-Precession: Breaking the Black Hole--Neutron Star Degeneracy}",
    eprint = "1402.3581",
    archivePrefix = "arXiv",
    primaryClass = "gr-qc",
    doi = "10.1088/2041-8205/798/1/L17",
    journal = "Astrophys. J. Lett.",
    volume = "798",
    number = "1",
    pages = "L17",
    year = "2015"
}

@article{Pratten:2020igi,
    author = "Pratten, Geraint and Schmidt, Patricia and Buscicchio, Riccardo and Thomas, Lucy M.",
    title = "{Measuring precession in asymmetric compact binaries}",
    eprint = "2006.16153",
    archivePrefix = "arXiv",
    primaryClass = "gr-qc",
    reportNumber = "LIGO-DCC P2000224",
    doi = "10.1103/PhysRevResearch.2.043096",
    journal = "Phys. Rev. Res.",
    volume = "2",
    number = "4",
    pages = "043096",
    year = "2020"
}

@article{Morras:2025nlp,
    author = "Morras, Gonzalo and Pratten, Geraint and Schmidt, Patricia",
    title = "{Improved post-Newtonian waveform model for inspiralling precessing-eccentric compact binaries}",
    eprint = "2502.03929",
    archivePrefix = "arXiv",
    primaryClass = "gr-qc",
    reportNumber = "IFT-UAM/CSIC-25-12",
    doi = "10.1103/PhysRevD.111.084052",
    journal = "Phys. Rev. D",
    volume = "111",
    number = "8",
    pages = "084052",
    year = "2025"
}

@article{Morras:2025xfu,
    author = "Morras, Gonzalo and Pratten, Geraint and Schmidt, Patricia",
    title = "{Orbital eccentricity in a neutron star - black hole binary merger}",
    eprint = "2503.15393",
    archivePrefix = "arXiv",
    primaryClass = "astro-ph.HE",
    reportNumber = "LIGO-DCC P2500105",
    doi = "10.3847/2041-8213/ae474c",
    journal = "Astrophys. J. Lett.",
    volume = "1000",
    number = "1",
    pages = "L2",
    year = "2026"
}

@article{Romero-Shaw:2022fbf,
    author = "Romero-Shaw, Isobel M. and Gerosa, Davide and Loutrel, Nicholas",
    title = "{Eccentricity or spin precession? Distinguishing subdominant effects in gravitational-wave data}",
    eprint = "2211.07528",
    archivePrefix = "arXiv",
    primaryClass = "astro-ph.HE",
    doi = "10.1093/mnras/stad031",
    journal = "Mon. Not. Roy. Astron. Soc.",
    volume = "519",
    number = "4",
    pages = "5352--5357",
    year = "2023"
}

@article{Madau:2014bja,
    author = "Madau, Piero and Dickinson, Mark",
    title = "{Cosmic Star Formation History}",
    eprint = "1403.0007",
    archivePrefix = "arXiv",
    primaryClass = "astro-ph.CO",
    doi = "10.1146/annurev-astro-081811-125615",
    journal = "Ann. Rev. Astron. Astrophys.",
    volume = "52",
    pages = "415--486",
    year = "2014"
}

@book{Trefethen:1997,
  author    = {Trefethen, Lloyd N. and Bau, David, III},
  title     = {Numerical Linear Algebra},
  series    = {Other Titles in Applied Mathematics},
  volume    = {50},
  edition   = {Illustrated},
  publisher = {SIAM},
  year      = {1997},
  isbn      = {9780898713619},
  pages     = {373}
}

@article{Thorne:1980ru,
    author = "Thorne, K. S.",
    title = "{Multipole Expansions of Gravitational Radiation}",
    doi = "10.1103/RevModPhys.52.299",
    journal = "Rev. Mod. Phys.",
    volume = "52",
    pages = "299--339",
    year = "1980"
}

@article{Moore:2015sza,
    author = "Moore, Christopher J. and Berry, Christopher P. L. and Chua, Alvin J. K. and Gair, Jonathan R.",
    title = "{Improving gravitational-wave parameter estimation using Gaussian process regression}",
    eprint = "1509.04066",
    archivePrefix = "arXiv",
    primaryClass = "gr-qc",
    reportNumber = "LIGO-P1500162",
    doi = "10.1103/PhysRevD.93.064001",
    journal = "Phys. Rev. D",
    volume = "93",
    number = "6",
    pages = "064001",
    year = "2016"
}

@article{Nee:2025nmh,
    author = "Nee, Peter James and others",
    title = "{Eccentric binary black holes: A new framework for numerical relativity waveform surrogates}",
    eprint = "2510.00106",
    archivePrefix = "arXiv",
    primaryClass = "gr-qc",
    month = "9",
    year = "2025",
    journal = ""
}

@article{Maurya:2025shc,
    author = "Maurya, Akash and Kumar, Prayush and Field, Scott E. and Mishra, Chandra Kant and Nee, Peter James and Paul, Kaushik and Pfeiffer, Harald P. and Ravichandran, Adhrit and Varma, Vijay",
    title = "{Chase Orbits, not Time: A Scalable Paradigm for Long-Duration Eccentric Gravitational-Wave Surrogates}",
    eprint = "2510.00116",
    archivePrefix = "arXiv",
    primaryClass = "gr-qc",
    month = "9",
    year = "2025",
    journal = ""
}

@article{Theodoropoulos:2026wkj,
    author = "Theodoropoulos, Anastasios and Villanueva, Nino and Gramaxo Freitas, Osvaldo and Fernandes, Tiago and Nunes, Solange and Torres-Forne, Alejandro and Font, Jose A. and Onofre, Antonio and Martin-Guerrero, Jose D.",
    title = "{An autoencoder-based surrogate waveform model for quasi-circular binary-black-hole mergers}",
    eprint = "2602.00203",
    archivePrefix = "arXiv",
    primaryClass = "astro-ph.IM",
    month = "1",
    year = "2026",
    journal = ""
}

@article{Tissino:2022thn,
    author = "Tissino, Jacopo and Carullo, Gregorio and Breschi, Matteo and Gamba, Rossella and Schmidt, Stefano and Bernuzzi, Sebastiano",
    title = "{Combining effective-one-body accuracy and reduced-order-quadrature speed for binary neutron star merger parameter estimation with machine learning}",
    eprint = "2210.15684",
    archivePrefix = "arXiv",
    primaryClass = "gr-qc",
    doi = "10.1103/PhysRevD.107.084037",
    journal = "Phys. Rev. D",
    volume = "107",
    number = "8",
    pages = "084037",
    year = "2023"
}

\end{document}